\documentclass{article}%
\usepackage{amssymb}
\usepackage{amsmath}
\usepackage{amsfonts}%
\usepackage{graphicx}

\begin{document}
Causal Models for Estimating the Effects of Weight Gain on Mortality

James Robins, MD

\bigskip

Correspondence: James Robins, MD, Mitchell L. and Robin LaFoley Dong Professor
of Epidemiology, Departments of Epidemiology and Biostatistics, Harvard School
of Public Health, Kresge Building, Room 821, 677 Huntington Avenue, Boston, MA
02115; Tel: 617-432-0207; Fax;617-432-1884; robins@hsph.harvard.edu

Running Head: Causal Models

\bigskip

\textbf{Abstract: }Suppose, contrary to fact, in 1950, we had put the cohort
of 18 year old non-smoking American men on a stringent mandatory diet that
guaranteed that no one would ever weigh more than their baseline weight
established at age 18. How would the counter-factual mortality of these 18
year olds have compared to their actual observed mortality through 2007? We
describe in detail how this counterfactual contrast could be estimated from
longitudinal epidemiologic data similiar to that stored in the electronic
medical records of a large HMO by applying g-estimation to a novel structural
nested model. Our analytic approach differs from any alternative approach in
that in that, in the abscence of model misspecification, it can successfully
adjust for (i) measured time-varying confounders such as exercise,
hypertension and diabetes that are simultaneously intermediate variables on
the causal pathway from weight gain to death and determinants of future weight
gain, (ii) unmeasured confounding by undiagnosed preclinical disease (i.e
reverse causation) that can cause both poor weight gain and premature
mortality [provided an upper bound can be specified for the maximum length of
time a subject may suffer from a subclinical illness severe enough to affect
his weight without the illness becomes clinically manifest], and (iii) the
prescence of particular identifiable subgroups, such as those suffering from
serious renal, liver, pulmonary, and/or cardiac disease, in whom confounding
by unmeasured prognostic factors so severe as to render useless any attempt at
direct analytic adjustment. However (ii) and (iii) limit the ability to
empirically test whether the structural nested model is misspecified. The
other two g-methods - parametric g-computation algorithm and inverse
probability of treatment weighted (IPTW) estimation of maginal structural
models (MSMs) can adjust for potential bias due to (i) but not due to (ii) or (iii).


\section{Introduction}

Suppose, contrary to fact, in 1950, we had put the cohort of 18 year old
non-smoking American men on a stringent mandatory diet that guaranteed that no
one would ever weigh more than their baseline weight established at age 18.
Specifically, each subject was weighed every day starting on the day before
his 18th birthday. Whenever his weight was greater than or equal to this
baseline weight, the subject's caloric intake was restricted, without changing
his usual mix of calorie sources and micronutrients, until the time (usually
within 1-3 days) that the subject fell to below baseline weight. (I restrict
to men solely to avoid the complicating issue of how much weight gain to allow
during pregnancy.)\ Thus, ignoring errors of a pound or two, no subject would
ever weigh more than his baseline weight. No instructions or restrictions were
given concerning exercise at any time or the amount or nature of what the
subject ate during non-calorie restricted periods. How would the
counter-factual mortality of these 18 year olds have compared to the actual
observed mortality through 2007.

Factually, a substantial fraction of 18 year old American male gains more than
30 lbs from age 18 to 74. Thus if the counterfactual mortality were much less
than the observed mortality, then, it would make sense for individuals to
maintain their baseline body weight by restricting caloric intake (regardless
of whether or not a practical, non-mandatory public health intervention exists
that would successfully maintain the baseline weight of most of the
(non-smoking) US population.) Here and throught we use the phrase " maintain
their age $x$ bodyweight" to mean that after age $x$ a subject's weight never
exceeds his weight at age $x,$ although it may drop below that weight.

The difference between the counterfactual mortality were no one to exceed
their age 18 body weight and the actual observed mortality of the non-smoking
US population has been discussed by Willett et al (1) as a useful way to
conceptualize the effect of weight on mortality. A major goal of this paper is
to show that g-estimation of a novel structural nested model (SNM) can be used
to directly estimate this difference from longitudinal observational data. A
SNM is a model that takes as input a subject's observed outcome in their
observed exposure (here, weight) history, and an unknown parameter and outputs
the response that would have been observed if, contrary to fact, the subject
to follow the stringent mandatory diet described above. The unknown parameter
vector of a SNM is estimated via the g-estimation procedure introduced in
Robins et al (12). Previous analytic approaches to the estimation of the
effect of weight on mortality do not provide a direct estimate of this
difference. In addition, previous approaches have suffered from one or more of
the following sources of bias : (i) failure to adequately control for measured
confounding due to time-varying exercise, blood lipids, blood pressure,
diabetes, and other chronic diseases (once diagnosed) because of concerns that
one will thereby be controlling for intermediate variables on the causal
pathway from overweight to death, (ii) failure to adequately control for
unmeasured confounding due to undiagnosed chronic disease such as cancer
(i.e., reverse causation) and (iii) failure to update the weight of a subject
whose weight changes after start of follow-up, because of concerns about
reverse causation and measurement error.

Bias due to confounding by measured time-varying confounders that are also
intermediate variables can be controlled by the use so-called of g-methods. G
- methods are statistical methods specifically designed to control bias
attributable to time-varying confounders affected by previous exposure. In
addition to g-estimation of structural nested models, g-methods include the
parametric g-formula estimator and inverse probability of treatment weighted
(IPTW) estimators (7,15). As yet g-methods have not been used to estimate the
effect of overweight on obesity with the exception of Ref. (2), where the
parametric g-formula estimator was used. In this paper we concentrate on
g-estimation of SNMs, because as discussed below, of the three g-methods, only
g-estimation of SNMs can adjust for unmeasured confounding due to undiagnosed
chronic disease\textbf{.}

Finally g-estimation of SNMs allows one to update the weight of a subject
whose weight changes after start of follow-up without introducing any bias due
to reverse causation. However issues of measurement error are more tricky and
will be discussed in the final section of the paper.

Even if maintenance of age 18 weight improves mortality, perhaps a mandatory
intervention that allowed weight gain of 0.3/12 pounds per month (i.e., 3
pounds per decade) would produce an even lower mortality. Perhaps the
mandatory intervention that would produce the lowest mortality ( i.e., the
optimal intervention among all "weight-gain" interventions) is one that allows
a weight gain of 0.3/12 pounds pounds per month in subjects free of
hypertension, diabetes, hyperlipidemia, or clinical CHD, but of only 0.1/12
pounds per two month (i.e., 1 pound per decade) once a subject developed one
of these risk factors.

To decide which mandatory intervention is optimal, we require a well-defined
numerical measure of overall mortality that can be used to rank interventions.
For example, one might use the total years of life (or quality adjusted life)
experienced by the cohort from 1950-2007 as a measure. Use of this measure is
mathematically equivalent to the use of "years (or quality-adjusted years) of
life lived from 1950-2007" as the (subject -specific) utility function in a
decision problem whose goal is to maximize expected utility. "Years (or
quality-adjusted years) of life lived" measures have a much more natural and
useful public health and policy interpretation than the rate ratio,
attributable fraction, and attributable risk measures routinely reported in
epidemiologic studies.

However even "years of life lived from 1950-2007" is an inadequate utility
function when follow-up of the cohort is not to extinction. This function
inappropriately assigns the same utility not only to all subjects alive at age
74 on Jan 1, 2008 regardless of their state of health, but also to a subject
who dies on Dec. 31, 2007 at 11:59 pm. Clearly among survivors in 2008, the
healthier ones (according to some agreed on standard measure of current health
) have a greater post-study expected survival (and thus warrant a higher
utility) than the less healthy survivors and a much greater expected survival
(and thus warrant a much higher utility) than the non-survivors who died in
late December 2007. We will not discuss further precisely how to decide on an
appropriate utility measure for the survivors, except to remark that such a
discussion is necessary. Rather, we will simply assume that, at the end of
follow-up, each cohort member has been given a utility measure $Y.\ $

Note that the benefit of any of the above counterfactual interventions is an
overall effect of the intervention. For example it is conceivable that the
mortality benefit of the intervention that maintained baseline weight was
wholly due to changes in exercise. Perhaps maintenance of baseline weight
makes individuals feel so much better that they exercise more.

In section 2, we assume we have observational retrospective follow-up data
through 2007 on a random sample of the cohort of US males who were non-smokers
and 18 in 1950. The data includes detailed medical records, analogous to those
currently available on subscribers to a comprehensive HMO. In Sec. 2.2, I
discuss three major sources of potential bias that complicate any attempt to
estimate the overall effect of the mandatory intervention "maintain baseline
weight" on the expected utility of our cohort: (i) measured time-varying
confounders such as exercise, hypertension and diabetes that are potentially
intermediate variables, (ii) unmeasured confounding by undiagnosed preclinical
disease (i.e reverse causation) that can cause both poor weight gain and
premature mortality, and (iii) the prescence of particular identifiable
subgroups, such as those suffering from serious renal, liver, pulmonary,
and/or cardiac disease, in whom confounding by unmeasured prognostic factors
is so severe as to render useless direct analytic adjustment for confounding.
In Section 3, I describe how g-estimation of a correctly specified SNM can
appropriately adjust for these potential sources of bias, [provided an upper
bound can be specified for the maximum length of time a subject may suffer
from a subclinical illness severe enough to affect his weight before the
illness becomes clinically manifest]. The SNM required for this adjustment is
novel in two ways. First it is a joint SNM, combining a structural nested
failure time model (SNFTM)\ for the counterfactual time to the earlier of
death or the diagnosis of a chronic illness and a conditional structural
nested mean model (SNMM) for the counterfactual mean of a subject's
counterfactual utility given his counterfactual time to death or a diagnosed
chronic illness. Second our SNM only models the causal effect of an any
increase in BMI between month $m$ and $m+1$ over a subject's maximum previous
BMI$.$ In particular, it does not model and thus is agnostic about the causal
effect a) of any decrease in BMI or b) of any increase in BMI between $m$ and
$m+1$ that fails to attain the previous maximum$.$ As a consequence, our SNM
is more robust than standard SNMs that also model a) and b), because our model
makes fewer asumptions than such alternative models, and thus is less likely
to be mispecified. However, the small number of assumptions made by our SNM
are sufficient to consistently estimate our parameter of interest $E\left[
Y_{0}\right]  .$

In Sections 3.2.4-3.2.5, however, I show that (ii) and (iii) limit the ability
to empirically test whether the joint structural nested model is misspecified.
I also show that, somewhat remarkably, to adjust for bias due to reverse
causation one need not assume a deterministic rank-preserving SNM.\ This is
important since a deterministing rank-preserving SNM assumes that the effect
of weight-gain on mortality is the same for different subjects, an assumption
that is clearly biologically implausible. In Section 4, I consider how to
account for censoring by administative end of follow-up. In Section 5, I
consider the estimation of the expected utility under alternative dietary
interventions. In Section 6, I discuss the consequences of measurement error
in BMI. Proofs and statements of several new theorems are collected in
Appendices 1and 2. Finally, estimation of the optimal
"weight-gain"intervention is discussed in the Appendix 3.

\section{Estimation of an overall effect}

\subsection{The Data}

I first describe the observational data that is supposed to be available.
First, I suppose that a subject's BMI is recorded at the end of each month $t
$, $t=0,1,...,K$, where $time$ t is in months since age 18 and $K+1=\left(
2007-1950\right)  \times12$ is the duration of follow-up. Let $A^{\ast}(t)$ be
the difference between BMI at the end of month $t$ and at the end of month
$t-1.$ Let $L\left(  t\right)  $ be the vector of covariates values recorded
in month $t$ and suppose $L\left(  t\right)  $ precedes $A^{\ast}\left(
t\right)  $ temporally. $L(t)$ includes blood pressure, HDL and LDL measures
of cholesterol, any diagnoses of and clinical and laboratory characteristics
of any chronic disease such as cancer, CAD, diabetes, asthma, COPD, liver,
renal disease, etc., level of exercise, measures of mobility and disability,
etc. The vector $L\left(  t\right)  $ also includes $BMI\left(  t\right)  ,$
the BMI just before $t$ rounded to the nearest pound. Thus
\begin{equation}
A^{\ast}\left(  t\right)  =BMI\left(  t+1\right)  -BMI\left(  t\right)  .
\end{equation}
$L\left(  t\right)  $ also includes the indicator $I\left(  T>t\right)  $ of
vital status at the beginning of month $t$ with $T$ the death time of a
subject and, for any proposition $B,$ $I\left(  B\right)  $ is the indicator
function that take the value $1$ if $B$ is true and zero otherwise Thus
$I\left(  T>t\right)  =1$ if a subject is alive at $t$ and zero if dead at
$t.$ If $I\left(  T>t\right)  =0,$ I include in $L\left(  t\right)  $ the
exact day of death and, by convention, assign the value zero to all other
components of $L\left(  t\right)  $.

By convention, set $A^{\ast}\left(  t\right)  $ and the remaining components
of $L\left(  t\right)  $ to zero once a subject has died.

The baseline covariates $L\left(  0\right)  $ include covariate and BMI data
on a subject before follow up starts at age $18.$ Specifically, let
$BMI\left(  0\right)  $ denote BMI at (just before) age $18$ (i.e., time 0)$.$
Our inclusion of BMI just before age $18$ as a covariate rather than a
treatment reflects the fact that "change" in BMI since 18 is our exposure. In
particular, note that $A^{\ast}\left(  0\right)  $ is the difference between
$BMI$ recorded just before $18$ yrs and 1 month and BMI recorded just before
18 years. As is standard in the literature, I have taken change in BMI rather
than in change in weight in pounds as the exposure variable. Let
$\overline{A^{\ast}}\left(  t\right)  $ and $\overline{L}\left(  t\right)  $
be change in BMI and covariate history through time $t$ and $\overline
{A^{\ast}}$=$\overline{A^{\ast}}\left(  K\right)  $ be a subjects (change in)
BMI history through month $K$ and $\overline{L}$=$\overline{L}\left(
K+1\right)  $ be $L$ history through the end of the study. A subject's utility
$Y,$ a measure of quality-adjusted survival, is calculated from $\overline{L}%
$=$\overline{L}\left(  K+1\right)  $ since $\overline{L}$ includes the
survival time of nonsurvivors, health status measures for survivors at end of
follow-up, and time-varying health status factors.

\subsection{Potential for Measured and Unmeasured Confounding:}

\subsubsection{\textbf{Reverse Causation and Unmeasured Confounding by
Subclinical Disease:}}

In the literature on the effect of BMI on mortality, a controversy has arisen
as whether and how to modify standard analytic methods to account for "reverse
causation". Reverse causation refers to the well-accepted fact that
preclinical (i.e., undiagnosed) chronic disease, such as preclinical cancer,
can cause both weight loss (or diminished weight gain) and death. It follows
that among subjects with identical BMI history $\left(  \overline{A^{\ast}%
}\left(  t-1\right)  ,BMI\left(  0\right)  \right)  $ and measured covariate
history $\overline{L}\left(  t\right)  $ before age $t$, the subset whose
monthly change $A^{\ast}\left(  t\right)  $ in BMI is negative \textit{are not
comparable }with regard to mortality risk to the subset with positive
$A^{\ast}\left(  t\right)  ,$ even if $BMI$ has no causal effect on mortality.
That is, reverse causation implies unmeasured confounding by undiagnosed
chronic disease. In fact, by an analogous argument, even among the subset with
$A^{\ast}\left(  t\right)  $ positive, there will be unmeasured confounding,
because those with a small gain in BMI are more likely to have preclinical
disease than those with a substantial gain.

It follows that one requires an analytic method that can adjust for unmeasured
confounding due to the prescence of preclinical disease. I will present a
method that is appropriate under the additional assumption that we are able to
specify an upper bound on the length of time a subject may have a subclinical
illness severe enough to affect his weight, before that illness becomes
clinically manifest.

\subsubsection{\textbf{Measured Confounders that are also Intermediate
Variables}}

I next turn to the issue of confounding by measured factors, i.e., by
components of the covariate vector $L\left(  t\right)  .$ For pedagogic
purposes, in this subsection, it will be simpler to imagine that the
unmeasured confounding due to reverse causation discussed above is not
present. Now it is fairly well accepted that obesity causes increased blood
pressure (BP), increased low density lipoproteins (LDL), diabetes (Db), and
decreased exercise and these four factors may in turn cause increased
mortality. Thus these four variables are intermediate variables on the causal
pathway from BMI to mortality. In order to prevent underestimation of the
overall effect of BMI on mortality due to adjusting for intermediate
variables, many analyses of the effect of BMI on mortality have failed to
adjust for BP, LDL , Db, or exercise in the analysis. However such a decison
can only be justified if these potential intermediate variables do not also
confound the BMI- mortality relationship.

A sufficient conditon for these intermediate variables to also be confounders
is that, among subjects with identical BMI history $\left(  BMI\left(
0\right)  ,\overline{A^{\ast}}\left(  t-1\right)  \right)  $ until $t$, the
subset whose monthly change $A^{\ast}\left(  t\right)  $ in BMI is negative
\textit{are not comparable }with regard to past BP, LDL, Db, and exercise
history to the subset with positive $A^{\ast}\left(  t\right)  .$ Such
non-comparability implies that, if data on time-varying BP, LDL, Db, and
exercise history are not used in the analysis, their will exist a non-causal
association between an increase of $A^{\ast}\left(  t\right)  $ in BMI during
month $t$ and subsequent adverse $mortality,$ even under the null hypothesis
of no overall effect of BMI on mortality. Such non-comparability can occur
whenever some or all of these intermediate variables are either a cause of a
change in BMI or are correlated with an unmeasured cause of a change in BMI

For example, it is likely that lack of exercise causes weight gain. In that
case, if regular exercise causes decreased mortality, then, in an analysis
that fails to adjust for exercise history prior to $t,$ the association
between an increase of $A^{\ast}\left(  t\right)  $ in BMI during month $t$
and subsequent adverse $mortality$ will be an overestimate of the true causal
effect of $A^{\ast}\left(  t\right)  $ on mortality, due to uncontrolled
confounding by exercise.

Similarly, suppose that chronic emotional stress and low grade depression not
only cause weight gain by inducing overeating as a soothing, self-medicating
behavior, but also directly cause elevated BP, elevated LDL, and Db
independently via various stress-induced metabolic, immune, and sympathetic
nervous system effects. If, as is true in most observational data bases, data
on chronic emotional stress and low grade depression are not recorded (i.e.,
measured), then, even under the null hypothesis of no overall effect of BMI on
mortality, the association between an increase $A^{\ast}\left(  t\right)  $ in
BMI and subsequent adverse $mortality$ will tend to be positive, whether or
not one adjusts for elevated BP, elevated LDL, and Db in the analysis, due to
uncontrolled confounding by chronic emotional stress and low grade depression.
However, these variables should be adjusted for in the analysis, because the
magnitude of positive overestimation will often be much less if they are
adjusted for, because of their correlation with the unmeasured causal
confounder - chronic emotional stress and depression.

In contrast with the last paragraph, suppose there is no confounding by
chronic emotional stress and low grade depression; rather, in the
observational data base, most indivivduals who developed an elevated BP,
elevated LDL, or Db became concerned about their health and instituted a diet
that resulted in their gaining less weight than those without these
conditions. Then the association found between an increase $A^{\ast}\left(
t\right)  $ in BMI and subsequent adverse $mortality$ in an analysis that
fails to adjust for these variables at $t$ would tend to underestimate the
true causal effect of $BMI$ on mortality due to negative confounding. I
conclude that elevated BP, elevated LDL, and Db could confound the association
between increase in BMI and subsequent adverse $mortality$ in either a
negative or positive direction, depending on which of the mechanisms described
in this paragraph and the last predominates.

In summary, time-dependent covariates such as exercise (i.e., physical
activity), BP, LDL, or Db that are recorded in $L\left(  t\right)  $ may be
both intermediate variables on the causal pathway from BMI to death and
confounders of the BMI-death relationship. It follows that one requires an
analytic method that can appropriately adjust for the effects of measured
time-varying covariates that are simultaneously intermediate variables and
time-dependent confounders.

\subsubsection{\textbf{Intractable Unmeasured Confounding in Subgroups}}

There may be subgroups defined by measured variables in whom confounding by
unmeasured factors is intractable. For example, among persons with diagnosed
chronic renal, liver, pulmonary or cardiac disease, rapid weight gain can
indicate increasing edema (water retention) due to unmeasured disease
progression rather than increasing fat stores; as a consequence, among chronic
disease patients with identical pasts, comparability would not hold because
individuals experiencing rapid weight gain may be at increased risk of death
due to unmeasured progression of disease compared to those with lesser weight
gain. In such a case unmeasured confounding by disease progression may be intractable.

Using other arguments, various investigators have argued that in both the
subgroup of subjects over age 70 and the subgroup with BMI less than 21,
subjects gaining weight at different rates are not comparable owing to
unmeasured confounding factors, even when data has been collected on many
potential confounders. .

Therefore one needs an analytic method that can remain valid even when there
exists intractable confounding among subjects with a diagnosed chronic
disease, an age of greater than 70, or a BMI below 21. In the next section, I
describe an analytic method that satisfies the requirements of this and the
two previous subsections.

\subsection{A Simplified Description of G-estimation of Structural Nested
Models (SNMs):}

In this subsection I give a nontechnical, conceptual description of how, even
in the prescence of the measured and unmeasured confounding described in
Section 2.2, g-estimation of structural nested models can be used to estimate
the expected utility had, contrary to fact, all non-smoking 18 year old
American men in 1950 been put on a stringent mandatory diet that guaranteed
that no one would ever weigh more than their weight at age 18. In order to
avoid technical digressions and thereby keep the description centered on
important conceptual issues, this nontechnical description is neither complete
nor fully accurate. Section 3 onwards provides a complete and accurate
description. This completeness and accuracy unfortunately place greater
technical demands on the reader.

A locally rank preserving SNM for $Y$ is a rule that takes as input a
subject's observed utility $Y,$ their observed BMI and covariate history
through the end of the study, and an unknown parameter $\beta^{\ast}$ and
outputs the utility $Y_{0\ }$ that would have been observed if, possibly
contrary to fact, the subject had followed the dietary intervention of the
first paragraph of the Introduction. If the rule is correct and we knew the
value of $\beta^{\ast},$ then we could calculate $Y_{0\ }$ for each study
subject. The average of these $Y_{0}$ in the cohort of all non-smoking 18 year
old American men in 1950 is our quantity of interest: the expected (i.e.
average) utility had one implemented a dietary intervention that guaranteed
that no one would ever weigh more than they did at age 18. However we do not
know the value of $\beta^{\ast}.$ Thus the challenge is to estimate
$\beta^{\ast}$ from the data. When, as in section 2.2.2, all confounding is
due to measured variables, Robins (12) proposed a method of estimation called
g-estimation that is described next.

If the only confounding is due to measured factors, then among subjects with
the same BMI and covariate history prior to time t with nonnegative $A\left(
t\right)  $, the increase $A(t)$ in BMI between $t$ and $t+1$ will be
conditionally uncorrelated with $Y_{0}.$ Thus to estimate $\beta^{\ast},$ we
simply try many different guesses $\beta.$ If a particular guess $\beta$ were
the true $\beta^{\ast},$ then the output of the rule would be uncorrelated
with $A(t).$ Thus I choose as our estimate $\widehat{\beta}$ of $\beta^{\ast
},$ the guess $\beta$ which results in an output that has smallest conditional
correlation with $A(t)$ when we combine the information across all months $t$
from 0 to end of follow-up at $K+1.$

When as in Section 2.2.3, there are certain identifiable subgroups in whom
confounding is intractable, bias can result because the output of the rule
will be conditionally correlated with $A\left(  t\right)  $ even when
$\beta=\beta^{\ast}$. To eliminate this bias, it suffices to search for lack
of correlation with $A\left(  t\right)  $ only among the subset of subjects
who are not members of these intractable confounded subgroups at time $t.$That
is, we simply restrict our g-estimation procedure at a given time $t$ to
subjects who are not currently members of these subgroups.

When there is unmeasured confounding by subclinical disease such as in Section
2.2.1, I must modify our g-estimation procedure. Suppose one can specify an
upper bound, say 6 years, on the length of time a subject may have a
subclinical illness severe enough to affect weight gain, before that illness
becomes clinically manifest. Then one can still validly estimate $\beta^{\ast
}$ if one restricts the g-estimation procedure at a given time $t$ to those
subjects who would have remained alive and free of a diagnosed chronic (i.e.,
of clinical) disease for the six years following $t$ had, possibly contrary to
fact, they followed a diet that prevented any further weight gain over those 6
years; by our assumption of a 6-year upper bound, such subjects did not have
their weight gain affected by an undiagnosed chronic disease. [It does not
suffice to restrict to subjects who actually remained alive and free of
clinical disease for the six years following $t,$ because if BMI change
$A\left(  t\right)  $ at $t$ causally effects the onset of clinical disease
and/or survival in the following six years, the variable 'survival without
clinical disease for six years after t' is a response affected by the exposure
$A\left(  t\right)  $ and thus cannot be adjusted for without introducing
selection bias as explained in Hernan et al{\Large .} (13){\Large . }Thus to
validly estimate $\beta^{\ast}$ using g-estimation, one must be able to
determine those "subjects who would have remained alive and free of clinical
disease for the six years following $t$ had, possibly contrary to fact, they
followed a diet that prevented any further weight gain over those 6 years."

One can do so by specfying a second SNM, called a locally rank preserving
structural nested failure time model (SNFTM), for the effect of change in BMI
on the time $X$ to the diagnosis of chronic disease or death (whichever comes
first). A locally rank preserving SNFTM is a rule that takes as input a
subject's observed time $X$ to (the earlier of) death or a diagnosed chronic
disease$,$ their observed BMI and covariate history through the end of the
study, and an unknown parameter $\psi^{\ast},$ and a time $t$ and outputs the
time $X_{t\ }$ that would have been observed if, possibly contrary to fact,
the subject had followed a dietary intervention in which no further weight was
gained after time $t.$ If $\psi^{\ast}$ were known or well-estimated, we could
compute $X_{t\ }$ for each subject, determine which subjects' $X_{t\ }$ failed
to exceed $t\ $by more than 6 years, and exclude such subjects from the
g-estimation procedure used to estimate $\beta^{\ast}.$

Thus it only remain to estimate the parameter $\psi^{\ast}$ of our locally
rank preserving SNFTM in the prescence of unmeasured confounding by
subclinical disease. Now among subjects with the same BMI and covariate
history prior to time t with nonnegative $A\left(  t\right)  $ who are not
members of an identifiable subgroup with intractable confounding, the change
$A(t)$ in BMI between $t$ and $t+1$ will be uncorrelated with $X_{t}$ if we
restrict to subjects with $X_{t}$ exceeding $t\ $by more than 6 years$.$ Thus
to estimate $\psi^{\ast},$ I simply try many different guesses $\psi. $ If a
particular guess $\psi$ were the true $\psi^{\ast},$ the output of the SNFTM
rule would be uncorrelated with $A(t)$ when I restrict the g-estimation
procedure to subjects whose output exceeds $t\ $by more than 6 years$.$ Thus I
choose as the estimate $\widehat{\psi}$ of $\psi^{\ast},$ the guess $\psi$
which results in an output that, under this restricted g-estimation procedure,
has the smallest conditional correlation with $A(t)$ when I combine the
information across all times $t$ from 0 to $end$ of follow-up at $K+1.$

Before proceeding to the more technical part of the paper, I provide a brief
non-technical discussion of several important but subtle points about SNMs.
First, locally rank preserving SNMs assume that the effect of a given increase
in BMI on the utility $Y$ and on $X$ is the same for any two subjects with the
same past measured covariate history. This assumption is biologically
implausible since unmeasured genetic and enviromental factors will clearly
modify the magnitude of the effect of weight gain on the responses $Y$ and
$X$. Fortunately, we prove in Section 3 that our g-estimator of the mean of
$Y_{0}$ remains valid even if we allow the magnitude of the effect of weight
gain on $Y$ and $X$ to be modified in an arbitrary manner by unmeasured
genetic and enviornmental factors.

The description of g-estimation of the parameters $\psi^{\ast}$ of our SNFTM
model for $X$ assumed that the time $X$ to death or diagnosed chronic disease
was available for every study subject. However, by end of follow-up, a number
of study subjects will remain alive and free of chronic disease. Such subjects
are said to be censored. In Section 4, I show how our g-estimation procedures
can be modified to approprately account for these censored observations.

The estimate of the mean of $Y_{0}$ will be biased if either the SNMM for $Y$
or the SNFTM for $X$ are misspecified. I discuss below how to construct tests
for misspecification. However, I also show that the power of such tests to
detect model misspecification can be quite limited in the prescence of reverse
causation by subclinical disease and intractable confounding in identifiable
subgroups. In Section 3.3, I offer some suggestions on how the impact of this
limited power on the quality of one's inferences can be lessened if one is
willing to change the parameter that is being estimated.

\section{\textbf{Estimation of the effect of the "maintain baseline weight
intervention"}}

In this section I describe how we can use G-estimation of structural nested
nodels (SNMs) to estimate the the expected utility had one put all non-smoking
18 year old American men on a stringent mandatory diet that guaranteed that no
one would ever weigh more than their baseline weight established at age 18.
For pedagogic reasons I first consider the simpler setting in which there is
no unmeasured confounding by preclinical disease.

\subsection{Case 1: No unmeasured confounding by preclinical disease.}

\subsubsection{A Locally Rank Preserving SNM.}

An SNM is a model for counterfactual variables $Y_{m}$ that denote a subject's
utility measured at end of follow-up under the following counterfactual
dietary intervention:

\textbf{Time }$m$\textbf{\ Dietary Intervention: }The subject follows his
observed diet up to month $m\ $following his 18th birthday and, from month $m
$ onwards, the subject is weighed every day: (i) whenever his weight is
greater than or equal to his maximum monthly BMI\ up to $m\ $ $[$i.e.,
$BMI_{\max}\left(  m\right)  \equiv\max\left\{  BMI\left(  0\right)
,....,BMI\left(  m\right)  \right\}  ]$, the subject's caloric intake is
restricted until the subject's BMI falls to below $BMI_{\max}\left(  m\right)
$; (ii) whenever his weight is less than $BMI_{\max}\left(  m\right)  ,$ the
subject is allowed to eat as he pleases without any intervention.

A subject's responses had, possibly contrary to fact, he been made to follow a
time $m$\ dietary intervention are referred to as counterfactual responses. We
assume that $Y_{m}$ is well-define in the sense that its value is insensitive
to the unspecified details of exactly how the subject's calories are to be
restricted in (i). We also assume a subject's counterfactual responses are
observed only for those $m$ for which a subject's actual BMI history was
consistent with his having followed the time $m$ dietary intervention. For
other values of $m$, the time $m$-specific counterfactuals remain unobserved.

The time $0$ dietary intervention is the dietary intervention in the first
paragraph of the Introduction. The counterfactual $Y_{0\ }$ is the utility
corresponding to this regime. Thus the expected value $E\left[  Y_{0}\right]
$ of $Y_{0}$ is our parameter of interest: the expected utility had we placed
in 1950 all non-smoking 18 year old American men on a diet that guaranteed
that no one would ever weigh more than they did at age 18.

Note that $Y_{K+1}$=$Y:$ if one were to follow his actual observed diet up to
the time $K+1$ at which the study ends, then no dietary intervention would
have occurred. Hence the counterfactual $Y_{K+1}$ must be the observed (i.e.,
actual) $Y.$

By definition, a subject's observed data through $k$ (but before $k+1)$ is
inconsistent or incompatible with following the "time m dietary intervention"
if and only if $BMI\left(  k+1\right)  >BMI_{\max}\left(  m\right)  $ for some
$k>m.$

Let us define $A\left(  t\right)  $ to be the difference between a subject's
observed BMI, $BMI\left(  t+1\right)  $, just prior to month $t+1$ and his
maximum value $BMI_{\max}\left(  t\right)  $ of BMI prior to month $t$,
whenever that difference is nonnegative. When the difference is negative, we
simply set $A\left(  t\right)  $ to be zero. Formally then
\begin{align}
A\left(  t\right)   & =BMI\left(  t+1\right)  -BMI_{\max}\left(  t\right)
\text{ if }BMI\left(  t+1\right)  \geq BMI_{\max}\left(  t\right) \\
\text{ }A\left(  t\right)   & =0\ if\text{ }BMI\left(  t+1\right)  <BMI_{\max
}\left(  t\right)  .
\end{align}
$A\left(  t\right)  $ is nonnegative. It follows that it is only when the
individual's observed data is incompatible with the "time m dietary
intervention" through time $m$ is $A\left(  m\right)  \neq0.$ If an
individual's observed data is consistent with his having followed the "time m
dietary intervention", it is consistent with his having followed the "time t
dietary intervention" for $t>m.$

Note that $Y_{m+1}-Y_{m}=0$ whenever $A\left(  m\right)  =0$. If $A\left(
m\right)  \neq0,Y_{m+1}-Y_{m}$ is the difference between (i) a subject's
utility when he has his observed $\overline{BMI}\left(  m+1\right)  $ history
and thereafter, possibly contrary to fact, the subject follows the dietary
intervention that guarantees his $BMI\left(  k\right)  $ for $k>m+1$ never
again exceeds $BMI\left(  m+1\right)  $ and (ii) his utility when he has his
observed $\overline{BMI}\left(  m\right)  \ $history and thereafter, possibly
contrary to fact, the subject follows the dietary intervention that guarantees
his $BMI\ $at $m+1$ equals his observed $BMI_{\max}\left(  m\right)  \ \left[
\text{rather than his observed }BMI\ \text{at }m+1\right]  $ and that his
$BMI\left(  k\right)  $ for $k>m+1$ never again exceeds $BMI_{\max}\left(
m\right)  .$ As a kind of shorthand for the previous sentence, whenever
$A\left(  m\right)  \neq0,$ we will refer to $Y_{m+1}-Y_{m} $ as the causal
effect of final blip of exposure of magnitude $A\left(  m\right)  $ on the
subject's utility.

An additive locally rank preserving SNM is a deterministic model for the
magnitude of the effect of a treatment $A(m)$ on $Y_{m+1}-Y_{m}.$
Mathematically an additive locally rank preserving SNM assumes that for each
time $m=0,...,K,$%
\begin{equation}
Y_{m+1}-Y_{m}=\gamma_{m}\left[  A(m),\overline{A}\left(  m-1\right)
,\overline{L}\left(  m\right)  ,\beta^{\ast}\right] \label{generalsnmm}%
\end{equation}
where (i) $\beta^{\ast}$ is the unknown true parameter vector, and (ii)
$\gamma_{m}\left[  A(m),\overline{A}\left(  m-1\right)  ,\overline{L}\left(
m\right)  ,\beta\right]  $ is a known function [such as $\left\{  \beta
_{0}+\beta_{1}m+\beta_{2}^{T}L\left(  m\right)  \right\}  A\left(  m\right)
]$ satisfying the restrictions $\gamma_{m}\left[  A(m),\overline{A}\left(
m-1\right)  ,\overline{L}\left(  m\right)  ,\beta\right]  =0$ if $A\left(
m\right)  =0\ or$ $\beta=0.$ Here $\theta_{2}$ is a column vector of length
equal to that of the vector $L\left(  m\right)  $. Furthermore,$^{{}}$ $T$ as
a superscript denotes the transpose of a matrix or vector. The first
restriction must logically hold because, by definition, if $A\left(  m\right)
=0,$ $Y_{m+1}=Y_{m}.$ We now show that the second restriction guarantees that
$\beta^{\ast}=0$ encodes the sharp null hypothesis that "following a diet that
prevents one's BMI from ever exceeding the baseline BMI" has no effect on any
subject's utility.

Recalling that $Y_{K+1}=Y,$ the model $\left(  \ref{generalsnmm}\right)  $ is
seen to be equivalent to the model
\begin{equation}
Y_{m}=Y-\sum_{m}^{K}\gamma_{m}\left[  A(m),\overline{A}\left(  m-1\right)
,\overline{L}\left(  m\right)  ,\beta^{\ast}\right]
\end{equation}
for m=0,1,...,K. To help understand equation (5) consider first the special
case $m=K$. Then equation (5) says that to calculate $Y_{K}$ from $Y,$ we
remove the causal effect $\gamma_{K}\left[  A(K),\overline{A}\left(
K-1\right)  ,\overline{L}\left(  K\right)  ,\beta^{\ast}\right]  $ of exposure
$A(K)$ at the last time $K.$ Next consider the special case $m=0$. Then
equation (5) says that to calculate $Y_{0},$ one successively removes the
effect of exposure at times $K,K-1,...,0.$ It follows from the restriction
$\gamma_{m}\left[  A(m),\overline{A}\left(  m-1\right)  ,\overline{L}\left(
m\right)  ,0\right]  =0$ that $\beta^{\ast}=0$ implies $\gamma_{m}\left[
A(m),\overline{A}\left(  m-1\right)  ,\overline{L}\left(  m\right)
,\beta^{\ast}\right]  =0$ for each $m.$ Thus $\beta^{\ast}=0$ encodes the
sharp null hypothesis that $Y_{0}=Y_{m}=Y$ for all subjects and all $m.$In
other words, one's utility at the end of the study will be the same regardless
of whether or not one follows any "time $m$ dietary intervention".

Since, by Eq. (5), a locally rank preserving SNM directly maps an individuals
observed utility $Y$ to the utility an individual would have under the "time m
dietary intervention", it is a model for individual causal effects.

Possible choices of $\gamma_{m}\left[  a(m),\overline{a}\left(  m-1\right)
,\overline{l}\left(  m\right)  ,\beta\right]  $ include (i) $\beta a\left(
m\right)  $,

(ii) $\left(  \beta_{0}+\beta_{1}m\right)  a\left(  m\right)  $, (iii)
$\left\{  \beta_{0}+\beta_{1}m+\beta_{2}^{T}l\left(  m\right)  \right\}
a\left(  m\right)  $. In model (i),$\ $the effect of a change of $A(m)$ in BMI
is the same for all $m$. Under model (ii),$\ $the effect varies linearly with
time $m$ . Under model (iii), the causal effect of $A\left(  m\right)  $ is
modified by the most recent covariate history.

In the following we assume the observed data $O$ on each subject is $O=\left(
Y,\overline{L},\overline{A}\right)  \equiv\left(  Y,\overline{L}\left(
K+1\right)  ,\overline{A}\left(  K\right)  \right)  $. That is $O$ consists of
a subject's utility $Y$ and his covariate and treatment histories through the
end of the study. The inclusion of $\overline{A}\left(  K\right)  $ is
actually redundant, since the $A-$history $\overline{A}(K)$ is determined by
$\overline{BMI}\left(  K+1\right)  ,$ and $\overline{BMI}\left(  K+1\right)  $
is a component of $\overline{L}(K+1).$ Thus we could write the observed data
as simply $\left(  Y,\overline{L}\left(  K+1\right)  \right)  .$ However
because we wish to use results on g-estimation of SNMs that were derived in
previous papers in which $\overline{A}(K)$ was not determined by $\overline
{L}(K+1)$, we will continue to write $O=\left(  Y,\overline{L},\overline
{A}\right)  $ and accept some redundancy in the notation. Let%

\begin{equation}
Y_{m}\left(  \beta\right)  =Y-\sum_{j=m}^{K}\gamma_{m}\left[  A(j),\overline
{A}\left(  j-1\right)  ,\overline{L}\left(  j\right)  ,\beta\right]
\end{equation}
so, under our model, $Y_{m}=Y_{m}\left(  \beta^{\ast}\right)  .$ Note that,
for each $\beta$, $Y_{m}\left(  \beta\right)  $ can be computed from the
observed data $\left(  Y,\overline{L},\overline{A}\right)  $. Suppose we had a
consistent estimate $\widehat{\beta}$ of $\beta^{\ast}.$ Then $Y_{0}\left(
\widehat{\beta}\right)  $ would be a consistent estimate of $Y_{0}%
=Y_{0}\left(  \beta^{\ast}\right)  .$ Thus the average $\sum_{i=1}^{n}%
Y_{0i}\left(  \widehat{\beta}\right)  /n$ over the $n$ study subjects would be
a consistent estimate of the parameter of interest $E\left[  Y_{0}\right]  .$
Further $\sum_{i=1}^{n}Y_{0i}\left(  \widehat{\beta}\right)  /n-\sum_{i=1}%
^{n}Y_{i}/n$ would be a consistent estimate of the difference $E\left[
Y_{0}\right]  -E\left[  Y\right]  $ between the expected utility $E\left[
Y_{0}\right]  $ under a dietary intervention guaranteeing BMI never excedes
the baseline BMI and the expected utility $E\left[  Y\right]  $ in the
abscence of any dietary intervention. Below we show how one can obtain a
consistent estimate $\widehat{\beta}$ by g-estimation if a certain
comparability assumption holds.

\textbf{The Innovative Aspect of our SNM: }\ The most important and innovative
aspect of our model is that it models the causal effect on the utility of an
increase in BMI of $A\left(  m\right)  $ over a subject's maximum past BMI,
$BMI_{\max}\left(  m\right)  .$ It does not model and thus is agnostic about
the causal effect a) of any decrease in BMI or b) of any increase in BMI
between $m$ and $m+1$ that fails to result in one's BMI exceeding $BMI_{\max
}\left(  m\right)  .$ Our model (4) is thus more robust than alternative
models that would also model a) or b). However, the small number of
assumptions made by our model are sufficient for our purposes; if we can
consistently estimate the parameter $\beta^{\ast},$ we can consistently
estimate our parameter of interest $E\left[  Y_{0}\right]  .$

Thus it only remains to estimate $\beta^{\ast}.$ In this section will do so
under the following assumption, which will be weakened in later sections.
Define the indicator variable $\Xi\left(  m\right)  $ taking values in the
two-element set $\left\{  0,1\right\}  $ by
\begin{equation}
\Xi\left(  m\right)  =1\Leftrightarrow BMI\left(  m+1\right)  \geq BMI_{\max
}\left(  m\right)  \text{ }%
\end{equation}
That is $\Xi\left(  m\right)  $ takes the value $1$ if a subject's BMI just
before $m+1$ is at least as great as his maximum BMI up to time $m$. Otherwise
$\Xi\left(  m\right)  $ takes the value $0$.

\textbf{Comparability Assumption} (CO): Among subjects with the same
$\overline{A}(m-1)$ history and covariate history $\overline{L}(m)\ ($which
includes BMI history $\overline{BMI}\left(  m\right)  )$ and with $\Xi\left(
m\right)  =1$, $A\left(  m\right)  $ is statistically independent of the
counterfactual $Y_{m}.\ $Formally, conditional on $\left(  \overline
{A}(m-1),\overline{L}(m),\Xi\left(  m\right)  =1\right)  ,$ $A\left(
m\right)  $ is independent of $Y_{m}.$ [Since past $A-$history $\overline
{A}(m-1)$ is determined by $\overline{BMI}\left(  m\right)  $, $\overline
{A}(m-1)$ in the conditioning event $\left(  \overline{A}(m-1),\overline
{L}(m)\right)  $ is redundant; nonetheless we shall retain the $\overline
{A}(m-1).]$

A comparability assumption such as CO is often referred to as an asssumption
of no confounding by unmeasured factors or as an assumption of sequential randomization.

\textbf{Remark}: To understand why we conditioned on $\Xi\left(  m\right)  =1$
in the CO assumption, imagine we had instead assumed that $A\left(  m\right)
$ is independent of $Y_{m}$ conditional on $\left(  \overline{A}%
(m-1),\overline{L}(m)\right)  .$ That would have implied that among the subset
of subjects with a given $\left(  \overline{A}(m-1),\overline{L}(m)\right)  ,$
the subgroup with $A\left(  m\right)  \neq0$ would have the same distribution
of the utility $Y_{m}$ under the time $m-$ dietary intervention as the
subgroup with $A\left(  m\right)  =0.$ But, under the time $m-$ intervention,
all subjects in the $A\left(  m\right)  \neq0$ subgroup would have BMI$\left(
m+1\right)  $ equal to their common $BMI_{\max}\left(  m\right)  \in
\overline{L}(m),$ while many subjects with $A\left(  m\right)  =0$
(specifically, those with $\Xi\left(  m\right)  =0)$ would have BMI$\left(
m+1\right)  <BMI_{\max}\left(  m\right)  .$ Thus, the $A\left(  m\right)  =0$
subgroup will have lower BMI at $m+1$ than the $A\left(  m\right)  \neq0$
subgroup under the time $m-$ intervention. Suppose the null hypothesis of no
biological BMI effect is false. Then, for an individual with $A\left(
m\right)  =0$, their utility $Y_{m}$ should depend on their BMI at $m+1$. As
such, it extremely unlikely that the $A\left(  m\right)  =0$ and $A\left(
m\right)  \neq0$ subgroups would be comparable. In contrast, if, as in
assumption $CO,$ we restrict the $A\left(  m\right)  =0$ subgroup to a subset
of the subjects with $\Xi\left(  m\right)  =1,$ then given $\overline{L}(m),$
this restricted $A\left(  m\right)  =0$ subgroup, like the $A\left(  m\right)
\neq0$ subgroup, will have BMI$\left(  m+1\right)  $ equal to the common
$BMI_{\max}\left(  m\right)  $ under the intervention, so the assumption of
noncomparability is plausible.

It is interesting to note that if we had used the coding convention that
vector $L\left(  k\right)  $ includes $\Xi\left(  k\right)  $ as a component,
we could then have stated our comparability assumption as $A\left(  m\right)
$ is independent of $Y_{m}$ conditional on $\left(  \overline{A}%
(m-1),\overline{L}(m)\right)  $ because, under this coding, $pr\left(
A\left(  k\right)  =0|\bar{A}\left(  k-1\right)  =\bar{0}\left(  k-1\right)
,\bar{L}\left(  k\right)  \right)  $ is one whenever $\Xi\left(  k\right)  $
takes the value zero. However, we will not use this coding convention.

Under the CO assumption, we can obtain a consistent estimator of $\beta^{\ast
}$ by g-estimation as follows. We specify a linear regression model

$E\left[  A\left(  m\right)  |\overline{L}(m),\overline{A}(m-1),\Xi\left(
m\right)  =1\right]  =\alpha^{T}W(m)$ for $m=0,...,K$. Here $W(m)=w_{m}\left[
\overline{L}(m),\overline{A}(m-1)\right]  $ is a vector of covariates
calculated from a subject's past data, $\alpha^{T}\ $is a row vector of
unknown parameters, and each person-month is treated as an independent
observation, so each person contributes up to $K+1$ observations. However,
person months for which $\Xi\left(  m\right)  \neq1$ are excluded from the
regression. Examples of $W(m)=w_{m}\left[  \overline{L}(m),\overline
{A}(m-1)\right]  $ would be the transpose of the row vector ($m,L^{T}%
(m),L^{T}(m-1))$. Let $\widehat{\alpha}$ be the OLS estimator of $\alpha$
computed using a standard statistical package.

For the moment assume $\beta$ is one dimensional. Let $\beta_{low}$ and
$\beta_{up}$ be much smaller and larger, respectively, than any substantively
plausible value of $\beta^{\ast}$.

Then, separately, for each $\beta$ on a grid from $\beta_{low}$ to $\beta
_{up}$, say $\beta_{low},\beta_{low}+0.1,\beta_{low}+0.2,...,\beta_{up}$,
perform the score test of the hypothesis $\theta=0$ in the extended linear
model
\begin{equation}
E\left[  A\left(  m\right)  |\overline{L}(m),\overline{A}(m-1),Y_{m}\left(
\beta\right)  ,\Xi\left(  m\right)  =1\right]  =\alpha^{T}W(m)+\theta
Y_{m}\left(  \beta\right)
\end{equation}
that adds the covariate $Y_{m}\left(  \beta\right)  $ at each time $m$ to the
above (pooled over persons and time) linear model. A $95\%$ confidence
interval for $\beta^{\ast}$ is the set of $\beta$ for which an $\alpha=0.05$
two-sided score test of the hypothesis $\theta=0$ does not reject. The
g-estimate $\widehat{\beta}$ of $\beta^{\ast}$ is the value of $\beta$ for
which the score test takes the value zero (i.e., the p-value is one).

The validity of g-estimation is proved as follows. By our comparability
assumption $Y_{m}\left(  \beta^{\ast}\right)  $ and $A(m)$ are conditionally
independent given

$\left(  \overline{L}(m),\overline{A}(m-1),\Xi\left(  m\right)  =1\right)  $.
That is, $Y_{m}\left(  \beta^{\ast}\right)  $ is not a predictor of $A(m)$
given $\left(  \overline{L}(m),\overline{A}(m-1),\Xi\left(  m\right)
=1\right)  $, which implies that the coefficient $\theta\ $of $Y_{m}\left(
\beta\right)  \ $must be zero in the extended model when $\beta=\beta^{\ast}$,
provided the model

$E\left[  A\left(  m\right)  |\overline{L}(m),\overline{A}(m-1),\Xi\left(
m\right)  =1\right]  =\alpha^{T}W(m)$ is correctly specified.

Now, we do not know the true value of $\beta$. Therefore, any value $\beta$
for which the data are consistent with the parameter $\theta$ of the term
$\theta Y_{m}\left(  \beta\right)  $ being zero might be the true $\beta
^{\ast}$, and thus belongs in our confidence interval. If consistency with the
data is defined at the $0.05$ level, then our confidence interval will have
coverage of $95\%$. Furthermore, the g-estimate $\widehat{\beta}$ of
$\beta^{\ast}$ is that $\beta$ for which adding the term $\theta Y_{m}\left(
\beta\right)  $ does not help to predict $A\left(  m\right)  $ whatsoever,
which is the $\beta$ for which the score test of $\theta=0$ is precisely zero.
The g-estimate $\widehat{\beta}$ is also the value of $\beta$ for which the
OLS estimator of $\theta$ is precisely zero.

It may appear peculiar that a function $Y_{m}\left(  \beta\right)  $ of the
response $Y$ measured at end of follow-up is being used to predict $A\left(
m\right)  $ at earlier times. However, this peculiarity evaporates when one
recalls that, for each $\beta$ on our grid, we are testing the null hypothesis
that $\beta=\beta^{\ast}$, and, under this null, $Y_{m}\left(  \beta\right)  $
is the counterfactual $Y_{m}$, which we can view as already existing at time
$m$ (although we cannot observe its value until time $K+1$ and then only if
A$\left(  t\right)  $ in the observed data is zero from $m$ onwards).

Suppose next that the parameter $\beta$ is a vector. To be concrete suppose we
consider the model with

$\gamma_{m}\left[  a\left(  m\right)  ,\overline{a}(m-1),\overline{l}\left(
m\right)  ,\beta\right]  =a\left(  m\right)  \left\{  \beta_{0}+\beta
_{1}m+\beta_{3}^{T}l\left(  m\right)  \right\}  \ $so $\beta$ is of dimension
dim$\left(  l\left(  m\right)  \right)  +2$ where dim$\left(  l\left(
m\right)  \right)  $ is the dimension of $l\left(  m\right)  $ which, for
concreteness, we take to be $3$. $\ $Hence $\beta$ is 5 dimensional. Then we
would use a $5$ dimensional grid, one dimension for each component of $\beta$.
So if we had $20$ grid points for each component, we would have $20^{5}%
\ $different values of $\beta$ on our $5$ dimensional grid. Now to estimate
$5$ parameters one requires $5$ additional covariates. Specifically, let
$Q_{m}\left(  \beta\right)  =q_{m}\left[  \bar{L}(m),\bar{A}(m-1),Y_{m}\left(
\beta\right)  \right]  $ be a $5$ dimensional vector of functions of $\left(
\bar{L}(m),\bar{A}(m-1),Y_{m}\left(  \beta\right)  \right)  $, such as
$Q_{m}\left(  \beta\right)  =\left[  1,m,L^{T}\left(  m\right)  \right]
\left(  Y_{m}\left(  \beta\right)  \right)  ^{2}$. We use the extended model
\[
E\left[  A\left(  m\right)  |\overline{L}(m),\overline{A}(m-1),Y_{m}\left(
\beta\right)  ,\Xi\left(  m\right)  =1\right]  =\alpha^{T}W(m)+\theta^{T}%
Q_{m}\left(  \beta\right)  .
\]

Our g-estimate $\widehat{\beta}$ is the $\beta$ for which the 5 degree of
freedom score test that all 5 components of $\theta$ equal zero is precisely
zero. The particular choice of the functions $q_{m}$ does not affect the
consistency of the point estimate, but it affects the width of the confidence interval.

When $\gamma_{m}\left[  a\left(  m\right)  ,\overline{a}(m-1),\overline
{l}\left(  m\right)  ,\beta\right]  =a\left(  m\right)  \beta^{T}R_{m}$ is
linear in $\beta$ with $R_{m}=r_{m}\left(  \bar{L}(m),\overline{A}%
(m-1)\right)  $ being a vector of known functions and we choose $Q_{m}\left(
\beta\right)  =Q_{m}^{\ast}Y_{m}\left(  \beta\right)  $ linear in
$Y_{m}\left(  \beta\right)  $, then, given the OLS estimator $\widehat{\alpha
}^{T}$of $\alpha^{T}$ in the model $E\left[  A\left(  m\right)  |\overline
{L}(m),\overline{A}(m-1)\right]  =\alpha^{T}W(m))$, there is an explicit
closed form expression for $\widehat{\beta}$ given by
\begin{equation}
\widehat{\beta}=\left\{  \sum_{i=1,m=0}^{i=n,m=K}\Xi_{i}\left(  m\right)
A_{i}\left(  m\right)  G_{im}\left(  \widehat{\alpha}\right)  Q_{im}^{\ast
}S_{im}^{T}\right\}  ^{-1}\left\{  \sum_{i=1,m=0}^{i=n,m=K}\Xi_{i}\left(
m\right)  Y_{i}G_{im}\left(  \widehat{\alpha}\right)  Q_{im}^{\ast}\right\}
\end{equation}
with $G_{im}\left(  \widehat{\alpha}\right)  =\left[  A_{i}\left(  m\right)
-\widehat{\alpha}^{T}W_{i}(m)\right]  $, $S_{im}=\sum_{i=1,j=m}^{i=n,j=K}%
R_{ij}$.

\textbf{Identification :}\ Suppose that two different values of $\beta,$ say
$\widehat{\beta}$ and $\widehat{\widehat{\beta}},$ both make the 5 degree of
freedom score test precisely zero and yet the two CI for $\beta^{\ast}$
centered at $\widehat{\beta}$ and $\widehat{\widehat{\beta}}$ do not overlap.
How should we choose between the estimates? [In such a case, the matrix whose
inverse is required in (9) will not be invertible and so (9) will fail.] Since
we can use any 5 vector $Q_{m}\left(  \beta\right)  =q_{m}\left[  \bar
{L}(m),\bar{A}(m-1),Y_{m}\left(  \beta\right)  \right]  $ in our procedure,
one simple approach is to try other choices of $Q_{m}\left(  \beta\right)  $
until we find a $Q_{m}\left(  \beta\right)  $ for which our CI for
$\beta^{\ast}$ includes only one of the $\widehat{\beta}$ and $\widehat
{\widehat{\beta}}$ , declare the one included to be our point estimate of
$\beta^{\ast}$ and ignore the excluded one. Will this approach always
succeeed? In general this approach should succeed in rather quickly excluding
all but one of the values of $\beta$ that originally made the score test zero,
provided that the model $\gamma_{m}\left[  a\left(  m\right)  ,\overline
{a}(m-1),\overline{l}\left(  m\right)  ,\beta\right]  $ is correct, except
when $\beta^{\ast}$ is not identified. By definition $\beta^{\ast} $ is not
identified if there is a $\beta^{\ast\ast}$ different from the true parameter
$\beta^{\ast}$ such that, with an infinite sample size, $\beta^{\ast\ast},$
like $\beta^{\ast},$ makes the 5 degree of freedom score test precisely zero
for all choices of $Q_{m}\left(  \beta\right)  .$ In our model, it follows
from Robins (3, 4) that under the positivity assumption that
\begin{equation}
\Pr\left[  A\left(  m\right)  =0|\overline{A}\left(  m-1\right)  ,\overline
{L}\left(  m\right)  ,\Xi\left(  m\right)  =1\right]  \neq0
\end{equation}
for all subjects and $m=0,...,K,$ $\beta^{\ast}$ is identified. In our context
the positivity assumption is a very weak assumption, that is almost certainly
true. Hence, for the remainder of the paper, we will silently assume that it holds.

\textbf{Remark:} We considered a linear regression model for

$E\left[  A\left(  m\right)  |\overline{L}(m),\overline{A}(m-1),Y_{m}\left(
\beta\right)  ,\Xi\left(  m\right)  =1\right]  $ in the above for expositonal
simplicity. In practice since $A\left(  m\right)  \geq0,$ we might use a log
linear model that specifies\hfill\linebreak\ $E\left[  A\left(  m\right)
|\overline{L}(m),\overline{A}(m-1),Y_{m}\left(  \beta\right)  ,\Xi\left(
m\right)  =1\right]  =\exp\left\{  \alpha^{T}W(m)+\theta^{T}Q_{m}\left(
\beta\right)  \right\}  $ and fit by non-linear least squares. In that case,
in the last display, $G_{im}\left(  \widehat{\alpha}\right)  =\left[
A_{i}\left(  m\right)  -\exp\left\{  \widehat{\alpha}^{T}W_{i}(m)\right\}
\right]  .$ Alternatively we could replace the repsonse variable $A\left(
m\right)  $ in the linear regression by $ln\left(  A_{m}+0.1\right)  $ where
the $0.1$ is added to insure the logarithm remains finite even when $A_{m}=0.
$ In that case, $G_{im}\left(  \widehat{\alpha}\right)  =\left[  ln\left\{
A_{i}\left(  m\right)  +0.1\right\}  -\widehat{\alpha}^{T}W_{i}(m)\right]  $

\subsubsection{An additive structural nested mean model (SNMM)}

An additive locally rank preserving SNM (4) implies that if two subjects have
the same observed data $O=\left(  Y,\overline{L},\overline{A}\right)  $ they
will have the same value of $Y_{0}$ under the "time 0 dietary intervention" of
the introduction. That is the model implies that for these subjects, the
effect of the "time 0 dietary intervention" will be identical. This assumption
is clearly biologically implausible in view of between-subject heterogeneity
in unmeasured genetic and environmental factors. To overcome this limitation,
we consider an additive structural nested mean model (SNMM)\
\begin{equation}
E\left[  Y_{m+1}-Y_{m}|\overline{A}\left(  m\right)  ,\overline{L}\left(
m\right)  \right]  =\gamma_{m}\left[  A(m),\overline{A}\left(  m-1\right)
,\overline{L}\left(  m\right)  ,\beta^{\ast}\right]
\end{equation}
that models the conditional mean of $Y_{m+1}-Y_{m}$ given $\left(
\overline{A}\left(  m\right)  ,\overline{L}\left(  m\right)  \right)  $ rather
than the individual differences $Y_{m+1}-Y_{m},$ and thus does not impose
local rank preservation. In particular $Y_{m}$ no longer is equal to
$Y_{m}\left(  \beta^{\ast}\right)  .$ However, Robins (4, 5) proved the
additive SNMM implies (and, in fact is equivalent to ) the assumption that
$Y_{m}$ and $Y_{m}\left(  \beta^{\ast}\right)  $ have the same mean given
$\overline{A}\left(  m\right)  ,\overline{L}\left(  m\right)  ,\Xi\left(
m\right)  =1.$ That is
\begin{equation}
E\left[  Y_{m}|\overline{A}\left(  m\right)  ,\overline{L}\left(  m\right)
\right]  =E\left[  Y_{m}\left(  \beta^{\ast}\right)  |\overline{A}\left(
m\right)  ,\overline{L}\left(  m\right)  \right]
\end{equation}
for each $m.$ Further he proved that, under the CO assumption, g-estimation of
$\beta^{\ast}$ retains all the properties described above, even in the absence
of local rank preservation, except now the function $Q_{m}\left(
\beta\right)  $ must be chosen linear in $Y\left(  \beta\right)  ,$ i.e.,
$Q_{m}\left(  \beta\right)  =Q_{m}^{\ast}Y_{m}\left(  \beta\right)  $ as above.

As a consequence, the definition of non-indentifiability must be modifed as
follows: the parameter $\beta^{\ast}$ of an SNMM $\beta^{\ast}$ is not
identified if there is a $\beta^{\ast\ast}$ different from the true parameter
$\beta^{\ast}$ such that, with an infinite sample size, $\beta^{\ast\ast},$
like $\beta^{\ast},$ makes the 5 degree of freedom score test precisely zero
for all choices of $Q_{m}\left(  \beta\right)  $ that are linear in
$Y_{m}\left(  \beta\right)  .$ A fuller discussion of rank preserving versus
non-rankpreservng models occurs in Section 3.2.4.

\subparagraph{Alternative Approaches:}

Under the CO assumption, it is shown in ref $\left[  12\right]  $ that the
$E\left[  Y_{m}\right]  $ are nonparametrically identified for $m=0,...,K$
from data $O=\left(  Y,\overline{L},\overline{A}\right)  $ $\ $by the IPTW
formula%
\begin{equation}
E\left[  YI\left\{  \underline{A}\left(  m\right)  =\underline{0}\left(
m\right)  \right\}  \mathbb{W}\left(  m\right)  \right]  ,
\end{equation}

where $\underline{A}\left(  m\right)  =(A\left(  m\right)  ,...,A\left(
K\right)  )$ and the IPTW weight
\[
\mathbb{W}\left(  m\right)  =1/%
{\displaystyle\prod\limits_{k=m}^{K}}
\left\{  pr\left(  A\left(  k\right)  =0|\overline{A}\left(  k-1\right)
,\overline{L}\left(  k\right)  \right)  \right\}  ^{\Xi\left(  k\right)  }%
\]
is the inverse of the conditional probability that a subject had his observed
treatment $\underline{A}\left(  m\right)  =\underline{0}\left(  m\right)  $.
That is, $E\left[  Y_{m}\right]  $ is the weighted mean of the observed
$utility$ $Y$ among subjects who observed data was consistent with following
the time $m$\ dietary intervention with weights given by the inverse of the
conditional probability of having data consistent with following the
intervention. Thus one could, in principle, consider estimating $E\left[
Y_{0}\right]  $ nonparametrically by the weighted average of $Y$ among
subjects whose weight never exceeded their baseline weight at age 18 with
weights proportional to an estimate $\widehat{\mathbb{W}}\left(  0\right)  $
of $\mathbb{W}\left(  0\right)  .$ That is, by $\left[
{\displaystyle\sum\limits_{\left\{  i;\underline{A}_{i}\left(  0\right)
=\underline{0}\left(  0\right)  \right\}  }}
\widehat{\mathbb{W}_{i}}\left(  0\right)  Y_{i}\right]  /\left[
{\displaystyle\sum\limits_{\left\{  i;\underline{A}_{i}\left(  0\right)
=\underline{0}\left(  0\right)  \right\}  }}
\widehat{\mathbb{W}_{i}}\left(  0\right)  \right]  .$ The problem with this
approach is that only the utility $Y$ of the rare person whose weight never
exceeds his age 18 weight contributes to the analysis. In contrast by
specifying a SNMM, data on the utility $Y$ of every subject contributes to the
estimate of $E\left[  Y_{0}\right]  .$ The price paid for the greater
efficiency of a SNMM is the possibility of bias if the SNMM (11) is misspecified.

However, under the CO, $E\left[  Y_{m}|\overline{A}\left(  m\right)
,\overline{L}\left(  m\right)  \right]  =E\left[  Y_{m}|\overline{A}\left(
m-1\right)  ,\overline{L}\left(  m\right)  \right]  $ and is nonparametrically
identified by the formula

$E\left[  YI\left\{  \underline{A}\left(  m\right)  =\underline{0}\left(
m\right)  \right\}  /\mathbb{W}\left(  m\right)  |\overline{L}\left(
m\right)  ,\overline{A}\left(  m-1\right)  \right]  $. Thus $E\left[
Y_{m+1}-Y_{m}|\overline{A}\left(  m\right)  ,\overline{L}\left(  m\right)
\right]  $ is nonparametrically identified for m=0,...,K. Hence, given a
sufficiently large sample size, one could in principle construct
misspecification tests of the model (11) that have power against all
alternatives when the model is incorrect. In practice, the available sample
size may greatly limit the power to detect model misspecification.

IPTW estimation of marginal structural models and the parametric g-formula are
alternative approaches to model-based estimation of $E\left[  Y_{0}\right]  $
that also use data on every subject's utility $Y$. See Appendix 2 for further
discussion of the latter approach.

\textbf{Remark: }The reader familiar with IPTW expects $\mathbb{W}\left(
m\right)  $ to be defined as $\mathbb{W}\left(  m\right)  =1/\underset
{k=m}{\overset{k}{\Pi}}\left\{  pr\left(  A\left(  k\right)  =0|\bar{A}\left(
k-1\right)  =0\left(  k-1\right)  ,\bar{L}\left(  k\right)  \right)  \right\}
$ rather than as \newline$1/\underset{k=m}{\overset{K}{\Pi}}\left\{  pr\left(
A\left(  k\right)  =0|\bar{A}\left(  k-1\right)  =\bar{0}\left(  k-1\right)
,\bar{L}\left(  k\right)  \right)  \right\}  ^{\Xi\left(  k\right)  }$. In
fact, the two expression would have been equal had we used the coding
convention that the vector $L\left(  k\right)  $ includes $\Xi\left(
k\right)  $ as a component because, under this coding, $pr\left(  A\left(
k\right)  =0|\bar{A}\left(  k-1\right)  =\bar{0}\left(  k-1\right)  ,\bar
{L}\left(  k\right)  \right)  $ is one whenever $\Xi\left(  k\right)  $ takes
the value zero. However, we do not use this convention.

\subsection{Case 2: Unmeasured confounding by preclinical disease.}

In this section we no longer assume A$\left(  m\right)  $ is statistically
independent of $Y_{m}$ given $\overline{A}\left(  m-1\right)  ,\overline
{L}(m),\Xi\left(  m\right)  =1$. To describe our new comparability assumption
we need to introduce some further notation. Let $X=\min\left(  T,\mathcal{D}%
\right)  $ be the minimum of the time $T$ to death and the time $\mathcal{D}$
to the diagnosis of a chronic disease, such as cancer, severe emphysema, liver
or renal disease, or any other chronic condition that would be severe enough
to affect weight gain. At each time $m,$ the indicator $I\left(  X\leq
m\right)  $ is a component of $L\left(  m\right)  .$ Further if $I\left(
X\leq m\right)  =1,$ the exact time $X$ is observed and included in $L\left(
m\right)  .$ Thus $X$ is observed if $X$ is less than K+1. However $X$ is
censored (i.e. not observed) on subjects whose $X$ exceeds $K+1,$ the end of
follow up time. For the present we shall avoid the additional complications
that arise from censoring by assuming that $X$ is less than $K+1$ for all
subjects, so that the data $O=\left(  Y,X,\overline{L},\overline{A}\right)  $
is observed on each subject. In Section 4, we relax this assumption and allow
for censoring.

Let $X_{m}=\min\left(  T_{m},\mathcal{D}_{m}\right)  $ be the counterfactual
version of $X=\min\left(  T,\mathcal{D}\right)  $ had "the time m dietary
intervention" been carried out. Then we make the following more realistic assumption.

\textbf{Realistic Comparability (RC)\ Assumption :} $A\left(  m\right)  $ is
statistically independent of $\left(  Y_{m},X_{m}\right)  $ given $\Xi\left(
m\right)  =1,\overline{L}(m),\overline{A}\left(  m-1\right)  \ $and
$\overline{U}\left(  m\right)  =\overline{0}\left(  m\right)  ,$ where
$U\left(  m\right)  =1$ if a subject has at $m$ or had prior to $m,$ an
undiagnosed chronic disease that was sufficiently advanced to interfere with
his normal weight trajectory. Otherwise $U\left(  m\right)  =0.$ We also
define $U\left(  m\right)  =1$ for subject's alive at $m$ with $X<m$ under the
assumption that there was probably a subclinical period prior to the time $X$
of clinical diagnosis in which weight gain may have been altered. Note that
$U\left(  m\right)  =0$ implies $\overline{U}\left(  m\right)  =\left(
U\left(  0\right)  ,...,U\left(  m\right)  \right)  =\overline{0}\left(
m\right)  $ is also zero.

\textbf{Remark:} The RC assumption cannot be recast as $\left(  Y_{m}%
,X_{m}\right)  $ independent of $A\left(  m\right)  $ given $\left(
\overline{L}(m),\overline{A}\left(  m-1\right)  ,\overline{U}\left(  m\right)
\right)  $ even had we used the coding convention that vector $L\left(
k\right)  $ includes $\Xi\left(  k\right)  $ as a component, because, even
under this coding, $pr\left(  A\left(  k\right)  =0|\bar{A}\left(  k-1\right)
=\bar{0}\left(  k-1\right)  ,\bar{L}\left(  k\right)  ,\overline{U}\left(
m\right)  ,\left(  Y_{m},X_{m}\right)  \right)  $ would be neither zero nor
one and could , under the RC assumption, depend on $\left(  Y_{m}%
,X_{m}\right)  $ whenever $\overline{U}\left(  m\right)  \neq0$ and
$\Xi\left(  k\right)  =1.$ For this reason it would perhaps be more precise to
refer to the RC as a selective comparability assumption as it only implies
comparability for a selected subset of the population.

We observe $\left(  Y,X,\overline{L},\overline{A}\right)  $ but $\overline
{U}\left(  m\right)  =\left(  U\left(  0\right)  ,...,U\left(  m\right)
\right)  $ is, of course, generally unobserved when $X>m$. Thus $\overline{U}$
is an unmeasured confounder. The most crucial of several assumptions needed to
allow consistent estimation of the parameter of interest $E\left[
Y_{0}\right]  $ in this setting is the following.

\textbf{Clinical Detection (CD) Assumption:} Any subject who has $U\left(
m\right)  =1$ [ie a sufficently advanced undiagnosed chronic disease at ( or
before) $m]$ and thereafter follows "the time m dietary intervention" will
either have died or been diagnosed with clinical chronic disease by time
$m+\zeta,$ where $\zeta$ is assumed known. Formally
\begin{equation}
U\left(  m\right)  =1\ \Rightarrow X_{m}\leq m+\zeta
\end{equation}
or equivalently
\begin{equation}
X_{m}>m+\zeta\Rightarrow\overline{U}\left(  m\right)  =\overline{0}\left(
m\right)  \
\end{equation}
Here $\Rightarrow$ is translated as 'implies'. A typical choice for $\zeta$
might be 72 months. It is useful to choose $\zeta$ to be the minimal time for
which $\left(  15\right)  $ holds as this increases both the effciency of
g-estimators and the power of goodness of fit tests to detect misspecification
of a structural nested model and decreases the likelihood that an SNM is
nonidentified. However if the chosen $\zeta$ is less then the true minimum
time for which (15) holds bias will result. As a consequence one should
routinely include a table that shows how one's estimate of $E\left[
Y_{0}\right]  $ changes as $\zeta$ is varied.

The RC and CD assumptions require that one record in X the minimum time of
clinical onset among the set of clinical conditions whose preclinical phase
could affect BMI. The exact clinical conditions that belong in this subset is
a substantive question, about which subject matter experts should be consulted.

\textbf{Remark:} We will later consider the effect of replacing the
counterfactual $X_{m}$ by the observed $X$ in the CD assumption.

\subsubsection{Estimation under a Rank Preserving SNM for $Y_{m}|X_{m}$ with
$X_{m}$ known}

To consistently estimate $E\left[  Y_{0}\right]  $ under $RC$ and $CD$ we must
replace our SNMM\ model with an additive SNMM model for $Y_{m}|X_{m}$ that
also conditions on and allows effect modification by the counterfactual
$X_{m}.$ For pedagogic purposes in this subsection we return to locally rank
preserving models. A locally rank preserving SNM for $Y_{m}|X_{m}$ states
that
\begin{equation}
Y_{m+1}-Y_{m}=\gamma_{m}\left[  A(m),\overline{A}\left(  m-1\right)
,\overline{L}\left(  m\right)  ,X_{m},\beta^{\ast}\right]
\end{equation}
where $\beta^{\ast}$ is an unknown parameter and $\gamma_{m}\left[
A(m),\overline{A}\left(  m-1\right)  ,\overline{L}\left(  m\right)
,X_{m},\beta\right]  $ is a known function that can now depend on $X_{m}$ that
takes the value zero if either $A(m)=0$ or $\beta=0.$ [We emphasize that it is
$X_{m}$ and not $\overline{X}_{m}$ that occurs in the last display$.].$ This
model is equivalent to assuming
\begin{equation}
Y_{m}=Y_{m}\left(  \beta^{\ast}\right)
\end{equation}
$\ $for each subject with $Y_{m}\left(  \beta\right)  $ now redefined as
\begin{equation}
Y_{m}\left(  \beta\right)  =Y-\sum_{j=m}^{K}\gamma_{m}\left[  A(j),\overline
{A}\left(  j-1\right)  ,\overline{L}\left(  j\right)  ,X_{j},\beta\right]
\end{equation}

Now, of course the counterfactual variable $X_{m}$ is itself unobserved.
However for pedagogic purposes in this subsection we unrealisticly assume that
in addition to the observed data $\left(  Y,X,\overline{L},\overline
{A}\right)  $, data on the counterfactuals $X_{m}$ are available.

\textbf{Remark:} We do not actually require a locally rank preserving SNM for
$Y_{m}|X_{m}.$ A locally rank preserving SNM for $Y_{m}|c\left(  X_{m}\right)
$ for a certain known function $c\left(  x\right)  $ could be used instead.
This remark is explored further in the Appendix.

Redefine $Q_{m}\left(  \beta\right)  =q_{m}\left[  \bar{L}(m),\bar
{A}(m-1),X_{m},Y_{m}\left(  \beta\right)  \right]  $ to possibly be a function
of $X_{m}.$ Consider again the g-estimator $\widehat{\beta}$ that is equal to
the $\beta$ for which the 5 degree of freedom score test of $\theta=0$ is
precisely zero in the model
\[
E\left[  A\left(  m\right)  |\overline{L}(m),\overline{A}(m-1),\ X_{m}%
,\Xi\left(  m\right)  =1,Y_{m}\left(  \beta\right)  \right]  =\alpha
^{T}W(m)+\theta^{T}Q_{m}\left(  \beta\right)  .
\]
$\widehat{\beta}$ would be a CAN estimator of $\beta^{\ast}$ under the CO
assumption$,\ $but not under the RC assumption. Under $RC$, the independence
needed to make $\theta=0$ when $\beta$=$\beta^{\ast}$ only holds when we also
condition on $\overline{U}\left(  m\right)  .$

However, consider the estimator $\widetilde{\beta}$ obtained when, for each
time $m,$ we only fit the previous model to subjects for whom $X_{m}>m+\zeta,$
excluding all subjects with $X_{m}\leq m+\zeta.$ This exclusion can be
expressed by saying that we now fit the model%
\begin{equation}
E\left[  A\left(  m\right)  |\overline{L}(m),\overline{A}(m-1),X_{m}%
,\Xi\left(  m\right)  =1,Y_{m}\left(  \beta\right)  ,X_{m}>m+\zeta\right]
=\alpha^{T}W(m)+\theta^{T}Q_{m}\left(  \beta\right)  .
\end{equation}

Then the estimator $\widetilde{\beta}$ is the $\beta$ for which the 5 degree
of freedom score test of the hypothesis $\theta=0$ is precisely zero in this
latter model. When $X_{m}>m+\zeta,$ $\overline{U}\left(  m\right)
=\overline{0}\left(  m\right)  ,$ by assumption CD. Hence we can rewrite the
last display as
\begin{align}
& E\left[  A\left(  m\right)  |\overline{L}(m),\overline{A}(m-1),\ X_{m}%
,\Xi\left(  m\right)  =1,\ Y_{m}\left(  \beta\right)  ,X_{m}>m+\zeta
,\overline{U}\left(  m\right)  =\overline{0}\left(  m\right)  \right]
\nonumber\\
& =\alpha^{T}W(m)+\theta^{T}Q_{m}\left(  \beta\right)
\end{align}
showing that we have succeeded in conditioning on $\overline{U}\left(
m\right)  =\overline{0}\left(  m\right)  ,$ even though $\overline{U}\left(
m\right)  $ is unmeasured! It follows that, when the parameter $\beta^{\ast} $
is identified, the estimator $\widetilde{\beta}$ is a consistent and
asymptotically normal (CAN) estimator of $\beta^{\ast}$ under the RC and
CD\ assumptions, since these assumptions imply the coefficient $\theta=0$ if
$\beta=\beta^{\ast}.$ However as discussed further below, under the RC and
CD\ assumptions, the positivity assumption no longer suffices to guarantee identification.

In summary, all that was required to produce a CAN estimator $\widetilde
{\beta}$ of the parameter $\beta^{\ast}$ of our locally rank preserving SNM
(17) for $Y_{m}|X_{m}$under the RC and CD\ assumptions was to restrict the
earlier g-estimation procedure at each time $m$ to those subjects with
$X_{m}>m+\zeta.$

Thus if $\gamma_{m}\left[  A\left(  m\right)  ,\overline{A}(m-1),\overline
{L}\left(  m\right)  ,X_{m},\beta\right]  =A\left(  m\right)  \beta^{T}R_{m}$
is linear in $\beta$ with $R_{m}=r_{m}\left(  \bar{L}(m),X_{m}\right)  $ being
a vector of known functions that now can depend on $X_{m}$, then, given the
OLS estimator $\widehat{\alpha}^{T}$of $\alpha^{T}$ in the model\newline%
$E\left[  A\left(  m\right)  |\overline{L}(m),\overline{A}(m-1),\Xi\left(
m\right)  =1\right]  =\alpha^{T}W(m)$ and $Q_{m}\left(  \beta\right)
=Q_{m}^{\ast}Y_{m}\left(  \beta\right)  $ linear in $Y_{m}\left(
\beta\right)  $, the CAN estimator $\widetilde{\beta}$ exists in closed form
as
\begin{align}
\widetilde{\beta}  & =\left\{  \sum_{i=1,m=0}^{i=n,m=K}I\left[  X_{im}%
>m+\zeta\right]  \Xi_{i}\left(  m\right)  A_{i}\left(  m\right)  G_{im}\left(
\widehat{\alpha}\right)  Q_{im}^{\ast}S_{im}^{T}\right\}  ^{-1}\\
& \times\left\{  \sum_{i=1,m=0}^{i=n,m=K}I\left[  X_{im}>m+\zeta\right]
\Xi_{i}\left(  m\right)  Y_{i}G_{im}\left(  \widehat{\alpha}\right)
Q_{im}^{\ast}\right\} \nonumber
\end{align}
with $G_{im}\left(  \widehat{\alpha}\right)  =\left[  A_{i}\left(  m\right)
-\widehat{\alpha}^{T}W_{i}(m)\right]  $, $S_{im}=\sum_{i=1,j=m}^{i=n,j=K}%
R_{ij}$.

From the above, it follows that if, in addition to the observed data $\left(
Y,X,\overline{L},\overline{A}\right)  ,$ data on the counterfactuals $X_{m}$
are available for each m, the sample average $\sum_{i}^{n}Y_{0}\left(
\widetilde{\beta}\right)  /n$ is a CAN estimator of the parameter of interest
$E\left[  Y_{0}\right]  \ $under the RC and CD\ assumptions, provided
$\beta^{\ast}$ is identified and both our locally rank preserving SNM for
$Y_{m}|X_{m\text{ }}$and our model
\begin{equation}
E\left[  A\left(  m\right)  |\overline{L}(m),\overline{A}(m-1),\Xi\left(
m\right)  =1\right]  =\alpha^{T}W(m)
\end{equation}
are correct. Of course data on $X_{m}$ are unavailable. However in the next
subsection we prove an analogue of this result holds without data on $X_{m}$
under a locally rank preserving SNFTM for the $X_{m}$ which allows us to
replace $X_{m}$ by an estimate $X_{m}\left(  \widetilde{\psi}\right)  ,$ where
$\widetilde{\psi}$ estimates the parameter $\psi^{\ast}$ of our SNFTM.

Before preceding to the next subsection, several additional points need to be made.

\textbf{Can we replace X}$_{m}$\textbf{\ by }$\mathbf{X}:$ A natural question
that arises is the following. Suppose we replaced $X_{m}$ by the observed $X$
in the CD assumption, in our definition of $Y_{m}\left(  \beta\right)  ,$ and
wherever else $X_{m}\ $occurs in this subsection, with the exception of the RC
assumption (as the RC assumption with X replacing $X_{m}$ would clearly be
false if BMI is a cause of T and/or C and thus of X.). Do $\widetilde{\beta}$
and $\sum_{i}^{n}Y_{0}\left(  \widetilde{\beta}\right)  /n $ remain CAN
estimators of $\beta^{\ast}$ and $E\left[  Y_{0}\right]  ?$ This question is
natural in the sense that it is not obvious that the CD assumption and RP SNM
based on $X_{m}$ are more likely to be true than when based on $X.$ So were
the answer "yes" it would be simpler and more straightforward to use $X$ in
place of $X_{m}$. In particular, since $X,$ unlike $X_{m},$ is observed, we
would eliminate the need to replace $X_{m}$ with the estimator $X_{m}\left(
\widetilde{\psi}\right)  $, thereby greatly simplifying the analysis.

Unfortunately $\widetilde{\beta}$ and thus $\sum_{i}^{n}\ Y_{0}\left(
\widetilde{\beta}\right)  /n$ do not remain consistent when we use $X$ in
place of $X_{m}.$ To see why consider the model
\begin{equation}
E\left[  A\left(  m\right)  |\overline{L}(m),\Xi\left(  m\right)
=1,\ \overline{A}(m-1),X,\text{ }Y_{m}\left(  \beta\right)  ,X>m+\zeta\right]
=\alpha^{T}W(m)+\theta^{T}Q_{m}Y_{m}\left(  \beta\right)
\end{equation}
which has replaced $X_{m}$ in Eq (19) with{\Large \ }$X.$ Clearly
$\widetilde{\beta}$ will only be consistent for the parameter $\beta^{\ast} $
of our locally RP SNM if $\theta=0$ when $\beta=\beta^{\ast}.$ That is
$\widetilde{\beta}$ will only be consistent if $Y_{m}=Y_{m}\left(  \beta
^{\ast}\right)  $ is independent of $A\left(  m\right)  $ given $\left(
\overline{L}(m),\ \overline{A}(m-1),\Xi\left(  m\right)  =1,X,X>m+\zeta
\right)  .$ Now by the CD assumption with $X$ replacing $X_{m},$ $X>m+\zeta$
implies $\overline{U}\left(  m\right)  =\overline{0}\left(  m\right)  .$ Thus
consistency of $\widetilde{\beta}$ requires $Y_{m}\left(  \beta^{\ast}\right)
$ independent of $A\left(  m\right)  $ given

$\left(  \overline{L}(m),\ \overline{A}(m-1),\Xi\left(  m\right)
=1,\ X,X>m+\zeta,\overline{U}\left(  m\right)  =\overline{0}\left(  m\right)
\right)  .$ However, we show in the next paragraph that this independence
statement is not implied by the RC assumption and thus will generally be
false, unless $A\left(  k\right)  $ has no causal effect on $X$ for $k\geq m$
in which case $X=X_{m}$ for each subject and we are back to Eq. (20).

When $A\left(  m\right)  $ has a causal effect on $X$ (whether directly or
through $A\left(  k\right)  ,$ $k>m$) then $X$ is a common effect of two
causes $A\left(  m\right)  $ and $X_{m}$ that are independent conditional on
the event $\left(  \overline{L}(m),\ \overline{A}(m-1),\Xi\left(  m\right)
=1,\ \overline{U}\left(  m\right)  =\overline{0}\left(  m\right)
,Y_{m}\left(  \beta^{\ast}\right)  \right)  .$ Therefore, conditional on both
the previous event and $\left(  X,X>m+\zeta\right)  ,$ $A\left(  m\right)  $
and $X_{m}$ are dependent and thus so are $A\left(  m\right)  $ and
$Y_{m}\left(  \beta^{\ast}\right)  ,$ since $X_{m}\ $and $Y_{m}\left(
\beta^{\ast}\right)  \ $are highly correlated, as both are functions of
$T_{m}$.

However, even when $A\left(  m\right)  $ has a causal effect on $X,$ a slight
modification of the above estimation procedure can be used to obtain CAN
estimators of $\beta^{\ast}\ $in the special case in which $A\left(  m\right)
$ has a known minimal latent period $\chi$ for its effect on $X$ of at least
$\zeta$ months.

\textbf{Definition of Minimal Latent Period (MLP) for effect on }$X$\textbf{:}
$A\left(  m\right)  $ has a minimal latent period for its effect on $X$ of
$\chi$ months if, for every subject and each time $k>m,$ $X_{k}>m+\chi
\Leftrightarrow X_{m}>m+\chi$ and $X_{k}=X_{m}$ if $X_{m}<m+\chi.$ In
particular by taking $k=K+1,$ the last two statements become $X>m+\chi
\Leftrightarrow X_{m}>m+\chi$ and $X=X_{m}$ if $X_{m}<m+\chi.$

When a known minimal latent period $\chi$ exceeds $\zeta,$ we can obtain CAN
estimators of $\beta^{\ast}$ by simply replacing $X>m+\zeta$ by m+$\chi$%
$>$%
$X>m+\zeta$ in model (23) since then the event $\left(  X,m+\chi
>X>m+\zeta\right)  $ is the event $\left(  X_{m},m+\chi>X_{m}>m+\zeta\right)
$ and we are back in the setting of Eq. (19), except for the additional
restriction, $m+\chi>X_{m},$ which does not introduce bias. Thus the existence
of a minimal latent period of length $\chi$ greater than $\zeta$ allows us to
estimate $\beta^{\ast}$ and $E\left[  Y_{0}\right]  $ without the need to
specify a SNFTM for the $X_{m}.$

We now prove that under the RC and CD assumptions, a MLP of length $\chi$
greater than $\zeta$ implies that
\[
X\amalg A\left(  m\right)  |\overline{L}(m),\Xi\left(  m\right)
=1,\ \overline{A}(m-1),m+\chi>X>m+\zeta.
\]
$\ $It follows that taking the RC and CD assumptions as given, we can test the
hypothesis that a MLP of length $\chi$ greater than $\zeta$ exists by testing
whether the last display is true. In fact a test of the hypothesis that the
parameter $\psi^{\ast}$ of a SNFTM serves as a test of the previous display.
To prove our previous claim note that, by the MLP assumption, the event
$m+\chi>X>m+\zeta$ is the event $m+\chi>X_{m}>m+\zeta$ which, by the CD
assumption, is the event $m+\chi>X_{m}>m+\zeta,\overline{U}\left(  m\right)
=\overline{0}\left(  m\right)  .$ Thus the last display is under the MLP and
CD assumption equivalent to the statement "$X_{m}$ is independent of $A\left(
m\right)  $ given $\left(  \overline{L}(m),\Xi\left(  m\right)  =1,\ \overline
{A}(m-1),m+\chi>X_{m}>m+\zeta,\overline{U}\left(  m\right)  =\overline
{0}\left(  m\right)  \right)  "$ which is true by the RC assumption.

Most experts believe it to be substantively implausible that an increase in
BMI has a minimum latent period of more than 72 months, our default choice for
$\zeta.$ In contrast, in occupational cohort studies of the effect of a
chemical carcinogen on time to clinical cancer, minimum latent periods of up
to 10 years are commonly assumed.

\subsubsection{Estimation of $E\left[  Y_{0}\right]  $ under a Rank Preserving
SNFTM:}

As mentioned above, an analogue of the above results hold when data on X$_{m}
$ are unavailable under under a locally rank preserving SNFTM for $X_{m}.$ The
simplest locally rank preserving SNFTM specifies that%

\begin{align}
X_{m}  & =m+\int_{m}^{X}\exp\left(  \psi^{\ast}A\left(  t\right)  \right)
dt\text{ if }X>m\\
X_{m}  & =X\text{ if }X\leq m,
\end{align}
where $\psi^{\ast}$ is an unknown parameter and $A\left(  t\right)  $ is as
defined previously when $t$ is a whole number of months and $A\left(
t\right)  =A\left(  \left\lfloor t\right\rfloor \right)  $ when $t$ is not a
whole number where $\left\lfloor t\right\rfloor $ is the largest integer less
than or equal to $t$. Thus, by the definition of an intergral as the area
under a curve,
\[
\int_{m}^{X}\exp\left(  \psi^{\ast}A\left(  t\right)  \right)  dt=%
{\displaystyle\sum\limits_{j=m}^{j=\left\lfloor X-1\right\rfloor }}
\exp\left(  \psi^{\ast}A\left(  j\right)  \right)  +\left\{  X-\left\lfloor
X\right\rfloor \right\}  \exp\left(  \psi^{\ast}A\left(  \left\lfloor
X\right\rfloor \right)  \right)  \text{.}%
\]
A locally rank preserving SNFTM directly maps an individual's observed failure
time $X$ to the failure time $X_{m}$ the individual would have under the "time
m dietary intervention". Thus it is a model for individual causal effects. If
$\psi^{\ast}=0,$ $\exp\left(  \psi^{\ast}A\left(  t\right)  \right)  =1$ and
thus $X_{m}=m+\int_{m}^{X}dt=m+X-m=X$ for any $m$. Hence $\psi^{\ast}=0$
encodes the sharp null hypothesis that $X_{0}=X$ for all subjects, i.e., the
"time 0 dietary intervention" has no effect on any subject's $X=\min\left(
T,\mathcal{D}\right)  .$ It is useful to note that when $\psi^{\ast}\neq0,$
the SNFTM $(24)$-$\left(  25\right)  $ implies that there is no minimal latent
period for the effect of treatment on $X.$

A general class (although not the most general class ) of locally RP SNFTMs
that includes the above one parameter model assumes
\begin{align}
X_{m}  & =m+\int_{m}^{X}\exp\left\{  \omega\left(  \overline{A}\left(
t\right)  ,\overline{L}\left(  t\right)  ,\psi^{\ast}\right)  \right\}
dt\text{ if }X>m\\
X_{m}  & =X\text{ if }X\leq m
\end{align}

where $\omega\left(  \overline{A}\left(  t\right)  ,\overline{L}\left(
t\right)  ,\psi\right)  \equiv\omega\left(  A\left(  t\right)  ,\overline
{A}\left(  t^{-}\right)  ,\overline{L}\left(  t\right)  ,\psi\right)  $ is a
known function satisfying $\omega\left(  A\left(  t\right)  ,\overline
{A}\left(  t^{-}\right)  ,\overline{L}\left(  t\right)  ,\psi\right)  =0$ if
$A\left(  t\right)  =0$ or $\psi=0$ and $\overline{A}\left(  t^{-}\right)  $
is the A-history until just prior to time $t.$ For example, we might have
$\omega\left(  A\left(  t\right)  ,\overline{A}\left(  t^{-}\right)
,\overline{L}\left(  t\right)  ,\psi\right)  =A\left(  t\right)  \left\{
\psi_{0}+\psi_{1}^{T}L\left(  t\right)  \right\}  $ where $L\left(  t\right)
=L\left(  \left\lfloor t\right\rfloor \right)  $ and $L\left(  \left\lfloor
t\right\rfloor \right)  $ is as defined earlier.

We next turn to estimation of $\psi^{\ast}.$ For the moment, suppose the CO
assumption modified to have $\left(  Y_{m},X_{m}\right)  $ in place of Y$_{m}$
held and that $\left(  Y,X,\overline{L},\overline{A}\right)  $ was observed.
Then we could consistently estimate $\psi^{\ast}$ by g-estimation.
Specifically we define%

\begin{align}
X_{m}\left(  \psi\right)   & =m+\int_{m}^{X}\exp\left\{  \omega\left(
\overline{A}\left(  t\right)  ,\overline{L}\left(  t\right)  ,\psi\right)
\right\}  dt\text{ if }X>m\\
X_{m}\left(  \psi\right)   & =X\text{ if }X\leq m
\end{align}
so under our model, $X_{m}=X_{m}\left(  \psi^{\ast}\right)  .$ Note that, for
each $\psi$, $X_{m}\left(  \psi\right)  $ can be computed from the observed
data. Suppose, for concreteness, $\psi^{\ast}$ is 5 dimensional so we search
over a $5$ dimensional grid. We let $Q_{m}^{\ast\ast}\left(  \psi\right)
=q_{m}^{\ast\ast}\left[  \bar{L}(m),\overline{A}\left(  m-1\right)
,X_{m}\left(  \psi\right)  \right]  $ be a $5$ dimensional vector of functions
of $\left(  \bar{L}(m),\overline{A}\left(  m-1\right)  ,X_{m}\left(
\psi\right)  \right)  $ such as $Q_{m}^{\ast\ast}\left(  \psi\right)
=X_{m}\left(  \psi\right)  \left[  1,m,L^{T}\left(  m\right)  \right]  .$ We
use an extended linear model
\[
E\left[  A\left(  m\right)  |\overline{L}(m),\overline{A}\left(  m-1\right)
,\Xi\left(  m\right)  =1,\ X_{m}\left(  \psi\right)  \right]  =\alpha
^{T}W(m)+\theta^{T}Q_{m}^{\ast\ast}\left(  \psi\right)
\]

Our g-estimate $\widehat{\psi}$ is the $\psi$ for which the 5 degree of
freedom score test that all 5 components of $\theta$ equal zero is precisely
zero. Since, by the modified CO assumption, $\theta=0$ if $\psi=\psi^{\ast},$
the g-estimate $\widehat{\psi}$ is CAN for $\psi^{\ast}$. The particular
choice of the functions $Q_{m}^{\ast\ast}\left(  \psi\right)  $ does not
affect the consistency of the point estimate, but it determines the width of
its confidence interval. Because $X_{m}\left(  \psi\right)  $ is a nonlinear
function of $\psi,$ there is not a closed form expression for $\widehat{\psi
}.$ However the equation solved by $\widehat{\psi}$ is a smooth function of
$\psi,$ so standard methods for solving nonlinear equations such as the
Newton-Raphson algorithm can be used to compute $\widehat{\psi}.$

Next suppose the observed data is still $\left(  Y,X,\overline{L},\overline
{A}\right)  ,$ but the modified CO assumption does not hold. Rather, the CD
and RC assumptions hold. Define the estimator $\widetilde{\psi}$ as the $\psi$
for which the 5 degree of freedom score test of the hypothesis $\theta=0$ is
precisely zero in the model
\[
E\left[  A\left(  m\right)  |\overline{L}(m),\overline{A}(m-1),\Xi\left(
m\right)  =1,\ X_{m}\left(  \psi\right)  ,X_{m}\left(  \psi\right)
>m+\zeta\right]  =\alpha^{T}W(m)+\theta^{T}Q_{m}^{\ast\ast}\left(
\psi\right)  .
\]
Note the set of subjects who do not contribute to the score test of $\theta=0$
(i.e subjects with $X_{m}\left(  \psi\right)  \leq m+\zeta)$ depends on
$\psi.$ When $X_{m}=X_{m}\left(  \psi^{\ast}\right)  >m+\zeta,$ then
$\overline{U}\left(  m\right)  =\overline{0}\left(  m\right)  ,$ by assumption
CD. Hence, at $\psi=\psi^{\ast},$ our procedure conditions on $\overline
{U}\left(  m\right)  =\overline{0}\left(  m\right)  .$ It follows that,
provided $\psi^{\ast}$ is identified, the estimator $\widetilde{\psi}$ is a
CAN estimator of $\psi^{\ast}$ under the RC assumption, as that assumption
implies the coefficient $\theta=0$ if $\psi=\psi^{\ast}.$ However, under the
CD and RC assumptions, the positivity assumption does not guarantee identification.

Now let $\widetilde{\beta}\left(  \widetilde{\psi}\right)  $ be defined like
$\widetilde{\beta}$ except that everywhere $X_{m}\left(  \widetilde{\psi
}\right)  $ replaces $X_{m},\ $so that $\ \widetilde{\beta}\left(
\widetilde{\psi}\right)  $ is a function of the data $\left(  Y,X,\overline
{L},\overline{A}\right)  $ only. Next define
\begin{equation}
\ Y_{m}\left(  \beta,\psi\right)  =Y-\sum_{j=m}^{K}\gamma_{m}\left[
A(j),\overline{A}\left(  j-1\right)  ,\overline{L}\left(  j\right)
,X_{j}\left(  \psi\right)  ,\beta\right]
\end{equation}
Note, by both models (17) and (26-27) being locally rank preserving,
$Y_{m}\left(  \beta^{\ast},\psi^{\ast}\right)  =Y_{m}$. Thus, when $\psi
^{\ast}$ and $\beta^{\ast}$ are identified, the sample average $\sum_{i}%
^{n}Y_{0}\left[  \left(  \widetilde{\beta}\left(  \widetilde{\psi}\right)
,\widetilde{\psi}\right)  \right]  /n$ is a CAN estimator of the parameter of
interest $E\left[  Y_{0}\right]  \ $under the RC and CD\ assumptions, provided
both our locally rank preserving SNM (17) for $Y_{m}|X_{m},$ our locally rank
preserving SNFTM (26-27) for $X_{m},$ and our model (22) are all correctly specified.

\subsubsection{Estimation of $E\left[  Y_{0}\right]  $ under a SNMM and a
SNFTM without Rank Preservation:}

As discused earlier, the assumption of local rank preservation is biologically
implausible. Thus we will no longer assume that our locally rank preserving
models (17) and (26-27) are true. As a consequence we can no longer assume
that there exists some $\left(  \beta^{\ast},\psi^{\ast}\right)  $ such that
the unobserved counterfactuals $\left(  X_{m},Y_{m}\right)  $ equal the
observed $\left(  X_{m}\left(  \psi\right)  ,Y_{m}\left(  \beta,\psi\right)
\right)  $ when $\left(  \beta,\psi\right)  =\left(  \beta^{\ast},\psi^{\ast
}\right)  $. However, suppose with $\left(  X_{m}\left(  \psi\right)
,Y_{m}\left(  \beta,\psi\right)  \right)  $ still defined by (28-29), and
(30), we assume that , for each $m,$ there exists some $\left(  \beta^{\ast
},\psi^{\ast}\right)  $ such that

\textbf{Assumption (i):} when $\psi=\psi^{\ast},X_{m}$ and $X_{m}\left(
\psi\right)  $ have the same conditional distribution given $\left(  A\left(
m\right)  ,\overline{L}(m),\overline{A}(m-1)\right)  $ and

\textbf{Assumption (ii)}
\begin{equation}
E\left[  Y_{m}|A\left(  m\right)  ,\overline{L}(m),\overline{A}(m-1),X_{m}%
=x\right]  =E\left[  Y_{m}\left(  \beta^{\ast},\psi^{\ast}\right)  |A\left(
m\right)  ,\overline{L}(m),\overline{A}(m-1),X_{m}\left(  \psi^{\ast}\right)
=x\right]
\end{equation}

In contrast with the assumption of local RP, there is no apriori biological
reason to exclude the possibility that (i) and (ii) both hold.

When assumptions (i) and (ii) hold for each $m$, we say the SNMM
\begin{equation}
\gamma_{m}\left[  A(m),\overline{A}\left(  m-1\right)  ,\overline{L}\left(
m\right)  ,x,\beta\right]
\end{equation}
for $Y_{m}|X_{m}$ and the SNFTM $\left(  28\right)  -\left(  29)\right)  $ for
$X_{m}$ jointly hold with true parameter $\left(  \beta^{\ast},\psi^{\ast
}\right)  $. If the RC and CD\ assumptions, the model (22) and (i) and (ii)
all hold, then $\widetilde{\psi},\widetilde{\beta}\left(  \widetilde{\psi
}\right)  ,$ and $\sum_{i}^{n}Y_{0}\left[  \left(  \widetilde{\beta}\left(
\widetilde{\psi}\right)  ,\widetilde{\psi}\right)  \right]  /n$ as defined
previously are CAN for $\psi^{\ast},\beta^{\ast},$ and the parameter of
interest $E\left[  Y_{0}\right]  $ respectively, provided $\left(  \beta
^{\ast},\psi^{\ast}\right)  $ are identified and we choose $Q_{m}\left(
\beta\right)  $ linear in $Y_{m}\left(  \beta\right)  $. [ In contrast,
$Q_{m}^{\ast\ast}\left(  \psi\right)  $ need not be chosen linear in
$X_{m}\left(  \psi\right)  .$]. In summary, $\widetilde{\psi},\widetilde
{\beta}\left(  \widetilde{\psi}\right)  ,$ and $\sum_{i}^{n}Y_{0}\left[
\left(  \widetilde{\beta}\left(  \widetilde{\psi}\right)  ,\widetilde{\psi
}\right)  \right]  /n$ have the same statistical properties under our joint
SNMM model for $Y_{m}|X_{m}$ and SNFTM for $X_{m}$ when local rank
preservation does not hold as when it does.

\subsubsection{Are Remarkable Results due to Some Sleight of Hand}

The result summarized in the last sentence is striking for a number of
reasons. Our comparability assumption, i.e. the RC\ assumption, only assumes
no unmeasured confounding conditional on $\overline{U}\left(  m\right)  .$ Yet
neither the SNMM for $Y_{m}|X_{m}$ nor SNFTM for $X_{m}$ is a model for causal
effects conditional on the unmeasured $\overline{U}\left(  m\right)  $. Thus,
it is remarkable that these models can be used to estimate causal contrasts
such as $E\left[  Y_{0}\right]  -E\left[  Y\right]  \ $under the RC and CD
assumptions. Furthermore, even though $X_{m}>m+\zeta$ implies $\overline
{U}\left(  m\right)  =\overline{0}\left(  m\right)  $ by the CD assumption,
nonetheless, in the abscence of local rank preservation, $X_{m}\left(
\psi^{\ast}\right)  >m+\zeta\ $does not imply $\overline{U}\left(  m\right)
=\overline{0}\left(  m\right)  .\ $Hence when local rank preservation does not
hold, even though we condition on $X_{m}\left(  \psi\right)  >m+\zeta$ in
computing our g-estimates $\widetilde{\psi},\widetilde{\beta}\left(
\widetilde{\psi}\right)  ,$ we do not thereby restrict the analysis to a
subset of subjects all of whom have the same value of $\overline{U}\left(
m\right)  ;$ thus one might guess confounding by the unmeasured $\overline
{U}\left(  m\right)  $ has not been controlled and our estimates of
$\widetilde{\psi},\widetilde{\beta}\left(  \widetilde{\psi}\right)  ,$ and
$\sum_{i=1}^{n}Y_{0,i}\left[  \left(  \widetilde{\beta}\left(  \widetilde
{\psi}\right)  ,\widetilde{\psi}\right)  \right]  /n$ must be inconsistent.
Remarkably, such is not the case.

How did we pull off the seemingly remarkable 'magic' described in the
preceding paragraph? We shall investigate whether we used some subtle "sleight
of hand". We use a simple paradigmatic instance of our model that only
involves a single time-independent exposure to guide our investigation.
Specifically, we next provide an explicit proof that contains no "sleight of
hand" of our results in the case of a time-independent exposure. The general
case is treated in the appendix. The reader who is interested more in the
methodology and less interested in foundational issues may feel free to skip
ahead to section 4.

\textbf{Paradigmatic Instance of a Time-Independent Exposure:} We suppose that
$K+1=1$ so time 0 is the only time of exposure. Further we assume there are no
covariates. In this setting the $RC\ $assumption becomes $\left(  Y_{0}%
,X_{0}\right)  $ independent of $A\left(  0\right)  $ given the unmeasured
confounder $U\left(  0\right)  =0$. The CD assumption becomes $X_{0}%
>\varsigma$ implies $U\left(  0\right)  =0.$ Our SNFTM for $X_{0}$ becomes

\textbf{Assumption (i)}: $X_{0}\left(  \psi\right)  =X$exp$\left(  \psi
A\left(  0\right)  \right)  \ $and $X_{0}\ $have the same conditional
distribution given $A\left(  0\right)  $ at $\psi=\psi^{\ast},$\newline while
our SNMM for $Y_{m}|X_{m}$ becomes

\textbf{Assumption (ii):} $E\left[  Y_{0}|A\left(  0\right)  ,X_{0}=x\right]
=E\left[  Y_{0}\left(  \beta^{\ast},\psi^{\ast}\right)  |A\left(  0\right)
,X_{0}\left(  \psi^{\ast}\right)  =x\right]  $ where $Y_{0}\left(  \beta
,\psi\right)  =Y-\gamma_{0}\left[  A(0),X_{0}\left(  \psi\right)
,\beta\right]  .$

Neither model makes any reference to $U\left(  0\right)  \ $and thus neither
is a model for causal effects conditional on $U\left(  0\right)  .$
Furthermore, although $X_{0}>\zeta$ implies $U\left(  0\right)  =0$ by the CD
assumption, nonetheless $X_{0}\left(  \psi^{\ast}\right)  >\zeta$ does not
imply $U\left(  0\right)  =0.$ Now to prove our results.

\paragraph{Proofs Of Our Results:}

\textbf{Proof that }$\widetilde{\psi}$\textbf{\ is CAN for }$\psi^{\ast}%
$\textbf{:} By assumption (i), $pr\left[  X_{0}\left(  \psi^{\ast}\right)
>t|A\left(  0\right)  ,X_{0}\left(  \psi^{\ast}\right)  >\zeta\right]
=pr\left[  X_{0}>t|A\left(  0\right)  ,X_{0}>\zeta\right]  .$ But, by the CD
and then the RC assumptions, $\allowbreak pr\left[  X_{0}>t|A\left(  0\right)
,X_{0}>\zeta\right]  $=$pr\left[  X_{0}>t|A\left(  0\right)  ,X_{0}%
>\zeta,U\left(  0\right)  =0\right]  =pr\left[  X_{0}>t|X_{0}>\zeta,U\left(
0\right)  =0\right]  .$ Hence $pr\left[  X_{0}\left(  \psi^{\ast}\right)
>t|A\left(  0\right)  ,X_{0}\left(  \psi^{\ast}\right)  >\zeta\right]  $ is
not a function of $A\left(  0\right)  .$ We conclude that $A\left(  0\right)
$ and $X_{0}\left(  \psi^{\ast}\right)  $ are independent given $X_{0}\left(
\psi^{\ast}\right)  >\zeta.$ Thus $E\left[  A\left(  0\right)  |X_{0}\left(
\psi^{\ast}\right)  ,X_{0}\left(  \psi^{\ast}\right)  >\zeta\right]
=\alpha+\theta X_{0}\left(  \psi^{\ast}\right)  $ has coefficient $\theta=0$
so, when $\psi^{\ast}$ is identified, the $\widetilde{\psi}$ for which the the
score test of $\theta=0$ takes the value 0 is CAN for $\psi^{\ast}.$

\textbf{Proof that }$\widetilde{\beta}\left(  \widetilde{\psi}\right)
$\textbf{\ is CAN for }$\beta^{\ast}:$ By Assumption (ii),
\[
E\left[  Y_{0}\left(  \beta^{\ast},\psi^{\ast}\right)  |A\left(  0\right)
,X_{0}\left(  \psi^{\ast}\right)  =x,X_{0}\left(  \psi^{\ast}\right)
>\zeta\right]  =E\left[  Y_{0}|A\left(  0\right)  ,X_{0}=x,X_{0}>\zeta\right]
.
\]

But, by the CD and then the RC assumptions,

$E\left[  Y_{0}|A\left(  0\right)  ,X_{0}=x,X_{0}>\zeta\right]  =E\left[
Y_{0}|A\left(  0\right)  ,X_{0}=x,X_{0}>\zeta,U\left(  0\right)  =0\right]  $

$=E\left[  Y_{0}|\ X_{0}=x,X_{0}>\zeta,U\left(  0\right)  =0\right]  \ $is not
a function of $A\left(  0\right)  .$

Thus, $0=E\left[  Y_{0}\left(  \beta^{\ast},\psi^{\ast}\right)  \left\{
A\left(  0\right)  -E\left[  A\left(  0\right)  |X_{0}\left(  \psi^{\ast
}\right)  ,X_{0}\left(  \psi^{\ast}\right)  >\zeta\right]  \right\}  \right]
.$

Hence $0=E\left[  Y_{0}\left(  \beta^{\ast},\psi^{\ast}\right)  \left\{
A\left(  0\right)  -E\left[  A\left(  0\right)  |X_{0}\left(  \psi^{\ast
}\right)  >\zeta\right]  \right\}  \right]  .$As a consequence, the
$\widetilde{\beta}\left(  \widetilde{\psi}\right)  $ for which the the score
test of $\theta=0$ takes the value 0 in the model

$E\left[  A\left(  0\right)  |X_{0}\left(  \widetilde{\psi}\right)
>\zeta,Y_{0}\left(  \beta,\widetilde{\psi}\right)  \right]  =\alpha+\theta
Y_{0}\left(  \beta,\widetilde{\psi}\right)  $ is CAN for $\beta^{\ast}$, when
$\beta^{\ast}$ and $\psi^{\ast}$ are identified.

\textbf{Proof that }$\sum_{i=1}^{n}Y_{0,i}\left[  \left(  \widetilde{\beta
}\left(  \widetilde{\psi}\right)  ,\widetilde{\psi}\right)  \right]
/n$\textbf{\ is CAN for }$E\left[  Y_{0}\right]  :$ $E\left[  Y_{0}\right]  =%
{\displaystyle\int}
{\displaystyle\int}
E\left[  Y_{0}|A\left(  0\right)  ,X_{0}=x\right]  dF_{X_{0}}\left(
x|A_{0}\right)  dF\left(  A_{0}\right)  $

$=%
{\displaystyle\int}
{\displaystyle\int}
E\left[  Y_{0}\left(  \beta^{\ast},\psi^{\ast}\right)  |A\left(  0\right)
,X_{0}\left(  \psi^{\ast}\right)  =x\right]  dF_{X_{0}\left(  \psi^{\ast
}\right)  }\left(  x|A_{0}\right)  dF\left(  A_{0}\right)  =E\left[
Y_{0}\left(  \beta^{\ast},\psi^{\ast}\right)  \right]  \ $by assumptions (i)
and (ii). Hence , $\sum_{i=1}^{n}Y_{0,i}\left[  \left(  \widetilde{\beta
}\left(  \widetilde{\psi}\right)  ,\widetilde{\psi}\right)  \right]  /n$ is
CAN for $E\left[  Y_{0}\left(  \beta^{\ast},\psi^{\ast}\right)  \right]
=E\left[  Y_{0}\right]  ,$ when $\beta^{\ast}$ and $\psi^{\ast}$ are identified.

This completes the promised proof of our results in the time-independent case.
The proof in the appendix of the general time-dependent case is not much more
difficult when one proceeds by induction$.$ We conclude no sleight of hand
occured in the proof.

\paragraph{\textbf{Do Correctly Specified SNMMs for }$Y_{m}|X_{m}%
$\textbf{\ and SNFTMs for }$X_{m}$\textbf{\ Always Exist?}}

Perhaps the sleight of hand occurred right at the start, when we supposed that
there exist $\left(  \beta^{\ast},\psi^{\ast}\right)  $ such that assumptions
(i) and (ii) hold. We now prove that no such slight of hand is afoot.
Specifically we prove that there always exist correctly specified SNMMs for
$Y_{m}|X_{m}$ and SNFTMs for $X_{m}.$ [This result does not, of course, imply
that the particular SNMM and SNFTM we actually choose to analyze are correct.]
We actually prove this result for an alternative, more intuitive, definition
of a SNMM for $Y_{m}|X_{m}$ and a SNFTM for $X_{m}$ and then prove these
alternative definitions are logically equivalent to assumptions (i) and (ii).
This is done in this subsection for the special case of a time-independent
exposure and in the Appendix for a general time-varying exposure.

Consider again, for simplicity, our paradigmatic instance. Write $A\left(
0\right)  $ as $A.$ Suppose that $Y$,$Y_{0},X,X_{0}$ are all non-negative
continuous random variables with support on $\left(  0,\infty\right)  ,$
satisfying the consistency assumption $X=X_{0}$ and $Y=Y_{0}$ if $A=0.$ Let
$S\left(  x|A\right)  =pr\left(  X>x|A\right)  .$ Let $S_{0}\left(
x|A\right)  =pr\left(  X_{0}>x|A\right)  .\ $Let $S_{0}^{-1}\left(
x|A\right)  $ be the inverse of $S_{0}\left(  x|A\right)  $ wrt the $x$
argument. Define the function $x_{0}^{\dagger}\left(  x,A\right)  =S_{0}%
^{-1}\left[  \left\{  S\left(  x|A\right)  \right\}  |A\right]  $ $.$
Substituting 0 for $A$, we find $x_{0}^{\dagger}\left(  x,0\right)  =x,$ so
\begin{equation}
x_{0}^{\dagger}\left(  X,0\right)  =X\text{ wp1}%
\end{equation}
Define $X_{0}^{\dagger}=x_{0}^{\dagger}\left(  X,A\right)  .$ Then
$X_{0}^{\dagger}=x_{0}^{\dagger}\left(  X,0\right)  =X,\ $when $A=0.$ It is
well known that $X_{0}^{\dagger}=x_{0}^{\dagger}\left(  X,A\right)  $ and
$X_{0}$ have the same conditional distribution given $A.$

Define $S\left(  t|A,X_{0}^{\dagger}=x\right)  =pr\left(  Y>t|A,X_{0}%
^{\dagger}=x\right)  $ and $S_{0}\left(  t|A,X_{0}=x\right)  =pr\left(
Y_{0}>t|A,X_{0}=x\right)  .$ Let $S_{0}^{-1}\left(  t|A,X_{0}=x\right)  $ be
the inverse of $S_{0}\left(  t|A,X_{0}=x\right)  $ wrt the $t$ argument. Let
$y_{0}^{^{\dagger}}\left(  t,x,A\right)  =S_{0}^{-1}\left(  \left\{  S\left(
t|A,X_{0}^{\dagger}=x\right)  \right\}  |A,X_{0}^{\dagger}=x\right)  $ and
$Y_{0}^{^{^{\dagger}}}=y_{0}^{^{\dagger}}\left(  Y,X,A\right)  .$ Then
$Y_{0}^{^{^{\dagger}}}|A,X_{0}^{\dagger}=x$ and $Y_{0}|A,X_{0}=x$ have the
same conditional distribution. It follows that $\left(  Y_{0}^{^{^{\dagger}}%
},X_{0}^{\dagger}\right)  |A\ $and $\left(  Y_{0},X_{0}\right)  |A$ have the
same joint conditional distribution . Thus$,$
\begin{equation}
E\left[  Y_{0}^{^{^{\dagger}}}|X_{0}^{\dagger}=x,A\right]  =E\left[
Y_{0}|X_{0}=x,A\right]
\end{equation}
Define
\begin{equation}
\gamma^{^{\dagger}}\left(  A,x\right)  =E\left[  Y|X_{0}^{\dagger}=x,A\right]
-E\left[  Y_{0}^{^{^{\dagger}}}|X_{0}^{\dagger}=x,A\right]  \equiv E\left[
Y-Y_{0}^{^{^{\dagger}}}|X_{0}^{\dagger}=x,A\right]
\end{equation}
$\ $The last two displays imply that%
\begin{equation}
E\left[  Y-\gamma^{^{\dagger}}\left(  A,X_{0}^{\dagger}\right)  |X_{0}%
^{\dagger}=x,A\right]  =E\left[  Y_{0}|X_{0}=x,A\right]
\end{equation}
and
\begin{equation}
\gamma^{^{\dagger}}\left(  0,X\right)  =0\text{ wp1}%
\end{equation}

since, by $Y_{0}^{^{^{\dagger}}}=y_{0}^{^{\dagger}}\left(  Y,X,A\right)  $,
$Y_{0}^{^{^{\dagger}}}=Y$ when $A=0$.

Here are the alternative definitions of a SNFTM for $X_{0}$ and a SNMM for
$Y_{0}|X_{0}.$

\textbf{Definition a:} Let $x_{0}\left(  t,a,\psi\right)  $ be known function
montone increasing in $t$ for each $\left(  a,\psi\right)  $ satisfying
$x_{0}\left(  t,a,\psi\right)  =1$ if $a=0$ or $\psi=0.$We say $x_{0}\left(
t,a,\psi\right)  $ is a correctly specified SNFTM for $X_{0}$ if there exists
$\psi^{\ast}$ such that $X_{0}\left(  \psi^{\ast}\right)  \equiv x_{0}\left(
X,A,\psi^{\ast}\right)  \ $equals $X_{0}^{\dagger}$ with probability one.$\ $

\textbf{Definition b:} We say a known function $\gamma\left(  a,x,\beta
\right)  $ satisfying $\gamma\left(  a,x,\beta\right)  =0$ if $a=0$ or
$\beta=0$ is a correctly specified SNMM for $Y_{0}|X_{0}$ if, for some
$\beta^{\ast},$ $\gamma\left(  A,X,\beta^{\ast}\right)  =\gamma^{^{\dagger}%
}\left(  A,X\right)  $ with probability 1.

Define $Y_{0}\left(  \beta^{\ast},\psi^{\ast}\right)  =Y-\gamma\left(
A,X_{0}\left(  \psi^{\ast}\right)  ,\beta^{\ast}\right)  .$

It is obvious from definitions a and b that there always exist correctly
specified SNMMs for $Y_{0}|X_{0}$ under Definition \textbf{b} and correctly
specified SNFTMs for $X_{0}$ under definition \textbf{a} since $\gamma
^{^{\dagger}}\left(  A,X\right)  $ and $x_{0}^{\dagger}\left(  X,A\right)  $
are well defined functions of $\left(  F,F_{0}\right)  $ satisfying
$\gamma^{^{\dagger}}\left(  0,X\right)  =0$ and $x_{0}^{\dagger}\left(
X,0\right)  =X $ with probability one, where $F$ and $F_{0},$ respectively,
denote the joint distribution of $\left(  Y,X,A\right)  $ and of $\left(
Y_{0},X_{0},A\right)  $. Note $\gamma^{^{\dagger}}\left(  A,X\right)  $ and
$x_{0}^{\dagger}\left(  X,A\right)  $ do not depend on the conditional joint
distribution of $\left\{  \left(  Y,X\right)  ,\left(  Y_{0},X_{0}\right)
\right\}  \ $given $A.$ This is as desired as this joint is not
non-parametrically identified from data $\left(  Y,X,A\right)  $ even when A
is randomly assigned.

Thus it only remain to show the logical equivalence of the original and
alternative defintions of a SNFTM for $X_{0}$ and a SNMM for $Y_{0}|X_{0}.$

The following Lemma shows that the alternative defintions of a SNFTM for
$X_{0}$ and a SNMM for $Y_{0}|X_{0}$ imply the previous definitions.

\textbf{\ Lemma: }Suppose\textbf{\ }$x_{0}\left(  t,a,\psi\right)  $ is a
correctly specified SNFTM for $X_{0}$ as defined in definition \textbf{a}.
Then $X_{0}\left(  \psi^{\ast}\right)  |A$ has the same distribution as
$X_{0}|A.$ Further assume that $\gamma\left(  a,x,\beta\right)  $ is a
correctly specified SNMM for $Y_{0}|X_{0}$ as defined in definition
\textbf{b}$.$ Then $E\left[  Y_{0}\left(  \beta^{\ast},\psi^{\ast}\right)
|A,X_{0}\left(  \psi^{\ast}\right)  =x\right]  =E\left[  Y_{0}|X_{0}%
=x,A\right]  .$

Proof: The first result follows immediately from $X_{0}^{\dagger}$ and $X_{0}$
having the same conditional distribution given $A.$ The second result follows from

$E\left[  Y-\gamma^{^{\dagger}}\left(  A,x\right)  |X_{0}^{\dagger
}=x,A\right]  =E\left[  Y_{0}|X_{0}=x,A\right]  .$

Finally, the following Lemma shows that the original definitions imply the
alternative definitions.

\textbf{Lemma: }Suppose\textbf{\ }$x_{0}\left(  t,a,\psi\right)  $ is montone
increasing in $t$ for each $\left(  a,\psi\right)  $ satisfying $x_{0}\left(
t,a,\psi\right)  =1$ if $a=0$ or $\psi=0.\ $Further suppose that $X_{0}\left(
\psi^{\ast}\right)  |A$ has the same distribution as $X_{0}|A$ wp1 where
$X_{0}\left(  \psi\right)  =x_{0}\left(  X,A,\psi\right)  .$ Then
$X_{0}\left(  \psi\right)  $ is a correctly specified SNFTM for $X_{0}\ $under
definition \textbf{a}. In addition, suppose that $\gamma\left(  a,x,\beta
\right)  $ is a function satisfying $\gamma\left(  a,x,\beta\right)  =0$ if
$a=0$ or $\beta=0.$ Suppose $E\left[  Y-\gamma\left(  A,x,\beta^{\ast}\right)
|X^{\dagger}=x,A=a\right]  =E\left[  Y_{0}|X_{0}=x,A=a\right]  $ for all
$\left(  x,a\right)  $ in a set of probability 1 under the law of $\left(
X_{0},A\right)  .$ Then$,\gamma\left(  a,x,\beta\right)  $ is a correctly
specified SNMM for $Y_{0}|X_{0}\ $under definition \textbf{b}$.$

\textbf{Proof:} The proof of the first part follows from the well known result
that $X_{0}^{\dagger}=x_{0}^{\dagger}\left(  X,A\right)  $ is the only
function $h\left(  X,A\right)  $ of $\left(  X,A\right)  $ satisfying
$h\left(  X,A\right)  |A$ has the same distribution as $X_{0}|A$ wp1. The
second part is proved by showing that $\gamma^{\dagger}\left(  a,x\right)  $
is the unique function $h\left(  a,x\right)  $ satisfying $E\left[  Y-h\left(
A,x\right)  |X^{\dagger}=x,A=a\right]  =E\left[  Y_{0}|X_{0}=x,A=a\right]  $
for all $\left(  x,a\right)  $ in a set of probability 1 as in Refs (8,10).

\paragraph{\textbf{Are }$\gamma^{^{\dagger}}\left(  a,x\right)  $%
\textbf{,}$x_{0}^{^{\dagger}}\left(  x,a\right)  ,$ and $E\left[
Y_{0}\right]  $\textbf{\ nonparametrically identified from data }$\left(
Y,X,A\right)  $\textbf{\ under our assumptions?}}

In this subsection, we finally uncover some slight of hand that provided us
with such seemingly magical results. Although we restrict our discussion to
the special case of a time-independent exposure, similiar results apply in the
general case. Specifically, we will show that $\gamma^{^{\dagger}}\left(
a,x\right)  $\textbf{,}$x_{0}^{^{\dagger}}\left(  x,a\right)  ,$ and $E\left[
Y_{0}\right]  $ are not identified by the distribution of $\left(
Y,X,A\right)  $ under the RC and CD\ assumptions. Previously, we saw that
$\gamma^{^{\dagger}}\left(  a,x\right)  $\textbf{,}$x_{0}^{^{\dagger}}\left(
x,a\right)  ,$ and $E\left[  Y_{0}\right]  $ are identified and equal
$\gamma\left(  a,x,\beta^{\ast}\right)  ,x_{0}\left(  x,a,\psi^{\ast}\right)
,$ and $E\left[  Y-\gamma\left(  A,X,\beta^{\ast}\right)  \right]  ,$
respectively when we assume a correctly specified SNFTM$\ x_{0}\left(
x,a,\psi\right)  $ for $X_{0}$ and a $SNMM$ $\gamma\left(  a,x,\beta\right)  $
for $Y_{0}|X_{0}$ whose true parameters $\psi^{\ast}$ and $\beta^{\ast} $ are
identified (by g-estimation). It follows that identification of $\gamma
^{^{\dagger}}\left(  a,x\right)  $\textbf{, }$x_{0}^{^{\dagger}}\left(
x,a\right)  ,$ and $E\left[  Y_{0}\right]  $ must result from the functional
form restrictions encoded in our models $x_{0}\left(  x,a,\psi\right)  $ and
$\gamma\left(  a,x,\beta\right)  .$ It follows that if we make the
restrictions imposed by our models less rigid by adding additional parameters,
we can lose identification of $\gamma^{^{\dagger}}\left(  a,x\right)
$\textbf{, }$x_{0}^{^{\dagger}}\left(  x,a\right)  ,$ and $E\left[
Y_{0}\right]  .$ This loss of identification occurs when, in an infinite
sample size, more than one combination of parameters, say the true parameters
$\left(  \psi^{\ast},\beta^{\ast}\right)  $ and the false parameters $\left(
\psi^{\ast\ast},\beta^{\ast\ast}\right)  ,$ both make the score tests in our
g-estimation procedures exactly zero for all choices of $Q_{m}\left(
\beta\right)  $ linear in $Y_{m}\left(  \beta\right)  $ and all choices of
$Q_{m}^{\ast\ast}\left(  \beta\right)  $. This loss of identification can be
expressed by saying that the data (even were the sample size infinite) can not
be used to determine whether the true causal quanties are $\gamma\left(
a,x,\beta^{\ast}\right)  ,x_{0}\left(  x,a,\psi^{\ast}\right)  ,$ and
$E\left[  Y-\gamma\left(  A,X,\beta^{\ast}\right)  \right]  $ versus
$\gamma\left(  a,x,\beta^{\ast\ast}\right)  ,x_{0}\left(  x,a,\psi^{\ast\ast
}\right)  ,$ and $E\left[  Y-\gamma\left(  A,X,\beta^{\ast\ast}\right)
\right]  .$

In contrast, $\gamma^{^{\dagger}}\left(  a,x\right)  $\textbf{, }%
$x_{0}^{^{\dagger}}\left(  x,a\right)  ,$ and $E\left[  Y_{0}\right]  $ are
identifed under the comparability assumption that $\left(  Y_{0},X_{0}\right)
$ is independent of $A_{0\text{ }}$, without any reliance on the functional
form restrictions encoded in our models. However, in contrast with assumption
RC, this comparability assumptions contradicts our substantive knowledge, as
it implies no unmeasured confounding by undiagnosed chronic disease.

The problem of lack of identification under the RC and CD assumptions has
little to do with the question of local rank preservation. Suppose we have
assumed a correctly specified SNFTM$\ x_{0}\left(  x,a,\psi\right)  $ for
$X_{0}$ and we do not assume RP. Suppose in truth RP holds. Nonetheless, a
second investigator who assumes the RP version of the SNFTM\ model gains
nothing thereby in regard to the estimation of $x_{0}^{^{\dagger}}\left(
x,a\right)  $: the causal quantity $x_{0}^{^{\dagger}}\left(  x,a\right)
\ $is identified under the non-rank preserving SNFTM if and only it is
identified under the RP SNFTM. However, a small amount could be gained by
assuming rank preservation for a SNMM; rarely by assuming RP a
non-identifiable SNMM can become identifiable as one can then use non-linear
functions $Q_{m}\left(  \beta\right)  $ of $Y_{m}\left(  \beta\right)  $ in
g-estimation. But this advantage is not actually due to rank preservation.
Rather it is due to the fact that an RP\ SNMM is actually a special case of a
structural nested distribution model (SNDM)\ as defined in Refs (5) and (7).
Our model SNMM model $\gamma\left(  a,x,\beta\right)  $ for $Y_{0}|X_{0} $ is
a SNDM if $Y-\gamma\left(  A,X,\beta^{\ast}\right)  $ is independent (rather
than just mean independent) of $A$ given $X.$ It is this independence (rather
than rank preservation) that licences the use of non-linear functions
$Q_{m}\left(  \beta\right)  $ of $Y_{m}\left(  \beta\right)  $ in g-estimation.

\subparagraph{Non Identifiability of $\gamma^{^{\dagger}}\left(  a,x\right)
$\textbf{, }$x_{0}^{^{\dagger}}\left(  x,a\right)  ,$ and $E\left[
Y_{0}\right]  $}

Suppose we do not impose a SNFTM for $X_{0}$ or a $SNMM$ for $Y_{0}|X_{0}.$
Then, it is clear that all we can conclude under assumptions RC and CD is that
$X_{0}^{^{^{\dagger}}}=x_{0}^{^{^{\dagger}}}\left(  X,A\right)  $ and $A\equiv
A\left(  0\right)  $ are independent given $X_{0}^{^{^{\dagger}}}>\zeta
$\textbf{\ and }$E\left[  Y-\gamma_{0}^{^{^{\dagger}}}\left(  A,x\right)
|A\left(  0\right)  ,X_{0}^{^{^{\dagger}}}=x,X_{0}^{^{^{\dagger}}}%
>\zeta\right]  =E\left[  Y-\gamma_{0}^{^{^{\dagger}}}\left(  A,x\right)
|X_{0}^{^{^{\dagger}}}=x,X_{0}^{^{^{\dagger}}}>\zeta\right]  .$ As a
consequence, our parameter of interest $E\left[  Y_{0}\right]  $\textbf{\ }is
not identified\textbf{. }Specifically, under RC and CD, with $p=pr\left(
A=0\right)  $
\begin{gather}
E\left[  Y_{0}\right]  =\\
E\left[  Y|X>\zeta,A=0\ \right]  \left\{  pr\left[  X>\zeta|A=0\right]
p+\left\{  1-pr\left[  X^{\dagger}<\zeta|A\neq0\right]  \right\}  \right\}
\left(  1-p\right) \\
+E\left[  Y|X\leq\zeta,A=0\right]  pr\left[  X\leq\zeta|A=0\right]  p\\
+E\left[  \left\{  Y-\gamma^{^{\dagger}}\left(  A,X^{\dagger}\right)
\right\}  |X^{\dagger}\leq\zeta,A\neq0\right]  pr\left[  X^{\dagger}%
<\zeta|A\neq0\right]  \left(  1-p\right)  .
\end{gather}

However the quantities
\begin{align}
pr\left[  X^{\dagger}<\zeta|A\neq0\right]   & =pr\left[  X_{0}<\zeta
|A\neq0\right]  ,\\
E\left[  \left\{  Y-\gamma^{^{\dagger}}\left(  A,X^{\dagger}\right)  \right\}
|X^{\dagger}\leq\zeta,A\neq0\right]   & =E\left[  Y_{0}|X_{0}\leq\zeta
,A\neq0\right]
\end{align}
are not identified under the RC and CD assumptions. It suffices to show this
when RP holds. So, for the moment assume RP. Because both quantities (42) and
(43) refer to the distribution of the counterfactuals responses $\left(
Y_{0},X_{0}\right)  $ under no exposure (no weight gain) among those who
actually were exposed $\left(  A\neq0\right)  ,$ we need an assumption to
identify them under RP. But under RC, we only have comparability conditional
on a value of $U\left(  0\right)  ,$ which is unknown when $X_{0}<\zeta$, so
identification fails.

When we additionally assume a SNFTM for $X_{0}$ and a $SNMM$ for $Y_{0}%
|X_{0},$ we may or may not obtain identification of $E\left[  Y_{0}\right]  $
depending on whether the additional functional form restrictions encoded in
the models suffice to identify the quantities (42) and (43) by allowing us to
extrapolate from $X_{0}>\zeta$ where we have comparability (since, by CD,
U$\left(  0\right)  =0)$) to $X\leq\zeta\ $where we do not. To clarify this
last statement, consider the following RP SNM for $Y_{0}|X_{0}:Y_{0}%
=Y-\gamma\left(  A,X_{0},\beta^{\ast}\right)  $ with
\begin{equation}
\gamma\left(  A,X_{0},\beta\right)  =\beta_{0}AI\left(  X_{0}\leq\zeta\right)
+\beta_{1}AI\left(  X_{0}>\zeta\right)  \text{.}%
\end{equation}
Under assumptions RC and CD, even if we unrealistically suppose that data on
$X_{0}$ was available for all subjects$,$ we could not identify $\beta^{\ast
}=\left(  \beta_{0}^{\ast},\beta_{1}^{\ast}\right)  ^{T},$ because $\beta
_{0}^{\ast}$ would not be identified, although $\beta_{1}^{\ast}$ would be
identified. This follows from the fact that, under RC and CD, no subject with
$X_{0}\leq\zeta$ may contribute to g-estimation of $\beta^{\ast}$. As a
consequence we cannot identify $E\left[  Y_{0}\right]  $ because
$Y_{0}=Y-\beta_{0}^{\ast}$ is not estimable on the subset of exposed subjects
($A=1)$ with $X_{0}\leq\zeta.$

In contrast, were data on $X_{0}$ available, $\beta^{\ast}$ and $E\left[
Y_{0}\right]  $ are identified in the RP SNM$\ \gamma\left(  A,X_{0}%
,\beta\right)  =\beta_{0}A+\beta_{1}AX_{0}$ because both $\beta_{0}^{\ast}$
and $\beta_{1}^{\ast}$ can be estimated by g-estimation restricted to subects
with $X_{0}>\zeta.$ Thus $Y_{0}=Y-\beta_{0}^{\ast}A-\beta_{1}^{\ast}AX_{0}$
can be estimated for all subjects, including those with $A=1$ and $X_{0}%
\leq\zeta,$ because, by having the same parameters apply to subects with
$X_{0}\leq\zeta$ as to subjects with $X_{0}>\zeta,$ the model allows
extrapolation from subjects with $X_{0}>\zeta$ to subjects with $X_{0}%
\leq\zeta.$ One must weigh the benefit of extrapolation that comes with
assuming model $\gamma\left(  A,X_{0},\beta\right)  =\beta_{0}A+\beta
_{1}AX_{0}$ against the risk that the model is misspecified for subjects with
$X_{0}\leq\zeta,$ as would be the case were the true model: $\gamma\left(
A,X_{0},\beta^{\ast}\right)  =\beta_{0}^{\ast}AI\left(  X_{0}>\zeta\right)
+\beta_{1}^{\ast}AX_{0}I\left(  X_{0}>\zeta\right)  +\beta_{2}^{\ast}AI\left(
X_{0}\leq\zeta\right)  +\beta_{3}^{\ast}AX_{0}I\left(  X_{0}\leq\zeta\right)
$ with $\beta_{2}^{\ast}\ $very different from $\beta_{0}^{\ast}$ and with
$\beta_{3}^{\ast}\ $very different from $\beta_{1}^{\ast}.$ Then the
extrapolated value $Y-\beta_{0}^{\ast}A-\beta_{1}^{\ast}AX_{0}$ for $Y_{0}$
based on the misspecified model would be a badly biased estimate of the true
$Y_{0}\ $for subjects with $A=1$ and $X_{0}\leq\zeta.$ Yet, because the model
$\gamma\left(  A,X_{0},\beta\right)  =\beta_{0}A+\beta_{1}AX_{0}$ is correct
for subjects with $X_{0}>\zeta,$ there exists no valid test of model fit that
could detect the biased extrapolation when we only assume RC and CD.

Suppose now, as is true in practice, data on $X_{0}$ are unavailable for
subjects with $A=1.$ Then, under assumptions, RC and CD, without the help of a
correct RP\ SNFTM\ for $X_{0}$ whose functional form provides for
extrapolation, we can no longer identify any aspect of the distribution of
$Y_{0}$ for any identifiable subset of subjects with $A\neq0.$ This is
because, although we know that the identified quantity $E\left[
Y|X>\zeta,A=0\ \right]  $ equals $E\left[  Y_{0}|X_{0}>\zeta,A\neq0\ \right]
,$ we cannot identify which subjects with $A\neq0\ $have $X_{0}>\zeta.$

In summary, in the realistic setting of longitudinal time -dependent
exposures, the possibility of sensitivity of one's estimate of $E\left[
Y_{0}\right]  $ to model extrapolation should be examined by reestimating
$E\left[  Y_{0}\right]  $ under a variety of models that differ in both the
dimension of the parameter vectors and in functional form.

A final point is that no individual who has developed a chronic disease by
time $m$ is included in our g-estimation procedure at $m$ because
$X_{m}\left(  \psi\right)  =X<m+\varsigma$ for such subjects$.$ Thus our
estimate of the effect of exposure at time $m$ on a subject with a chronic
disease at $m$ is identified wholly by extrapolation from the effect on
subjects without chronic disease at $m.$ One approach to lessening the degree
of extrapolation is to require a subject to be rather ill before they meet the
definition of having a diagnosed chronic disease. For example, mild to
moderate diabetes or hypertension need not qualify as having a chronic
disease, especially if regular data on blood pressure and blood glucose have
been recorded in the data base, as unmeasured confounding by undiagnosed mild
to moderate diabetes or hypertension should then be minimal. If our definition
of a diagnosed chronic disease is sufficiently stringent, then few subjects
who meet the definition at $m$ will be observed to gain weight subsequent to
$m.$ In that case, model-based extrapolation must be minimal - any model-based
extrapolation is restricted to those gaining weight at $m,$ because our models
are models for the causal effect of weight gain (not loss) at $m.$ In Section
3.3 we offer a different appproach to lessening our reliance on model misspecification.

\subsubsection{\textbf{Can we replace X}$_{m}$\textbf{\ by }$\mathbf{X}$
Revisited:}

We revisit the issue of whether we could have replaced $X_{m}$ by the observed
$X$ in the CD assumption if we are willing to assume a SNFTM for $X_{m}$ so as
to link the distribution of $X$ with that of $X_{m}.$ We take the observed
data to be $\left(  \overline{A}\left(  K\right)  ,\overline{L}\left(
K+1\right)  ,Y,X\right)  .$ We will study the implications of 2 different
SNFTMs. The first SNFTM is the model discussed above that assumes
$X_{m}\left(  \psi^{\ast}\right)  $ and $X_{m}$ have the same conditional
distribution given $\left(  \overline{L}(m),\overline{A}(m)\right)  .$ The
second assumes $X_{m}\left(  \psi^{\ast}\right)  $ and $X_{m}$ have the same
conditional distribution given $\left(  \overline{L}(m),\overline
{A}(m),\overline{U}\left(  m\right)  =0\right)  $. In both cases $X_{m}\left(
\psi\right)  $ is defined by Eqs $\left(  28)-(29\right)  .$ Note a locally RP
SNFTM\ implies $X_{m}\left(  \psi^{\ast}\right)  =X_{m}$ and thus both models
are true. When rank preservation does not hold, the truth of one model does
not imply the truth of the other. We first show that when rank preservation
does not hold, under the RC assumption and the modified CD assumption in which
$X_{m}$ is replaced by the observed $X$, the parameter $\psi^{\ast}$ of the
first SNFTM may not be identifiable; however, the parameter of the second
model is estimable by g-estimation. Thus one might assume we might impose the
modified CD assumption and the second model in lieu of the unmodified CD
assumption and the first model. However we shall see this approach has a
drawback: knowledge of the parameter $\psi^{\ast}$ of the second model in
contrast to that of the first model does not help identify the parameter of
interest $E\left[  Y_{0}\right]  .$

We now show that $\psi^{\ast}$ is identifiable in the second SNFTM\ model
under RC assumption and the modified CD assumption. Note $X>\zeta+m$ is
equivalent to $X_{m}\left(  \psi\right)  =m+\int_{m}^{X}\exp\left\{
\omega\left(  \overline{A}\left(  t\right)  ,\overline{L}\left(  t\right)
,\psi\right)  \right\}  dt>m+\int_{m}^{\zeta+m}\exp\left\{  \omega\left(
\overline{A}\left(  t\right)  ,\overline{L}\left(  t\right)  ,\psi\right)
\right\}  dt.$ Thus, the modified CD assumption implies that whenever
$X_{m}\left(  \psi\right)  \geq m+\int_{m}^{\zeta+m}\exp\left\{  \omega\left(
\overline{A}\left(  t\right)  ,\overline{L}\left(  t\right)  ,\psi\right)
\right\}  dt,\ $we have $\overline{U}\left(  m\right)  =0.$ However, even if
we made the the rank preservation assumption that $X_{m}\left(  \psi^{\ast
}\right)  =X_{m},$ we cannot therefore conclude from the RC assumption that
$A\left(  m\right)  $ is independent of $X_{m}\left(  \psi^{\ast}\right)  $
given $\left(  \overline{L}(m),\overline{A}(m),X_{m}\left(  \psi^{\ast
}\right)  \geq m+\int_{m}^{\zeta+m}\exp\left\{  \omega\left(  \overline
{A}\left(  t\right)  ,\overline{L}\left(  t\right)  ,\psi\right)  \right\}
dt\right)  ;$ although this conditioning event indeed implies $\overline
{U}\left(  m\right)  =0,$ nonetheless, the conditioning event also depends on
$A\left(  t\right)  $ for $t>m,$ while the conditioning events in the
RC\ assumption do not.

However, if we let $d\left(  m,\psi,\zeta\right)  $ be the maximum value of
$X_{m}\left(  \psi\right)  $ among all subjects with $m<X<\zeta+m$ (i.e.,
subjects with $m<X_{m}\left(  \psi\right)  <m+\int_{m}^{\zeta+m}\exp\left\{
\omega\left(  \overline{A}\left(  t\right)  ,\overline{L}\left(  t\right)
,\psi\right)  \right\}  dt),\ $then $X_{m}\left(  \psi\right)  >d\left(
m,\psi,\zeta\right)  $ implies $X>\zeta+m$ and thus $\overline{U}\left(
m\right)  =0.$ Thus, we can conclude from the RC assumption that, under a rank
preserving model, $A\left(  m\right)  $ and $X_{m}\left(  \psi^{\ast}\right)
$ are independent given $\left(  \overline{L}(m),\overline{A}(m),X_{m}\left(
\psi^{\ast}\right)  \geq d\left(  m,\psi^{\ast},\zeta\right)  \right)  ,$
since $d\left(  m,\psi,\zeta\right)  $ does not vary among the subjects.
(Technically, this independence only holds if we replace $d\left(
m,\psi,\zeta\right)  $ by its probability limit. But this distinction is
unimportant for inference because $d\left(  m,\psi,\zeta\right)  $ converges
to its probability limit at a rate even faster than $n^{1/2}$ under mild
regularity conditions$.)$ Thus, given a rank preserving SNFTM, we can use
g-estimation to obtain a CAN estimate $\widetilde{\psi}$ of $\psi^{\ast}$
under the RC and modified CD assumption. Specifically, $\widetilde{\psi}$ is
the $\psi$ for which the 5 degree of freedom score test of the hypothesis
$\theta=0$ is precisely zero in the model
\begin{align*}
& E\left[  A\left(  m\right)  |\overline{L}(m),\overline{A}(m-1),\Xi\left(
m\right)  =1,\ X_{m}\left(  \psi\right)  ,X_{m}\left(  \psi\right)  >d\left(
m,\psi,\zeta\right)  \right] \\
& =\alpha^{T}W(m)+\theta^{T}Q_{m}^{\ast\ast}\left(  \psi\right)  .
\end{align*}
Suppose now rank preservation is absent. If we assume the second SNFTM, we
know $X_{m}\left(  \psi^{\ast}\right)  $ and $X_{m}$ have the same
distribution given $\overline{L}(m),\overline{A}(m),\overline{U}\left(
m\right)  =0.$ Thus, by the RC$\ $assumption $A\left(  m\right)  $ and
$X_{m}\left(  \psi^{\ast}\right)  $ are independent given $\left(
\overline{L}(m),\overline{A}(m),X_{m}\left(  \psi^{\ast}\right)  \geq d\left(
m,\psi^{\ast},\zeta\right)  \right)  ,\overline{U}\left(  m\right)  =0.$ Hence
$A\left(  m\right)  $ and $X_{m}\left(  \psi^{\ast}\right)  $ are independent
given $\left(  \overline{L}(m),\overline{A}(m),X_{m}\left(  \psi^{\ast
}\right)  \geq d\left(  m,\psi^{\ast},\zeta\right)  \right)  $ since the event

\bigskip$\overline{L}(m),\overline{A}(m),X_{m}\left(  \psi^{\ast}\right)
>d\left(  m,\psi^{\ast},\zeta\right)  $ is equivalent to the event
$\overline{L}(m),\overline{A}(m),X_{m}\left(  \psi^{\ast}\right)  >d\left(
m,\psi^{\ast},\zeta\right)  ,\overline{U}\left(  m\right)  =0$. So
$\widetilde{\psi}$ generally remains CAN for $\psi^{\ast}$.

We next show that $\psi^{\ast}$ is not identifiable in the first SNFTM\ model
under RC assumption and the modified CD assumption. Under the first SNFTM, we
only know $X_{m}\left(  \psi^{\ast}\right)  $ and $X_{m}$ have the same
distribution given $\overline{L}(m),\overline{A}(m).$ Thus $X_{m}\left(
\psi^{\ast}\right)  |\overline{L}(m),\overline{A}(m),X_{m}\left(  \psi^{\ast
}\right)  >d\left(  m,\psi^{\ast},\zeta\right)  $ has the same distribution as
$X_{m}|\overline{L}(m),\overline{A}(m),X_{m}>d\left(  m,\psi^{\ast}%
,\zeta\right)  .$

Thus, by equivalence of the conditioning events, both $X_{m}\left(  \psi
^{\ast}\right)  |\overline{L}(m),\overline{A}(m),X_{m}\left(  \psi^{\ast
}\right)  >d\left(  m,\psi^{\ast},\zeta\right)  $ and $X_{m}\left(  \psi
^{\ast}\right)  |\overline{L}(m),\overline{A}(m),X_{m}\left(  \psi^{\ast
}\right)  >d\left(  m,\psi^{\ast},\zeta\right)  ,\overline{U}\left(  m\right)
=0$ have the same distribution as $X_{m}|\overline{L}(m),\overline{A}%
(m),X_{m}>d\left(  m,\psi^{\ast},\zeta\right)  $. However, under the first
SNFTM and without rank preservation, this equality does not allow us to invoke
the RC assumption, since the the conditioning event $\overline{L}%
(m),\overline{A}(m),X_{m}>d\left(  m,\psi^{\ast},\zeta\right)  ,$
$\overline{U}\left(  m\right)  =0$ in that assumption differs from the
conditioning event $\overline{L}(m),\overline{A}(m),X_{m}>d\left(
m,\psi^{\ast},\zeta\right)  . $ Thus we cannot conclude $A\left(  m\right)  $
and $X_{m}\left(  \psi^{\ast}\right)  $ are independent given $\left(
\overline{L}(m),\overline{A}(m),X_{m}\left(  \psi^{\ast}\right)  \geq d\left(
m,\psi^{\ast},\zeta\right)  \right)  $ and so $\widetilde{\psi}$ will not be
CAN for $\psi^{\ast}$ under the first SNFTM. Indeed indentification is not possible.

Finally we argue that knowledge of the parameter $\psi^{\ast}$ of the second
model in contrast to that of the first model does not help identify $E\left[
Y_{0}\right]  $. Under the second model, we only learn the causal effect of
treatment at time $m$ among those with U$\left(  m\right)  =0.$ This does not
allow us to estimate the distributions of $X_{m}$ and thus $Y_{m}$ for all
subjects. In fact, the counterfactual distribution of $X_{m}$ and thus $Y_{m}$
are not even identified in those with $\overline{U}\left(  m\right)  =0$ for
$m<K,$ because the distributions of $X_{K}$ and thus $Y_{K}$ are not
identifiable in those with $\overline{U}\left(  m\right)  =0$ but
$\overline{U}\left(  K\right)  \neq0.$ One way to understand the difference is
that the second model does not allow for the extensive model-based
extrapolation that the first model does. Whether that is viewed as a drawback
of the second model clearly depends on one's faith in versus skepticism about
model-based extrapolation.

\subsection{Intractable Confounding In Subgroups:}

Our comparability assumption RC that $A\left(  m\right)  $ is statistically
independent of $\left(  Y_{m},X_{m}\right)  $ given both $\overline{L}%
(m)\ $and $\overline{U}\left(  m\right)  =\overline{0}\left(  m\right)  $ at
time $m$ may not be reasonable for particular, identifiable subgroups of the
study population. That is, there may be identifable subgroups in whom
confounding by unmeasured factors is intractable, where, by defintion, a
subgroup is identifable at time $m$ if membership in the subgroup is
determined by the measured variables $\overline{L}\left(  m\right)  $. In
Section 2.2.3, we noted that possible examples of such subgroups include
subjects with a diagnosed chronic disease, an age of greater than 70, or a BMI
below 21. In fact, since we have assumed $U\left(  m\right)  =1$ whenever
$X<m$,$\ $we have all along been assuming intractable confounding in the
identifiable subgroup consisting of those alive with a diagnosed chronic
disease at $m$ $\left(  X<m,T>m\right)  .$ We have therefore been excluding
them from our g-estimation procedure by requiring $X_{m}>m+\varsigma$ for
inclusion. Recall that if $X<m,$ then $X=X_{m}.$

Suppose therefore we wish to conduct an analysis where no comparability
assumption (neither CO nor RC)\ is assumed at time $m$ for subjects who, at
$m,$ have an age of greater than 70, or a BMI below 21. To do so, as described
in Ref. (16), we simply redefine $\Xi\left(  m\right)  \ $to be zero for such
subjects regardless of whether or not their $BMI\left(  m+1\right)  \geq
BMI_{\max}\left(  m\right)  $, so that they too are excluded from contributing
to g-estimation at time $m.$ In so doing, we do not change the models being
fit, the interventions under consideration, or the parameter of interest
$E\left[  Y_{0}\right]  .$ Rather we only change, by decreasing, the number of
person-time observations used to estimate our model parameters. We thereby
sacrifice some power and efficiency. As a consequence, even were willing to
make assumption CO for the remaining subjects with $\Xi\left(  m\right)  =1,$
$E\left[  Y_{0}\right]  $ would no longer be nonparametrically identified,
because model-based extrapolation is now being used for identification.

In contrast to g-estimation of SNMs, when confounding by unmeasured factors is
present in certain subgroups of the study population, neither IPTW estimation
nor the parametric g-formula estimator can be used to estimate $E\left[
Y_{0}\right]  .$

If a substantial fraction of the total person time is accrued by subjects in
identifiable subgroups with intractable confounding then either identification
will fail or, more often, the validity of one's estimate of $E\left[
Y_{0}\right]  $ will rely heavily on model extrapolation. One, albeit not
altogether satisfactory, way to decrease the reliance on model extrapolation
is to give up the attempt to estimate the parameter of interest $E\left[
Y_{0}\right]  .$ Instead, let $IN\left(  m\right)  \ $be the indicator of
intractable confounding in identifiable subgroups that takes the value 1 if at
time m a subject is in an identifiable subgroup with intractable confounding
and $0$ otherwise. Note that, based on the above discussion, subjects alive at
$m$ with $X<m$ have $IN\left(  m\right)  =1.$

Define $Y_{m}^{^{\intercal}}$ to be one's counterfactual outcome when
following the time $m^{\intercal}$dietary intervention in which a subjects
follows his observed diet up through month $m$ and is thereafter weighed
daily. On any day in month $k>m$ that his weight exceeds his previous maximum
monthly weight, the subject's caloric intake is restricted whenever $IN\left(
k\right)  =0.$ However, during months in which a subject is in an intractable
subgroup [$IN\left(  k\right)  =1]$, we place no restrictions on his diet or
weight gain, reflecting the fact that due to intractible confounding, we are
unable to estimate the effect of preventing weight gain among subjects with
$IN\left(  m\right)  =1,$ except by model extrapolation.

Our new goal becomes to estimate $E\left[  Y_{0}^{^{\intercal}}\right]  ,$ the
mean utility under an intervention in which, starting at age 18, each time $m$
a subject with $IN\left(  m\right)  =0$ exceeds his past maximum past BMI, we
calorie restrict him to prevent further weight gain. To estimate $E\left[
Y_{0}^{^{\intercal}}\right]  $ by g-estimation we proceed exactly as above
except (i) we define new variables $A^{\intercal}\left(  m\right)  $ and
$\Xi^{\intercal}\left(  m\right)  $ that equal $A\left(  m\right)  $ and
$\Xi\left(  m\right)  $ whenever $IN\left(  m\right)  =0$ but are zero
whenever $IN\left(  m\right)  =1,\ $and (ii) everywhere replace $A\left(
m\right)  $ and $\Xi\left(  m\right)  $ in our g-estimation procedure by
$A^{\intercal}\left(  m\right)  $ and $\Xi^{\intercal}\left(  m\right)  .$
Then, our algorithm that had estimated $E\left[  Y_{0}\right]  $ will now
output an estimator of $E\left[  Y_{0}^{^{\intercal}}\right]  .$ In summary,
at the cost of estimating a parameter $E\left[  Y_{0}^{^{\intercal}}\right]  $
of lesser interest than $E\left[  Y_{0}\right]  ,$ we have eliminated the
model extrapolation required to estimate the effect of weight gain among
subjects with $IN\left(  m\right)  =1.$

However, the procedure in the preceding paragraph has not eliminated the model
extrapolation required to estimate the effect of weight gain among the
intractably confounded nonidentifiable subgroup defined by $m<X_{m}%
<m+\varsigma$. As a consequence $E\left[  Y_{0}^{^{\intercal}}\right]  $, like
$E\left[  Y_{0}\right]  ,$ fails to be nonparametrically identified and must
rely on model extrapolation for identification. Specifically the subgroup with
$m<X_{m}<m+\varsigma$ is intractably confounded by $U\left(  m\right)  .$ It
is not identifiable because the observed data cannot determine membership. For
example, among subjects with $A^{\intercal}\left(  m\right)  >0,\ $we cannot
determine if a subject with $X$ observed to be between $m\ $and $m+\varsigma$
is a subject with $m<X_{m}<m+\varsigma$ versus a subject with $X_{m}%
>m+\varsigma,$ with $X$ occurring before $m+\varsigma$ owing to the causal
effect of his weight gain $A^{\intercal}\left(  m\right)  .$ As a consequence
it is not possible to assign all members of the intractably confounded
subgroup with $m<X_{m}<m+\varsigma$ the value $IN\left(  m\right)  =1$%
,$\ $while assigning all members of the unconfounded subgroup with
$m+\varsigma<$ $X_{m}$ the value $IN\left(  m\right)  =0.$ The latter subgroup
is unconfounded under the RC assumption because $m+\varsigma<$ $X_{m}$ implies
$\overline{U}\left(  m\right)  =\overline{0}\left(  m\right)  \ by$ the CD assumption.

In fact, a minimal latent period with length $\chi>\varsigma$ is required for
nonparametric identification of $E\left[  Y_{0}^{^{\intercal}}\right]  .$ For
the remainder of this subsection, assume such a MLP. Then subjects with
$m<X_{m}<m+\varsigma$ form an identifiable subgroup, as $m<X<m+\varsigma$ and
$m<X_{m}<m+\varsigma$ are equivalent$.$ Similiarly subjects with
$X_{m}>m+\varsigma$ now form an identifiable subgroup. Thus we can now assign
$IN\left(  m\right)  =1$ to all subjects in the confounded subgroup
$m<X_{m}<m+\varsigma$ and $IN\left(  m\right)  =0$ to all members of the
subgroup $X_{m}>m+\varsigma$ who were not already known to have $IN\left(
m\right)  =1$ by virtue of membership in some other intractably confounded
subgroup (eg age greater than 70.) Once we have assigned all members of the
subgroup $m<X_{m}<m+\varsigma$ the value $IN\left(  m\right)  =1$, our time
$m^{\intercal}$ dietary interventions no longer restrict the diet of any
subject of any intractably confounded subgroup. As a consequence $E\left[
Y_{0}^{^{\intercal}}\right]  $ is now nonparametrically identified. A formal
proof is given in the appendix where it is also shown that, owing to the
nonparametric identification, $E\left[  Y_{0}^{^{\intercal}}\right]  $ can be
estimated using the parametric g-formula estimator and the IPTW estimator, as
well as by g-estimation of structural nested models.

\section{Censoring:}

We now consider the realistic setting in which the available data are
$O=\overline{A}\left(  K\right)  ,\overline{L}\left(  K+1\right)  ,Y,XI\left(
X\leq K+1\right)  \ $indicating that $X$ is not observed in subjects for whom
$X$ exceeds the end of follow up time $K+1.$ For such censored subjects,
$X_{m}\left(  \psi\right)  $ is not observed. As a consequence g-estimation as
described above cannot be done. We will describe a modified estimation
procedure that can be validly applied to censored data. In the interest of
brevity, we only consider a procedure that is easy to describe. The down side
is that the procedure we describe is not as efficient as other more complex procedures.

Given a SNFTM for $X_{0}$ we can still use g-estimation to obtain CAN
estimates $\widetilde{\psi}$ of $\psi^{\ast}$ from censored data by everywhere
replacing $X_{m}\left(  \psi\right)  $ by $C_{m}\left(  \psi\right)
=\min\left(  X_{m}\left(  \psi\right)  ,K_{m}\left(  \psi\right)  \right)  ,$
in the g-estimation procedure, where
\begin{equation}
K_{m}\left(  \psi\right)  =m+\min_{\left\{  i;X_{i}>K+1\right\}  }\left\{
\int_{m}^{K+1}\exp\left\{  \omega\left(  \overline{A}_{i}\left(  t\right)
,\overline{L}_{i}\left(  t\right)  ,\psi\right)  \right\}  dt\right\}
\end{equation}
is the smallest possible value of $X_{m}\left(  \psi\right)  $ any censored
subject could possibly have (as $m+\left\{  \int_{m}^{K}\exp\left\{
\omega\left(  \overline{A}\left(  t\right)  ,\overline{L}\left(  t\right)
,\psi\right)  \right\}  dt\right\}  $ would be $X_{m}\left(  \psi\right)  $
for a given censored subject had he died, unbeknownst to us, immediately after
end of follow up.). Note $C_{m}\left(  \psi\right)  >m+\zeta$ implies
$X_{m}\left(  \psi\right)  >m+\zeta$ so our g-estimation procedures remain
restricted to subjects with $\overline{U}\left(  m\right)  =0.$

Similiarly, given a SNMM model we can still use g-estimation to obtain CAN
estimates $\widetilde{\beta}\left(  \widetilde{\psi}\right)  $ of $\psi^{\ast
}$ from censored data by replacing $X_{m}\left(  \psi\right)  $ by
$C_{m}\left(  \psi\right)  ,$ everywhere in the g-estimation procedure$.$
However there is a subtlety in interpretation. Specifically define the
function $c_{m}^{\dagger}\left(  x\right)  =\min\left(  x,K_{m}\left(
\psi^{\ast}\right)  \right)  ,$ so $c_{m}^{\dagger}\left(  X_{m}\left(
\psi^{\ast}\right)  \right)  =C_{m}\left(  \psi\right)  .$ Define $C_{m}%
=c_{m}^{\dagger}\left(  X_{m}\right)  .$ The correct definition of our SNMM
model is
\begin{align}
& E\left[  Y_{m}|A\left(  m\right)  ,\overline{L}(m),\overline{A}%
(m-1),C_{m}=x\right] \\
& =E\left[  Y_{m}\left(  \beta^{\ast},\psi^{\ast}\right)  |A\left(  m\right)
,\overline{L}(m),\overline{A}(m-1),C_{m}\left(  \psi\right)  =x\right]
\end{align}
where, now,%
\begin{equation}
Y_{m}\left(  \beta,\psi\right)  =Y-\sum_{j=m}^{K}\gamma_{m}\left[
A(j),\overline{A}\left(  j-1\right)  ,\overline{L}\left(  j\right)
,C_{j}\left(  \psi\right)  ,\beta\right]  .
\end{equation}
We refer to this model as a SNMM model for $Y_{m}|C_{m}$. Technical details
are given in the appendix. Finally a CAN estimator of $E\left[  Y_{0}\right]
$ from censored data is $\sum_{i=1}^{n}Y_{0,i}\left[  \left(  \widetilde
{\beta}\left(  \widetilde{\psi}\right)  ,\widetilde{\psi}\right)  \right]  /n$
as before with $\widetilde{\beta}\left(  \widetilde{\psi}\right)  $ and
$\widetilde{\psi}$ as redefined in this section.

\section{ Maximum Weight Gain Dietary Intervention Regimes}

We use \underline{$g$}$_{m}$ to denote a general maximum weight gain dietary
intervention regime beginning at time $m$. Mathematically \underline{$g$}%
$_{m}$ is a collection of functions \underline{$g$}$_{m}=\left\{  g_{k}\left[
\bar{a}(k-1),\bar{l}(k)\right]  ;k=m,...,K\right\}  $. Under a regime
\underline{$g$}$_{m}$ a subject follows his own observed diet history prior to
$m$ and then, for $K\geq k\geq m,$ $g_{k}\left[  \bar{a}(k-1),\bar
{l}(k)\right]  $ is a non-negative function that specifies the increase in
maximum BMI to be allowed at time $k$ for a subject with past exposure and
covariate history $\left[  \bar{a}(k-1),\bar{l}(k)\right]  .$ See the
definition in the following paragraph for a precise statement. We use $g$ as
shorthand for a regime \underline{$g$}$_{0}$ beginning at time $0.$ Note that
any regime $g=$\underline{$g$}$_{0}=\left\{  g_{k}\left[  \bar{a}(k-1),\bar
{l}(k)\right]  ;k=0,...,K\right\}  $ is naturally associated with a particular
regime \underline{$g$}$_{m}:$ the regime \underline{$g$}$_{m}=\left\{
g_{k}\left[  \bar{a}(k-1),\bar{l}(k)\right]  ;k=m,...,K\right\}  $ where one
follows his oberved diet up till time $m$ and then follows regime
\underline{$g$}$_{m}$ using functions $g_{k}\left[  \bar{a}(k-1),\bar
{l}(k)\right]  $ specified by $g $ for $k\geq m.$ Therefore, we can define the
following counterfactuals.

Let $Y_{m\ }^{g}$be a subject's utility measured at the end of follow-up when
the counterfactual intervention \underline{$g$}$_{m}\ $is followed.
Similiarly, let $\overline{BMI}_{m\ }^{g}\left(  k\right)  ,$ $\overline
{L}_{m}^{g}(k),BMI_{m,\max\ }^{g}\left(  k\right)  ,$ $\overline{A}_{m}%
^{g}(k)$ be a subject's BMI, covariate, maximum BMI and $A-$ history through
$k$ under \underline{$g$}$_{m}$. Note $\overline{BMI}_{m\ }^{g}\left(
k\right)  \in$ $\overline{L}_{m}^{g}(k).$ Then we have the following formal definition.

\textbf{Definition of a general time m maximum weight gain dietary
intervention regime }\underline{$g$}$_{m}$: The subject follows his observed
diet up to time $m\ $and from month $m$ onwards, the subject is weighed every
day: (i) if $A\left(  m\right)  =BMI\left(  m+1\right)  -BMI_{\max}\left(
m\right)  \geq g_{m}\left[  \overline{A}(k-1),\overline{L}(k)\right]  $, the
subject's caloric intake is restricted until the subject's BMI falls to below
$BMI_{\max}\left(  m\right)  +g_{m}\left[  \overline{A}(k-1),\overline
{L}(k)\right]  $; (ii) for m+1$\leq k\leq K$ if (a) $A_{m}^{g}(k)\equiv
BMI_{m\ }^{g}\left(  k+1\right)  -BMI_{m,\max\ }^{g}\left(  k\right)  \geq
g_{k}\left[  A_{m}^{g}(k-1),\overline{L}_{m}^{g}(k)\right]  ,$ the subject's
caloric intake is restricted until the subject's BMI falls to below
$BMI_{m,\max\ }^{g}\left(  k\right)  +g_{k}\left[  \overline{A}_{m}%
^{g}(k-1),\overline{L}_{m}^{g}(k)\right]  ;$ (b) if his BMI is less than
$BMI_{m,\max\ }^{g}\left(  k\right)  +g_{k}\left[  A_{m}^{g}(k-1),\overline
{L}_{m}^{g}(k)\right]  ,$ the subject is allowed to eat as he pleases without
any intervention.

Note, by definition, $\overline{L}_{m}^{g}(k)$ equals $\overline{L}_{m}(k)$
and $A_{m}^{g}(k-1)$ equals $A_{m}(k-1)$ for $k\leq m.$ Furthermore, given a
regime $g=$\underline{$g$}$_{0},$ we say a subject's observed data is
consistent with following the associated regime \underline{$g$}$_{m}$ if and
only if $A_{m}^{g}(k)\leq g_{k}\left[  \bar{A}_{m}^{g}(k-1),\overline{L}%
_{m}^{g}(k)\right]  $ for $k\geq m.$ It follows that if a subject's observed
data is consistent with following the associated regime \underline{$g$}$_{m},
$ then subject's observed data is consistent with following the associated
regime \underline{$g$}$_{k}$ for any $k>m.$

If for all $k\geq m,$ $g_{k}\left[  \bar{a}(k-1),\bar{l}(k)\right]  $ is a
constant $a\left(  k\right)  $ that does not depend on $\left(  \bar
{a}(k-1),\bar{l}(k)\right)  ,$ the regime \underline{$g$}$_{m}$ is said to be
non-dynamic or static and is written \underline{$g$}$_{m}=$\underline{$a$%
}$\left(  m\right)  .$ Otherwise it is dynamic. An intervention that allowed a
BMI gain of 0.1/12 per month (i.e., of 1 per decade) starting at time 0 (age
18) is the regime \underline{$g$}$_{0}=$\underline{$a$}$\left(  0\right)  $
with each $a\left(  m\right)  =0.1/12.$ A dynamic intervention starting at
time 0 that allows a BMI gain of 0.1/12 per month in subjects free of
hypertension, diabetes, hyperlipidemia, or clinical CHD, but of only 0.05/12
per month once a subject developed one of these risk factors is a dynamic
regime \underline{$g$}$_{0}\ $with has $g_{k}\left[  \bar{a}(k-1),\bar
{l}(k)\right]  =$0.1/12 if $\bar{l}(k)$ indicates a subject is free at $k$ of
hypertension, diabetes, hyperlipidemia, or clinical CHD and $g_{k}\left[
\bar{a}(k-1),\bar{l}(k)\right]  =.05/12$ otherwise.

The expected value $E\left[  Y_{0\ }^{g}\right]  $ is our parameter of
interest associated with the regime $g$: the expected utility had we placed in
1950 all 18 year old non-smoking American men on the maximum weight gain
intervention regime $g.$

Let $\left\lfloor t\right\rfloor $ denote the smallest integer less than or
equal to $t$ and define $b_{+}=b$ if $b\geq0$ and $b_{+}=0$ if $b<0.$ Note
because data is only obtained monthly, for any non-negative real number $t,$
$A\left(  t\right)  =A(\left\lfloor t\right\rfloor )$ and $L\left(  t\right)
=L(\left\lfloor t\right\rfloor ).$ Given a regime $g,$ let $A_{\Delta}%
^{g}\left(  t\right)  =\left[  A(\left\lfloor t\right\rfloor )-g_{\left\lfloor
t\right\rfloor }\left[  \overline{A}(\left\lfloor t\right\rfloor
-1),\overline{L}\left(  \left\lfloor t\right\rfloor \right)  \right]  \right]
_{+}$

$=\left[  BMI\left(  \left\lfloor t\right\rfloor +1\right)  -\left\{
BMI_{\max}\left(  \left\lfloor t\right\rfloor \right)  +g_{\left\lfloor
t\right\rfloor }\left[  \overline{A}(\left\lfloor t\right\rfloor
-1),\overline{L}\left(  \left\lfloor t\right\rfloor \right)  \right]
\right\}  \right]  _{+}$ so $A_{\Delta}^{g}\left(  t\right)  =0$ for all $t$
if and only if a subject's observed data is consistent with following regime
$g$ from time 0. When$\ A_{\Delta}^{g}\left(  t\right)  \neq0,A_{\Delta}%
^{g}\left(  t\right)  $ measures how much greater one's observed weight gain
is than the maximum prescribed by $g.$ Define
\begin{align}
X_{m}^{g}\left(  \psi\right)   & =m+\int_{m}^{X}\exp\left\{  \omega\left(
A_{\Delta}^{g}\left(  t\right)  ,\overline{A}\left(  t^{-}\right)
,\overline{L}\left(  t\right)  ,\psi\right)  \right\}  dt\text{ if }X>m\\
X_{m}^{g}\left(  \psi\right)   & =X\text{ if }X\leq m
\end{align}%
\begin{equation}
\ Y_{j}^{g}\left(  \beta,\psi\right)  =Y-\sum_{m=j}^{K}\gamma_{m}\left[
A_{\Delta}^{g}(m),\overline{A}\left(  m-1\right)  ,\overline{L}\left(
m\right)  ,X_{m}\left(  \psi\right)  ,\beta\right]
\end{equation}
where the functions $\omega\left(  a\left(  t\right)  ,\overline{a}\left(
t^{-}\right)  ,\overline{l}\left(  t\right)  ,\psi\right)  $ and $\gamma
_{m}\left(  a\left(  m\right)  ,\overline{a}\left(  m-1\right)  ,\overline
{l}\left(  m\right)  ,\psi\right)  $ are again known functions satisfying
$\omega\left(  a\left(  t\right)  ,\overline{a}\left(  t^{-}\right)
,\overline{l}\left(  t\right)  ,\psi\right)  =0$ if $a\left(  t\right)  =0$ or
$\psi=0\ $and $\gamma_{m}\left(  a\left(  m\right)  ,\overline{a}\left(
m-1\right)  ,\overline{l}\left(  m\right)  ,\beta\right)  =0$ if $a\left(
m\right)  =0$ or $\beta=0$.

Given a regime $g,$ we say that (49)-(50) is a correctly specified SNFTM for
$X_{m}^{g}$ and (51) is a correctly specified SNMM for $Y_{m}^{g}|X_{m}^{g}$
with true parameters $\left(  \beta^{\ast},\psi^{\ast}\right)  $ when there
exists some $\left(  \beta^{\ast},\psi^{\ast}\right)  $ such that, for each
$m,$

Assumption (i): $X_{m}^{g}$ and $X_{m}^{g}\left(  \psi^{\ast}\right)  $ have
the same conditional distribution given $\left(  A_{\Delta}^{g}(j),\overline
{A}\left(  j-1\right)  ,\overline{L}\left(  j\right)  \right)  $ and

Assumption (ii):
\begin{equation}
E\left[  Y_{m}^{g}|A_{\Delta}^{g}(m),\overline{A}(m-1),\overline{L}\left(
m\right)  ,X_{m}^{g}=x\right]  =E\left[  Y_{m}^{g}\left(  \beta^{\ast}%
,\psi^{\ast}\right)  |A_{\Delta}^{g}(m),\overline{A}(m-1),\overline{L}\left(
m\right)  ,X_{m}^{g}\left(  \psi^{\ast}\right)  =x\right]
\end{equation}

Recall $\overline{A}(m-1)$ is a function of $\overline{L}\left(  m\right)  $
and thus its appearance in the conditioning event is redundant. Define
\begin{equation}
\Xi^{g}\left(  m\right)  =1\Leftrightarrow BMI\left(  m+1\right)  \geq
BMI_{\max}\left(  m\right)  +g_{m}\left[  \overline{A}(m-1),\overline
{L}\left(  m\right)  \right]
\end{equation}
so $A_{\Delta}^{g}(m)>0$ implies $\Xi^{g}\left(  m\right)  =1.$

Given a regime $g,$ let the RC$^{g}$ assumption be the RC assumption but with
$X_{m}^{g},Y_{m}^{g},\Xi^{g}\left(  m\right)  $ replacing their counterparts
without $g$ and $A_{\Delta}^{g}\ $replacing $A.$ Let CD$^{g}$ be the
CD\ assumption but with $X_{m}^{g}$ replacing $X_{m}$ and "time m dietary
intervention" replaced by the "\underline{$g$}$_{m}$ dietary intervention".
Henceforth we assume the CD$^{g}$ and the RC$^{g}$ hold for all regimes $g.$

Suppose we carry out g-estimation as in section 3 except with $X_{m}%
^{g}\left(  \psi\right)  ,Y_{m}^{g}\left(  \beta,\psi\right)  ,\Xi^{g}\left(
m\right)  $ replacing replacing their counterparts without $g$ and $A_{\Delta
}^{g}\ $replacing $A.$ Then results of Robins (4) imply that, under the
RC$^{g}$\ and CD$^{g}$ assumptions$,\ $if the model
\[
E\left[  A_{\Delta}^{g}|\overline{L}(m),\overline{A}\left(  m-1\right)
,\Xi^{g}\left(  m\right)  =1\right]  =\alpha^{T}W(m)
\]
is correct, and our SNFTM for $X_{m}^{g}$ and SNMM for $Y_{m}^{g}|X_{m}^{g}$
are correctly specified, then $\widetilde{\psi},\widetilde{\beta}\left(
\widetilde{\psi}\right)  ,$ and $n^{-1}\sum_{i}^{n}Y_{0}^{g}\left[  \left(
\widetilde{\beta}\left(  \widetilde{\psi}\right)  ,\widetilde{\psi}\right)
\right]  $ are CAN for $\psi^{\ast},\beta^{\ast},$ and the parameter of
interest $E\left[  Y_{0}^{g}\right]  $ respectively, provided $\left(
\beta^{\ast},\psi^{\ast}\right)  $ are identified and we choose $Q_{m}\left(
\beta\right)  $ linear in $Y_{m}\left(  \beta\right)  .$

\section{Measurement Error}

In studies of the effect of a time-independent exposure, random exposure
measurement error generally leads to bias towards the null and loss of power.
However, the consequences of random exposure measurement error are much more
complex in longitudinal studies of a time-dependent exposure in the prescence
of time- varying counfounders. Specifically, in such a study, exposure history
prior to time $t$ needs to be considered as a potential confounder for the
effect of exposure at $t,$ even under the sharp null hypothesis of no causal
effect of exposure at any time on the outcome $Y$. Since random measurement
error in a confounder can cause bias in any direction, random error in
recorded BMI can, in principle, cause bias even under the null! See Ref (6).
Futhermore this random error should be seen as including not only errrors in
measurement of BMI but also short term flucuations in BMI due to illness, a
New Years resolution to loose weight, etc. These random fluctuations in BMI
may have little effect on eventual mortality, but they can easily obscure the
actual trend in someone's BMI for periods of up to a year. Thus if we use a
monthly scale of analysis as described above, the random fluctuations in BMI
may dominate any trend within a subject. Further given that past BMI must be
controlled for in the regression models for current BMI used in g-estimation,
the true correlation between past and present BMI trends within a person will
be obscured by random fluctuations, which can even result in bias away from
the null. This can occur when the confounding effect of past trends in BMI are
inadequately controlled due to the random mismeasurement in past BMI. What to do?

One approach would be to specify a complex statistical model for the
relationship between true and mismeasured BMI. At present. I tend to seriously
doubt the robustness of such an approach owing to inevitable model mispecification.

The alternative is to increase the "time\textquotedblright\ between
measurements used in the analysis from say 1 month up to as high as 5-6 years.
By increasing the time between measurements, the problem of random
fluctuations in BMI is markedly reduced, as the BMI signal (the true
difference betwen measurement occassions) is made much greater, while the
random fluctuations may not increase or may even decrease if the fluctuations
are autocorrelated on a time scale of a few to many months. The drawback of
increasing the "time\textquotedblright\ between measurements in the analysis
is that this can lead to poorer control of the confounding attributable to
evolving time-varying factors. As an example, because the temporal ordering of
events between the measurement times used in the analysis is lost; the
confounding effect of changes in exercise may be incorrectly attributed to a
causal effect of BMI.

At present I would recommend repeating one's analysis using a number of
different \ between measurements "times\textquotedblright\ and report all
results. In this way, the sensitivity of one's conclusions to the choice of
the "time\textquotedblright\ between measurements will be known. If important,
this sensitivity will stimulate further discusion and the development of
better analytic methods.

\section{Appendix 1:}

\subsection{A Formal Definition of a Joint SNFTM\ for $X_{m}$ and a SNMM for
$Y_{m}|X_{m}$}

The definition here is the alternative, more intuitive and more general
definition mentioned in the main text. The equivalence with the definitions in
the main text are proved below.

We first consider the uncensored case. The observed data is $O=\overline
{A}\left(  K\right)  ,\overline{L}\left(  K+1\right)  ,X,Y,$ where $X$ is a
continuous time to event variable and $Y$ is measured at $K+1.$ The
counterfactual data are $\left(  X_{m},Y_{m}\right)  ,$ m=0,...,K+1, denoting
$X$ and $Y$ under treatment regimes where one experiences his observed
treatment $\overline{A}\left(  m-1\right)  $ up to $m$ and then receives no
treatment (treatment level $0)$ thereafter. We make the assumption that
$X_{K+1}=X,Y_{K+1}=Y.$ The covariate $L\left(  k\right)  $ precedes $A\left(
k\right)  $ which precedes $L\left(  k+1\right)  .$

The function $x_{m}^{\dagger}\left(  x,\overline{L}\left(  m\right)
,\overline{A}\left(  m\right)  \right)  =S_{X_{m}|\overline{L}\left(
m\right)  ,\overline{A}\left(  m\right)  }^{-1}\left\{  S_{X_{m+1}%
|\overline{L}\left(  m\right)  ,\overline{A}\left(  m\right)  }\left(
x\right)  \right\}  $ is a counterfactual conditional quantile -quantile
function, where $S$ and $S^{-1} $ denote a survivor function and its inverse.
It is a standard result that $x_{m}^{\dagger}\left(  x,\overline{L}\left(
m\right)  ,\overline{A}\left(  m\right)  \right)  $ is the unique function for
which $X_{m}^{\ast}\equiv x_{m}^{\dagger}\left(  X_{m+1},\overline{L}\left(
m\right)  ,\overline{A}\left(  m\right)  \right)  $ and $X_{m}$ have the same
conditional distribution, i.e.,
\begin{equation}
X_{m}^{\ast}|\overline{L}\left(  m\right)  ,\overline{A}\left(  m\right)
\symbol{126}X_{m}|\overline{L}\left(  m\right)  ,\overline{A}\left(  m\right)
\end{equation}

Define $X_{K+1}^{\dagger}=X\ $and then recursively define $X_{m}^{\dagger
}=x_{m}^{\dagger}\left(  X_{m+1}^{\dagger},\overline{L}\left(  m\right)
,\overline{A}\left(  m\right)  \right)  .$ Robins and Wasseman (7) proved the following

\textbf{Theorem A1:}%

\begin{equation}
X_{m}|\overline{L}\left(  m\right)  ,\overline{A}\left(  m\right)
\symbol{126}X_{m}^{\dagger}|\overline{L}\left(  m\right)  ,\overline{A}\left(
m\right)
\end{equation}

where we silently take such displays to hold for all m=0,...,K.

Furthermore, Robins (8,10) and Lok (9) proved the function $x_{m}^{\dagger}$
is unique. That is if the above display holds for with $X_{m}^{\dagger}$
replaced by some $H_{m}=h_{m}\left(  H_{m+1},\overline{L}\left(  m\right)
,\overline{A}\left(  m\right)  \right)  $ and $H_{K+1}=X,$ then the function
$h_{m}$ must be the function $x_{m}^{\dagger}.$

A SNFTM for $X_{m}$ assumes
\begin{equation}
x_{m}\left(  X_{m+1},\overline{L}\left(  m\right)  ,\overline{A}\left(
m\right)  ;\psi^{^{\ast}}\right)  =x_{m}^{\dagger}\left(  X_{m+1},\overline
{L}\left(  m\right)  ,\overline{A}\left(  m\right)  \right)
\end{equation}
for a known function $x_{m}\left(  x,\overline{L}\left(  m\right)
,\overline{A}\left(  m\right)  ;\psi\right)  $ satisfying $x_{m}\left(
x,\overline{L}\left(  m\right)  ,\overline{A}\left(  m\right)  ,\psi\right)
=x$ if $\psi=0$ or $A\left(  m\right)  =0$ with $\psi^{^{\ast}}$ an unknown
parameter vector.

It follows immediately that
\begin{gather}
X_{m}\left(  \psi^{^{\ast}}\right)  |\overline{L}\left(  m\right)
,\overline{A}\left(  m\right)  \symbol{126}X_{m}^{\dagger}|\overline{L}\left(
m\right)  ,\overline{A}\left(  m\right)  \text{,}\\
with\text{ }X_{K+1}\left(  \psi^{^{\ast}}\right)  =X\text{ and }X_{m}\left(
\psi^{^{\ast}}\right)  \equiv x_{m}\left(  X_{m+1}\left(  \psi^{^{\ast}%
}\right)  ,\overline{L}\left(  m\right)  ,\overline{A}\left(  m\right)
;\psi^{^{\ast}}\right)  \text{ }%
\end{gather}

The uniqueness of $x_{m}^{\dagger}$ implies that SNFTMs as defined in the text
are also SNFTMs as defined here.

Recall $X_{m}^{\ast}\equiv x_{m}^{\dagger}\left(  X_{m+1},\overline{L}\left(
m\right)  ,\overline{A}\left(  m\right)  \right)  $ and define
\begin{equation}
\gamma_{m}^{\dagger}\left(  \overline{A}\left(  m\right)  ,\overline{L}\left(
m\right)  ,x\right)  \equiv E\left[  Y_{m+1}|\overline{A}\left(  m\right)
,\overline{L}\left(  m\right)  ,X_{m}^{\ast}=x\right]  -E\left[
Y_{m}|\overline{A}\left(  m\right)  ,\overline{L}\left(  m\right)
,X_{m}=x\right]
\end{equation}

which is equivalent to
\begin{equation}
E\left[  Y_{m+1}-\gamma_{m}^{\dagger}\left(  \overline{A}\left(  m\right)
,\overline{L}\left(  m\right)  ,X_{m}^{\ast}\right)  |\overline{A}\left(
m\right)  ,\overline{L}\left(  m\right)  ,X_{m}^{\ast}=x\right]  =E\left[
Y_{m}|\overline{A}\left(  m\right)  ,\overline{L}\left(  m\right)
,X_{m}=x\right]  .
\end{equation}

Define $Y_{K+1}^{\dagger}=Y\ $and then recursively define $Y_{m}^{\dagger
}=Y_{m+1}^{\dagger}-\gamma_{m}^{\dagger}\left(  \overline{A}\left(  m\right)
,\overline{L}\left(  m\right)  ,X_{m}^{\dagger}\right)  .$

Below we prove the following theorem.

\textbf{Theorem A2:}%

\begin{equation}
E\left[  Y_{m}^{\dagger}|\overline{L}\left(  m\right)  ,\overline{A}\left(
m\right)  ,X_{m}^{\dagger}=x\right]  =E\left[  Y_{m}|\overline{L}\left(
m\right)  ,\overline{A}\left(  m\right)  ,X_{m}=x\right]
\end{equation}

Furthermore the function $\gamma_{m}^{\dagger}$ is unique. That is if the
above display holds with $Y_{m}^{\dagger}$ replaced by some $H_{m}%
=H_{m+1}-h_{m}\left(  \overline{A}\left(  m\right)  ,\overline{L}\left(
m\right)  ,X_{m}^{\dagger}\right)  $ and $H_{K+1}=Y,$ then the function $h_{m}
$ must be the function $\gamma_{m}^{\dagger}.$

An additive SNMM for $Y_{m}|X_{m}$ assumes
\begin{equation}
\gamma_{m}\left(  \overline{A}\left(  m\right)  ,\overline{L}\left(  m\right)
,x;\beta^{^{\ast}}\right)  =\gamma_{m}^{\dagger}\left(  \overline{A}\left(
m\right)  ,\overline{L}\left(  m\right)  ,x\right)
\end{equation}
for a known function $\gamma_{m}\left(  \overline{A}\left(  m\right)
,\overline{L}\left(  m\right)  ,x;\beta\right)  $ satisfying $\gamma
_{m}\left(  \overline{A}\left(  m\right)  ,\overline{L}\left(  m\right)
,x;\beta\right)  =0$ if $\beta=0$ or $A\left(  m\right)  =0$ with $\beta
^{\ast}$ an unknown parameter vector.

It follows immediately that
\begin{gather}
E\left[  Y_{m}\left(  \beta^{^{\ast}},\psi^{^{\ast}}\right)  |\overline
{L}\left(  m\right)  ,\overline{A}\left(  m\right)  ,X_{m}\left(  \psi
^{^{\ast}}\right)  =x\right]  =E\left[  Y_{m}|\overline{L}\left(  m\right)
,\overline{A}\left(  m\right)  ,X_{m}=x\right]  \text{,}\\
\text{with }Y_{K+1}\left(  \beta^{^{\ast}},\psi^{^{\ast}}\right)  =Y\text{ and
}Y_{m}\left(  \beta^{^{\ast}},\psi^{^{\ast}}\right)  \equiv Y_{m+1}\left(
\beta^{^{\ast}},\psi^{^{\ast}}\right)  -\gamma_{m}\left(  \overline{A}\left(
m\right)  ,\overline{L}\left(  m\right)  ,X_{m}\left(  \psi^{^{\ast}}\right)
;\beta^{^{\ast}}\right)
\end{gather}

The uniqueness of $\gamma_{m}^{\dagger}$ implies that an additive SNMM for
$Y_{m}|X_{m}$ as defined in the text is equivalent to the additive SNMM for
$Y_{m}|X_{m}$ as defined here.

\textbf{Proof of Theorem A2:} By backward induction.

Case 1:m=K; $E\left[  Y_{K}^{\dagger}|\overline{L}\left(  K\right)
,\overline{A}\left(  K\right)  ,X_{K}^{\dagger}=x\right]  $

=$E\left[  Y_{K+1}-\gamma_{K}^{\dagger}\left(  \overline{A}\left(  K\right)
,\overline{L}\left(  K\right)  ,X_{K}^{\ast}\right)  |\overline{A}\left(
K\right)  ,\overline{L}\left(  K\right)  ,X_{K}^{\dagger}=x\right]  $

=$E\left[  Y_{K+1}-\gamma_{K}^{\dagger}\left(  \overline{A}\left(  K\right)
,\overline{L}\left(  K\right)  ,X_{K}^{\ast}\right)  |\overline{A}\left(
K\right)  ,\overline{L}\left(  K\right)  ,X_{K}^{\ast}=x\right]  =E\left[
Y_{K}|\overline{A}\left(  K\right)  ,\overline{L}\left(  K\right)
,X_{K}=x\right]  $

where the first equality uses the definition of $Y_{K}^{\dagger}$ and that
$Y_{K+1}=Y=Y_{K+1}^{\dagger}$, the second uses that $X_{K}^{\ast}%
=X_{K}^{\dagger}$ by $X_{K+1}=X=X_{K+1}^{\dagger},$ and the third is the
definition of $\gamma_{K}^{\dagger}\left(  \overline{A}\left(  K\right)
,\overline{L}\left(  K\right)  ,X_{K}^{\ast}\right)  .$

Case 2: Assume true for m. We prove true for m+1.

We will require the following Lemma

Lemma:
\[
f\left(  \overline{L}\left(  m+1\right)  ,\overline{A}\left(  m+1\right)
|\overline{L}\left(  m\right)  ,\overline{A}\left(  m\right)  ,X_{m}^{\dagger
}=x\right)  =f\left(  \overline{L}\left(  m+1\right)  ,\overline{A}\left(
m+1\right)  |\overline{L}\left(  m\right)  ,\overline{A}\left(  m\right)
,X_{m}^{\ast}=x\right)
\]

Proof: $f\left(  \overline{L}\left(  m+1\right)  ,\overline{A}\left(
m+1\right)  |\overline{L}\left(  m\right)  ,\overline{A}\left(  m\right)
,X_{m}^{\dagger}=x\right)  $

=$\left\{  f\left(  X_{m}^{\dagger}=x|\overline{L}\left(  m\right)
,\overline{A}\left(  m\right)  \right)  \right\}  ^{-1}f\left(  \overline
{L}\left(  m+1\right)  ,\overline{A}\left(  m+1\right)  ,X_{m}^{\dagger
}=x|\overline{L}\left(  m\right)  ,\overline{A}\left(  m\right)  \right)  $

=$\left\{  f\left(  X_{m}^{\ast}=x|\overline{L}\left(  m\right)  ,\overline
{A}\left(  m\right)  \right)  \right\}  ^{-1}\times$

$\frac{\partial x_{m}^{\dagger-1}\left(  x,\overline{L}\left(  m\right)
,\overline{A}\left(  m\right)  \right)  }{\partial x}f\left(  \overline
{L}\left(  m+1\right)  ,\overline{A}\left(  m+1\right)  ,X_{m+1}^{\dagger
}=x_{m}^{\dagger-1}\left(  x,\overline{L}\left(  m\right)  ,\overline
{A}\left(  m\right)  \right)  |\overline{L}\left(  m\right)  ,\overline
{A}\left(  m\right)  \right)  $

=$\left\{  f\left(  X_{m}^{\ast}=x|\overline{L}\left(  m\right)  ,\overline
{A}\left(  m\right)  \right)  \right\}  ^{-1}$

$\frac{\partial x_{m}^{\dagger-1}\left(  x,\overline{L}\left(  m\right)
,\overline{A}\left(  m\right)  \right)  }{\partial x}f\left(  \overline
{L}\left(  m+1\right)  ,\overline{A}\left(  m+1\right)  ,X_{m+1}%
=x_{m}^{\dagger-1}\left(  x,\overline{L}\left(  m\right)  ,\overline{A}\left(
m\right)  \right)  |\overline{L}\left(  m\right)  ,\overline{A}\left(
m\right)  \right)  $

=$\left\{  f\left(  X_{m}^{\ast}=x|\overline{L}\left(  m\right)  ,\overline
{A}\left(  m\right)  \right)  \right\}  ^{-1}f\left(  \overline{L}\left(
m+1\right)  ,\overline{A}\left(  m+1\right)  ,X_{m}^{\ast}=x|\overline
{L}\left(  m\right)  ,\overline{A}\left(  m\right)  \right)  $

=$f\left(  \overline{L}\left(  m+1\right)  ,\overline{A}\left(  m+1\right)
|\overline{L}\left(  m\right)  ,\overline{A}\left(  m\right)  ,X_{m}^{\dagger
}=x\right)  $

where the first equality is by Bayes rule, the second by $X_{m}^{\ast}$ and
$X_{m}^{\dagger}$ both having the same law as $X_{m}$ conditional on
$\overline{L}\left(  m\right)  ,\overline{A}\left(  m\right)  \ $and a change
of variables from $X_{m}^{\dagger}$ to $X_{m+1}^{\dagger},$ the third by
$\left(  X_{m+1}^{\dagger},\overline{L}\left(  m+1\right)  ,\overline
{A}\left(  m+1\right)  \right)  $ and $\left(  X_{m+1},\overline{L}\left(
m+1\right)  ,\overline{A}\left(  m+1\right)  \right)  $ having same joint
distribution, the fourth by the definition of $X_{m}^{\ast}$ and a change of
variables, and the 5th by Bayes rule.

Now to the proof of case 2:

$E\left[  Y_{m}^{\dagger}|\overline{L}\left(  m\right)  ,\overline{A}\left(
m\right)  ,X_{m}^{\dagger}=x\right]  $

$=E\left[  Y_{m+1}^{\dagger}-\gamma_{m}^{\dagger}\left(  \overline{A}\left(
m\right)  ,\overline{L}\left(  m\right)  ,X_{m}^{\dagger}\right)
|\overline{L}\left(  m\right)  ,\overline{A}\left(  m\right)  ,X_{m}^{\dagger
}=x\right]  $

=E$\left\{  E\left[  Y_{m+1}^{\dagger}|\overline{L}\left(  m+1\right)
,\overline{A}\left(  m+1\right)  ,X_{m}^{\dagger}|\overline{L}\left(
m\right)  ,\overline{A}\left(  m\right)  ,X_{m}^{\dagger}=x\right]  \right\}
-\gamma_{m}^{\dagger}\left(  \overline{A}\left(  m\right)  ,\overline
{L}\left(  m\right)  ,x\right)  $

$=E\left\{  E\left[  Y_{m+1}^{\dagger}|\overline{L}\left(  m+1\right)
,\overline{A}\left(  m+1\right)  ,X_{m+1}^{\dagger}=x_{m}^{\dagger-1}\left(
x,\overline{L}\left(  m\right)  ,\overline{A}\left(  m\right)  \right)
|\overline{L}\left(  m\right)  ,\overline{A}\left(  m\right)  ,X_{m}^{\dagger
}=x\right]  \right\}  $

$-\gamma_{m}^{\dagger}\left(  \overline{A}\left(  m\right)  ,\overline
{L}\left(  m\right)  ,x\right)  $

=$E\left\{  E\left[  Y_{m+1}|\overline{L}\left(  m+1\right)  ,\overline
{A}\left(  m+1\right)  ,X_{m+1}=x_{m}^{\dagger-1}\left(  x,\overline{L}\left(
m\right)  ,\overline{A}\left(  m\right)  \right)  \right]  |\overline
{L}\left(  m\right)  ,\overline{A}\left(  m\right)  ,X_{m}^{\dagger
}=x\right\}  $

$-\gamma_{m}^{\dagger}\left(  \overline{A}\left(  m\right)  ,\overline
{L}\left(  m\right)  ,x\right)  $

=$E\left\{  E\left[  Y_{m+1}|\overline{L}\left(  m+1\right)  ,\overline
{A}\left(  m+1\right)  ,X_{m+1}=x_{m}^{\dagger-1}\left(  x,\overline{L}\left(
m\right)  ,\overline{A}\left(  m\right)  \right)  \right]  |\overline
{L}\left(  m\right)  ,\overline{A}\left(  m\right)  ,X_{m}^{\ast}=x\right\}  $

$-\gamma_{m}^{\dagger}\left(  \overline{A}\left(  m\right)  ,\overline
{L}\left(  m\right)  ,x\right)  $

=$E\left\{  E\left[  Y_{m+1}|\overline{L}\left(  m+1\right)  ,\overline
{A}\left(  m+1\right)  ,X_{m}^{\ast}=x\right]  |\overline{L}\left(  m\right)
,\overline{A}\left(  m\right)  ,X_{m}^{\ast}=x\right\}  $

$-\gamma_{m}^{\dagger}\left(  \overline{A}\left(  m\right)  ,\overline
{L}\left(  m\right)  ,x\right)  $

=$E\left[  Y_{m+1}|\overline{L}\left(  m\right)  ,\overline{A}\left(
m\right)  ,X_{m}^{\ast}=x\right]  -\gamma_{m}^{\dagger}\left(  \overline
{A}\left(  m\right)  ,\overline{L}\left(  m\right)  ,x\right)  $

=$E\left[  Y_{m}|\overline{L}\left(  m\right)  ,\overline{A}\left(  m\right)
,X_{m}=x\right]  $,

where the first equality is by the definition of $Y_{m}^{\dagger},$ the second
by iterated expectations, the third by the definiton of $X_{m}^{\dagger},$ the
fourth by the induction hypothesis, the fifth by the preceding Lemma, the
sixth by the definition of $X_{m}^{\ast},$the 7th by the laws of probability,
and the eighth by the definition of $\gamma_{m}^{\dagger}\left(  \overline
{A}\left(  m\right)  ,\overline{L}\left(  m\right)  ,x\right)  .$ Uniqueness
is proved as in Refs (4,10) and is omitted.

\bigskip

Additive SNMM for $Y_{m}|X_{m}$ may not be not appropriate for analyzing
censored data due to adminstrative censoring of $X$ at time K as discussed in
the text. As indicated in Section 4, our approach requires that we consider a
broader class of SNMM models which we now describe.

Consider a collection of functions $c_{m}^{\dagger}\left(  x,\overline
{A}\left(  m\right)  ,\overline{L}\left(  m\right)  \right)  $ indexed by $m$
and define $C_{m}^{\ast}=c_{m}^{\dagger}\left(  X_{m}^{\ast},\overline
{A}\left(  m\right)  ,\overline{L}\left(  m\right)  \right)  ,C_{m}^{\dagger
}=c_{m}^{\dagger}\left(  X_{m}^{\dagger},\overline{A}\left(  m\right)
,\overline{L}\left(  m\right)  \right)  ,C_{m}=c_{m}^{\dagger}\left(
X_{m},\overline{A}\left(  m\right)  ,\overline{L}\left(  m\right)  \right)  ,$
and $C_{m}\left(  \psi\right)  =c_{m}^{\dagger}\left(  X_{m},\overline
{A}\left(  m\right)  ,\overline{L}\left(  m\right)  ,\psi\right)  .$ For fixed
$\overline{A}\left(  m\right)  ,\overline{L}\left(  m\right)  ,c_{m}^{\dagger
}\left(  x,\overline{A}\left(  m\right)  ,\overline{L}\left(  m\right)
\right)  $ need not be a 1-1 function of $x.$ The approach described in the
text for handling right censoring of $X$ at tiime $K+1$ amounts to the
selection of particular functions $c_{m}^{\dagger}$ that guarantee that
$C_{m}\left(  \psi\right)  $ is an observable (i.e. uncensored) random variable.

Let $c_{m}$ denote an arbitrary element in the range of $c_{m}^{\dagger}.$ Redefine%

\begin{equation}
\gamma_{m}^{\dagger}\left(  \overline{A}\left(  m\right)  ,\overline{L}\left(
m\right)  ,c_{m}\right)  \equiv E\left[  Y_{m+1}|\overline{A}\left(  m\right)
,\overline{L}\left(  m\right)  ,C_{m}^{\ast}=c_{m}\right]  -E\left[
Y_{m}|\overline{A}\left(  m\right)  ,\overline{L}\left(  m\right)
,C_{m}=c_{m}\right]
\end{equation}

which is equivalent to
\begin{equation}
E\left[  Y_{m+1}-\gamma_{m}^{\dagger}\left(  \overline{A}\left(  m\right)
,\overline{L}\left(  m\right)  ,C_{m}^{\ast}\right)  |\overline{A}\left(
m\right)  ,\overline{L}\left(  m\right)  ,C_{m}^{\ast}=c_{m}\right]  =E\left[
Y_{m}|\overline{A}\left(  m\right)  ,\overline{L}\left(  m\right)
,C_{m}=c_{m}\right]  .
\end{equation}

Define $Y_{K+1}^{\dagger}=Y\ $and then recursively redefine $Y_{m}^{\dagger
}=Y_{m+1}^{\dagger}-\gamma_{m}^{\dagger}\left(  \overline{A}\left(  m\right)
,\overline{L}\left(  m\right)  ,C_{m}^{\dagger}\right)  .$

Below we prove the following theorem.

\textbf{Theorem A3: }Suppose, for $m=0,....,K-1,$
\begin{equation}
c_{m+1}^{\dagger}\left(  x,\overline{A}\left(  m+1\right)  ,\overline
{L}\left(  m+1\right)  \right)  =d_{m+1}\left[  c_{m}^{\dagger}\left\{
x_{m}^{\dagger}\left(  x,\overline{A}\left(  m\right)  ,\overline{L}\left(
m\right)  \right)  ,\overline{A}\left(  m\right)  ,\overline{L}\left(
m\right)  \right\}  ,\overline{A}\left(  m+1\right)  ,\overline{L}\left(
m+1\right)  \right]
\end{equation}

for some function $d_{m+1}\left(  c_{m},\overline{A}\left(  m+1\right)
,\overline{L}\left(  m+1\right)  \right)  ,$ where $x_{m}^{\dagger}\left(
x,\overline{A}\left(  m\right)  ,\overline{L}\left(  m\right)  \right)  $ is
as defined previously. That is, the function $c_{m+1}^{\dagger}=d_{m+1}\circ
c_{m}^{\dagger}\circ x_{m}^{\dagger}.$ Then%

\begin{equation}
E\left[  Y_{m}^{\dagger}|\overline{L}\left(  m\right)  ,\overline{A}\left(
m\right)  ,C_{m}^{\dagger}=x\right]  =E\left[  Y_{m}|\overline{L}\left(
m\right)  ,\overline{A}\left(  m\right)  ,C_{m}=c_{m}\right]
\end{equation}

Furthermore the function $\gamma_{m}^{\dagger}$ is unique. That is, if the
above display holds with $Y_{m}^{\dagger}$ replaced by some $H_{m}%
=H_{m+1}-h_{m}\left(  \overline{A}\left(  m\right)  ,\overline{L}\left(
m\right)  ,C_{m}^{\dagger}\right)  $ and $H_{K+1}=Y,$ then the function $h_{m}
$ must be the function $\gamma_{m}^{\dagger}.$

Remark: The need for Eq 67 in the supposition to Theorem A.3 is because the
function $c_{m}^{\dagger}\left(  x,\overline{A}\left(  m\right)  ,\overline
{L}\left(  m\right)  \right)  $ need not be a 1-1 function of $x.$

An additive SNMM for $Y_{m}|C_{m}$ assumes
\begin{equation}
\gamma_{m}\left(  \overline{A}\left(  m\right)  ,\overline{L}\left(  m\right)
,c_{m};\beta^{^{\ast}}\right)  =\gamma_{m}^{\dagger}\left(  \overline
{A}\left(  m\right)  ,\overline{L}\left(  m\right)  ,c_{m}\right)
\end{equation}
for a known function $\gamma_{m}\left(  \overline{A}\left(  m\right)
,\overline{L}\left(  m\right)  ,c_{m};\beta\right)  $ satisfying $\gamma
_{m}\left(  \overline{A}\left(  m\right)  ,\overline{L}\left(  m\right)
,c_{m};\beta\right)  =0$ if $\beta=0$ or $A\left(  m\right)  =0$ with
$\beta^{\ast}$ an unknown parameter vector.

It follows immediately that
\begin{gather}
E\left[  Y_{m}\left(  \beta^{^{\ast}},\psi^{^{\ast}}\right)  |\overline
{L}\left(  m\right)  ,\overline{A}\left(  m\right)  ,C_{m}\left(  \psi
^{^{\ast}}\right)  =x\right]  =E\left[  Y_{m}|\overline{L}\left(  m\right)
,\overline{A}\left(  m\right)  ,C_{m}=x\right]  \text{,}\\
\text{with }Y_{K+1}\left(  \beta^{^{\ast}},\psi^{^{\ast}}\right)  =Y\text{ and
}Y_{m}\left(  \beta^{^{\ast}},\psi^{^{\ast}}\right)  \equiv Y_{m+1}\left(
\beta^{^{\ast}},\psi^{^{\ast}}\right)  -\gamma_{m}\left(  \overline{A}\left(
m\right)  ,\overline{L}\left(  m\right)  ,C_{m}\left(  \psi^{^{\ast}}\right)
;\beta^{^{\ast}}\right)
\end{gather}

Proof of $A.3:$ We only describe where the proof differs from that of its
special case Theorem A.2. The proof is essentially identical except for the
replacement of $X_{m}$ by $C_{m}$, $x$ by $c_{m},$ and $x_{m}^{\dagger
-1}\left(  x,\overline{L}\left(  m\right)  ,\overline{A}\left(  m\right)
\right)  $ by
\begin{equation}
c_{m+1}^{\dagger}\left\{  x_{m}^{\dagger-1}\left(  c_{m}^{\dagger-1}\left(
c_{m},\overline{L}\left(  m\right)  ,\overline{A}\left(  m\right)  \right)
,\overline{L}\left(  m\right)  ,\overline{A}\left(  m\right)  \right)
,\overline{L}\left(  m+1\right)  ,\overline{A}\left(  m+1\right)  \right\}
\end{equation}
$.$The only problematic point is that, now, since $x_{m}^{\dagger-1}\left(
c_{m}^{\dagger-1}\left(  c_{m},\overline{L}\left(  m\right)  ,\overline
{A}\left(  m\right)  \right)  ,\overline{L}\left(  m\right)  ,\overline
{A}\left(  m\right)  \right)  $ is the subset $\mathcal{X}\left(
c_{m},\overline{L}\left(  m\right)  ,\overline{A}\left(  m\right)  \right)
=\left\{  x:c_{m}^{\dagger}\left(  x_{m}^{\dagger}\left(  x,\overline
{L}\left(  m\right)  ,\overline{A}\left(  m\right)  \right)  ,\overline
{L}\left(  m\right)  ,\overline{A}\left(  m\right)  \right)  =c_{m}\right\}  $
of the nonnegative real line, a necessary condition for

$c_{m+1}^{\dagger}\left\{  x_{m}^{\dagger-1}\left(  c_{m}^{\dagger-1}\left(
c_{m},\overline{L}\left(  m\right)  ,\overline{A}\left(  m\right)  \right)
,\overline{L}\left(  m\right)  ,\overline{A}\left(  m\right)  \right)
,\overline{L}\left(  m+1\right)  ,\overline{A}\left(  m+1\right)  \right\}  $
to be a well defined function is that every element of the set $\mathcal{X}%
\left(  c_{m},\overline{L}\left(  m\right)  ,\overline{A}\left(  m\right)
\right)  $ has the same image under $c_{m+1}^{\dagger}\left(  \cdot
,\overline{L}\left(  m+1\right)  ,\overline{A}\left(  m+1\right)  \right)  $.
The choice

$c_{m+1}^{\dagger}\left(  \cdot,\overline{L}\left(  m+1\right)  ,\overline
{A}\left(  m+1\right)  \right)  \equiv c_{m}^{\dagger}\left\{  x_{m}^{\dagger
}\left(  \cdot,\overline{A}\left(  m\right)  ,\overline{L}\left(  m\right)
\right)  ,\overline{A}\left(  m\right)  ,\overline{L}\left(  m\right)
\right\}  $ satisfies this constraint with the image being $c_{m}$ itself.
More generally, the choice of $c_{m+1}^{\dagger}\left(  \cdot,\overline
{L}\left(  m+1\right)  ,\overline{A}\left(  m+1\right)  \right)  \ $given in
Eq 67 satisifies the constraint.

\section{Appendix 2: Estimation of Effects with the Parametric G-formula and
IPTW \textbf{When a Sufficiently Long Minimal Latent Period Exists:}}

In this section we show that the the parametric G-formula and IPTW can be used
to estimate certain causal effects when their exists a sufficiently long
minimal latent period. We begin with a preliminary discussion of these two
methods of estimation.

\subsection{Preliminaries:}

In this preliminary discussion, we assume that, as in Section 3.1.1, there is
neither confounding by pre-clinical disease nor a minimal latent period.
Specifically we assume, for each regime $g,$ the $CO^{g}$ assumption that, for
each $j,$ $\left(  Y_{0}^{g},X_{0}^{g}\right)  \amalg A_{\Delta}^{g}\left(
j\right)  |\overline{L}\left(  j\right)  ,\overline{A}_{\Delta}^{g}\left(
j-1\right)  =\overline{0}\left(  j-1\right)  ,\Xi^{g}\left(  j\right)  =1$
holds, with $\Xi^{g}\left(  m\right)  $ defined in Equation (53).

\textbf{Recoding:} Without loss of generality, we henceforth redefine (ie
recode) $\overline{L}\left(  j\right)  $ such that $\Xi^{g}\left(  j\right)  $
is now one of the components of $\overline{L}\left(  j\right)  $ but we remove
from $\overline{L}\left(  j\right)  $ the components corresponding to $X,$ ie
the components $\left(  XI\left(  X\leq j\right)  ,I\left(  X\leq j\right)
\right)  .$ Then we can write the $CO^{g}$ assumption as%
\begin{equation}
CO^{g}:\left(  Y_{j}^{g},X_{j}^{g}\right)  \amalg A_{\Delta}^{g}\left(
j\right)  |\overline{L}\left(  j\right)  ,\overline{A}_{\Delta}^{g}\left(
j-1\right)  =\overline{0}\left(  j-1\right)  ,\left(  XI\left(  X\leq
j\right)  ,I\left(  X\leq j\right)  \right)
\end{equation}
$\sin$ce, from their definitions, $\Xi^{g}\left(  j\right)  =0$ implies
$A_{\Delta}^{g}\left(  j\right)  =0.$ The $CO^{g}$ assumption implies $\ $%
\begin{equation}
\left(  Y_{0}^{g},X_{0}^{g}\right)  \amalg A_{\Delta}^{g}\left(  j\right)
|\overline{L}\left(  j\right)  ,\overline{A}_{\Delta}^{g}\left(  j-1\right)
=\overline{0}\left(  j-1\right)  ,\left(  XI\left(  X\leq j\right)  ,I\left(
X\leq j\right)  \right)
\end{equation}
since $\overline{A}_{\Delta}^{g}\left(  j-1\right)  =\overline{0}\left(
j-1\right)  $ implies $\left(  Y_{j}^{g},X_{j}^{g}\right)  =\left(  Y_{0}%
^{g},X_{0}^{g}\right)  $. This last display is the standard definiton of no
unmeasured confounding given $\left(  \overline{L}\left(  j\right)  ,\left(
XI\left(  X\leq j\right)  ,I\left(  X\leq j\right)  \right)  \right)  $ for
the effect of $A_{\Delta}^{g}\left(  j\right)  $ on the counterfactuals
$Y_{0}^{g},X_{0}^{g}.$ Let $\lambda\left(  u|\cdot\right)  =\lim
_{h\rightarrow0}pr\left[  u\leq X<u+h|\cdot,u\leq X\right]  /h$ be the
conditional hazard of $X$ given $\cdot.$

Robins (14,15) proves that Eq 74 implies that $S_{X_{0}^{g}}\left(  u\right)
\equiv pr\left(  X_{0}^{g}>u\right)  $ is identified via
\begin{align}
& S_{X_{0}^{g}}\left(  u\right) \nonumber\\
& =\int\cdot\cdot\cdot\int\exp\left\{  -\int_{0}^{%
\operatorname{u}%
}\lambda\left(  t|\overline{L}\left(  t\right)  ,\overline{A}_{\Delta}%
^{g}\left(  t\right)  =\overline{0}\right)  dt\right\}  \times\\
&
{\displaystyle\prod\limits_{m=0}^{m=\left\lfloor u\right\rfloor }}
dF\left[  L\left(  m\right)  |\overline{L}\left(  m-1\right)  ,\overline
{A}_{\Delta}^{g}\left(  m-1\right)  =\overline{0},X>m\right] \nonumber\\
& =E\left[  I\left\{  X>u\right\}  I\left\{  \overline{A}_{\Delta}^{g}\left(
u\right)  =\overline{0}\right\}  \mathbb{W}^{g,\ast}\left(  u\right)  \right]
\\
& with\nonumber\\
\mathbb{W}^{g,\ast}\left(  u\right)   & =1/%
{\displaystyle\prod\limits_{m=0}^{m=\left\lfloor u\right\rfloor }}
pr\left[  A_{\Delta}^{g}\left(  m\right)  =0|\overline{L}\left(  m\right)
,\overline{A}_{\Delta}^{g}\left(  m-1\right)  ,X>m\right]  ,
\end{align}
where the first formula for $S_{X_{0}^{g}}\left(  u\right)  $ is referred to
as the g-computation algorithm formula (g-formula, for short) and the second
formula as the IPTW formula. To shorten the formulae we have written
$\overline{0}$ as shorthand for $\overline{0}\left(  t\right)  $ when the time
$t$ is clear. In fact Robins $(14,15)$ shows that the assumption
\begin{equation}
X_{0}^{g}\amalg A_{\Delta}^{g}\left(  j\right)  |\overline{L}\left(  j\right)
,\overline{A}_{\Delta}^{g}\left(  j-1\right)  =\overline{0}\left(  j-1\right)
,X>j,
\end{equation}
which is implied by the assumption of Eq (74), suffices to establish the
identifying formulae. To estimate $S_{X_{0}^{g}}\left(  u\right)  $ we can
using either the parametric g-formula estimator that replaces the unknowns
$\lambda\left(  t|\overline{L}\left(  t\right)  ,\overline{A}_{\Delta}%
^{g}\left(  t\right)  =\overline{0}\right)  $ and $f\left[  L\left(  m\right)
|\overline{L}\left(  m-1\right)  ,\overline{A}_{\Delta}^{g}\left(  m-1\right)
=\overline{0},X>m\right]  $ in the first formula by estimates based on
parametric models or the IPTW estimator that replaces the unknown $pr\left[
A_{\Delta}^{g}\left(  m\right)  =0|\overline{L}\left(  m-1\right)
,\overline{A}_{\Delta}^{g}\left(  m-1\right)  =\overline{0},X>m\right]  $ in
the second formula with a parametric estimate and the unknown expectation with
a sample average. Both approaches are alternatives to g-estimation of
structural nested models.

Robins (14,15) proves $E\left[  Y_{0}^{g}\right]  $ is identified under the
assumption of Eq (74) by
\begin{gather*}
E\left[  Y_{0}^{g}\right]  =\int_{0}^{K+1}dx\times\\
\int\cdot\cdot\cdot\int\text{ }\lambda_{X}\left(  x|\overline{L}\left(
x\right)  ,\overline{A}_{\Delta}^{g}\left(  x\right)  =\overline{0}\right)
\exp\left\{  -\int_{0}^{x}\lambda_{X}\left(  t|\overline{L}\left(  t\right)
,\overline{A}_{\Delta}^{g}\left(  t\right)  =\overline{0}\right)  dt\right\}
\times\\%
{\displaystyle\prod\limits_{m=0}^{m=\left\lfloor x\right\rfloor }}
dF\left[  L\left(  m\right)  |\overline{L}\left(  m-1\right)  ,\overline
{A}_{\Delta}^{g}\left(  m-1\right)  =\overline{0},X>m\right]  \times\\%
{\displaystyle\prod\limits_{m=\left\lfloor x+1\right\rfloor }^{K+!}}
dF\left[  L\left(  m\right)  |\overline{L}\left(  m-1\right)  ,\overline
{A}_{\Delta}^{g}\left(  m-1\right)  =\overline{0},X=x\right]  \times\\
E\left[  Y|\overline{L}\left(  K+1\right)  ,\overline{A}_{\Delta}^{g}\left(
K\right)  =\overline{0},X=x\right] \\
=E\left[  YI\left\{  \overline{A}_{\Delta}^{g}\left(  K\right)  =\overline
{0}\right\}  \mathbb{W}^{g,\ast}\right] \\
with\\
\mathbb{W}^{g,\ast}=\mathbb{W}^{g,\ast}\left(  X\right)  \left\{  1/%
{\displaystyle\prod\limits_{m=\left\lfloor X+1\right\rfloor }^{K}}
pr\left[  A_{\Delta}^{g}\left(  m\right)  =0|\overline{L}\left(  m\right)
,\overline{A}_{\Delta}^{g}\left(  m-1\right)  =0,X,X<m\right]  \right\}
\end{gather*}

In the above formulae, we have assumed for simplicity that $X$ has support on
$(0.K+1)$ so censoring for $X$ is absent.

We next consider whether $S_{X_{0}^{g}}\left(  u\right)  $ and $E\left[
Y_{0}^{g}\right]  $ remain identified in the prescence of confounding by
pre-clinical disease and a sufficiently long minimal latent period (MLP).

\subsection{Identification and Estimation of $S_{X_{0}^{g}}\left(  u\right)
:$}

The following theorem establishes the identification of $S_{X_{0}^{g}}\left(
u\right)  .$ First note under our recoding, the RC$^{g}$ assumption becomes
\begin{equation}
RC^{g}:\left(  Y_{j}^{g},X_{j}^{g}\right)  \amalg A_{\Delta}^{g}\left(
j\right)  |\overline{L}\left(  j\right)  ,\overline{A}_{\Delta}^{g}\left(
j-1\right)  ,\overline{U}\left(  j\right)  =0,\left(  XI\left(  X\leq
j\right)  ,I\left(  X\leq j\right)  \right)  ,
\end{equation}

\textbf{Theorem A4:} Given a regime $g,$ let a g-specific MLP satisfy the
definition of a MLP\ of Sec. 3.2.1 except with $X_{k}$ and $X_{m}$ replaced by
by $X_{k}^{g}$ and $X_{m}^{g}\ $and $A\left(  m\right)  $ replaced by
$A_{\Delta}^{g}\left(  m\right)  .$ Suppose $A_{\Delta}^{g}\left(  m\right)  $
has a $g-specific$ MLP of $\chi$ months for its effect on $X$ where $\chi$
exceeds the time $\varsigma$ in the CD$^{g}$ assumption. Then, under the
CD$^{g}$ and RC$^{g}$\ assumptions, $S_{X_{0}^{g}}\left(  u\right)  $ remains
identified by both the g-formula and the IPTW\ formula when the recoded
$L\left(  t\right)  $ and $A_{\Delta}^{g}\left(  t\right)  $ are redefined as
$L^{\dagger}\left(  t\right)  $ and $A_{\Delta}^{g,\dagger}\left(  t\right)  $ where%

\begin{equation}
L^{\dagger}\left(  t\right)  =L\left(  t-\chi\right)  ,A_{\Delta}^{g,\dagger
}\left(  t\right)  =A_{\Delta}^{g,\dagger}\left(  t-\chi\right)
\end{equation}

The theorem thus states that the identifying formulas are the usual g-formula
and IPTW formula except we replace both the treatment variable $A_{\Delta}%
^{g}\left(  t\right)  $ and the covariate variable $L^{\dagger}\left(
t\right)  $ by their values $\chi$ time units earlier. [For the IPTW formula
the transformation is applied to $\mathbb{W}^{g,\ast}\left(  u\right)  .]$ It
is important to emphasize that a similiar transformation is not applied to
$X.$ Thus the conditioning event $\overline{L}\left(  m-1\right)
,\overline{A}_{\Delta}^{g}\left(  m-1\right)  =\overline{0},X>m$ transforms to
$\overline{L}\left(  m-1-\chi\right)  ,\overline{A}_{\Delta}^{g}\left(
m-1-\chi\right)  =\overline{0},X>m$.

\textbf{Proof of Theorem}: It suffices to show Eq. $\left(  78\right)  $ holds
when $L\left(  t\right)  $ and $A_{\Delta}^{g}\left(  t\right)  $ are replaced
by $L^{\dagger}\left(  t\right)  $ and $A_{\Delta}^{g,\dagger}\left(
t\right)  .$ By RC$^{g}$ $,$ $\left(  Y_{j}^{g},X_{j}^{g}\right)  \amalg
A_{\Delta}^{g}\left(  j\right)  |\overline{L}\left(  j\right)  ,\overline
{A}_{\Delta}^{g}\left(  j-1\right)  =\overline{0},\overline{U}\left(
j\right)  =0,X>j.$ Thus, $\left(  Y_{j}^{g},X_{j}^{g}\right)  \amalg
A_{\Delta}^{g}\left(  j\right)  |\overline{L}\left(  j\right)  ,\overline
{A}_{\Delta}^{g}\left(  j-1\right)  =\overline{0},\overline{U}\left(
j\right)  =\overline{0},X>j,X_{j}^{g}>j+\chi.$ By CD$^{g}$ and $\chi
>\varsigma,$ $\left(  Y_{j}^{g},X_{j}^{g}\right)  \amalg A_{\Delta}^{g}\left(
j\right)  |\overline{L}\left(  j\right)  ,\overline{A}_{\Delta}^{g}\left(
j-1\right)  =\overline{0},X_{j}^{g}>j+\chi,X>j.$

Thus $\left(  Y_{m-\chi}^{g},X_{m-\chi}^{g}\right)  \amalg A_{\Delta}%
^{g}\left(  m-\chi\right)  |\overline{L}\left(  m-\chi\right)  ,\overline
{A}_{\Delta}^{g}\left(  m-\chi-1\right)  =\overline{0},X>\left(
m-\chi\right)  ,X_{m-\chi}^{g}>m$ with $m\equiv\chi+j.$

Now the event $X>\left(  m-\chi\right)  $ is the event $X_{m-\chi}^{g}>\left(
m-\chi\right)  .$Further, by $\chi$ a g-specific minimal latent period we also
have the event $X_{m-\chi}^{g}>m$ is the event $X>m.$ Thus we have $\left(
Y_{m-\chi}^{g},X_{m-\chi}^{g}\right)  \amalg A_{\Delta}^{g}\left(
m-\chi\right)  |\overline{L}\left(  m-\chi\right)  ,\overline{A}_{\Delta}%
^{g}\left(  m-\chi\right)  =\overline{0}\left(  m-\chi\right)  ,X>m.$ Since,
given $\overline{A}_{\Delta}^{g}\left(  m-\chi\right)  =\overline{0}\left(
m-\chi-1\right)  ,$ we have $\left(  Y_{m-\chi}^{g},X_{m-\chi}^{g}\right)
=\left(  Y_{0}^{g},X_{0}^{g}\right)  ,$ we conclude $\left(  Y_{0}^{g}%
,X_{0}^{g}\right)  \amalg A_{\Delta}^{g}\left(  m-\chi\right)  |\overline
{L}\left(  m-\chi\right)  ,\overline{A}_{\Delta}^{g}\left(  m-\chi-1\right)
=\overline{0}\left(  m-\chi\right)  ,X>m,$ which is exactly Eq. $\left(
78\right)  $ with $L\left(  t\right)  $ and $A_{\Delta}^{g}\left(  t\right)  $
replaced by $L^{\dagger}\left(  t\right)  $ and $A_{\Delta}^{g,\dagger}\left(
t\right)  ,$ proving the theorem.

In contrast, under the conditions of the previous theorem, $E\left[  Y_{0}%
^{g}\right]  $ is not identified because Eq. $\left(  74\right)  ,$ in
contrast to Eq. $\left(  78\right)  ,$ fails to hold when $L\left(  t\right)
$ and $A_{\Delta}^{g}\left(  t\right)  $ are replaced by $L^{\dagger}\left(
t\right)  $ and $A_{\Delta}^{g,\dagger}\left(  t\right)  .$ Specifically, Eq.
$\left(  74\right)  $ can be written as the conjunction of Eq $\left(
78\right)  $,
\begin{align}
\left(  Y_{0}^{g},X_{0}^{g}\right)  \amalg A_{\Delta}^{g}\left(  m\right)
|\overline{L}\left(  m\right)  ,\overline{A}_{\Delta}^{g}\left(  m-1\right)
& =\overline{0},X,m>X>m-\chi+\varsigma\\
& and\nonumber\\
\left(  Y_{0}^{g},X_{0}^{g}\right)  \amalg A_{\Delta}^{g}\left(  m\right)
|\overline{L}\left(  m\right)  ,\overline{A}_{\Delta}^{g}\left(  m-1\right)
& =\overline{0},X,m-\chi+\varsigma>X
\end{align}

Below we show that under the conditions of the previous Theorem, Eq.$\left(
81\right)  $ holds but Eq $\left(  82\right)  $ does not when $L\left(
t\right)  $ and $A_{\Delta}^{g}\left(  t\right)  $ are replaced by
$L^{\dagger}\left(  t\right)  $ and $A_{\Delta}^{g,\dagger}\left(  t\right)
.$ To show (81) we modify slightly the proof of eq (78) as follows:

$\left(  Y_{j}^{g},X_{j}^{g}\right)  \amalg A_{\Delta}^{g}\left(  j\right)
|\overline{L}\left(  j\right)  ,\overline{A}_{\Delta}^{g}\left(  j-1\right)
=\overline{0},\overline{U}\left(  j\right)  =0,X>j$ (by RC$^{g})$

$\Rightarrow\left(  Y_{j}^{g},X_{j}^{g}\right)  \amalg A_{\Delta}^{g}\left(
j\right)  |\overline{L}\left(  j\right)  ,\overline{A}_{\Delta}^{g}\left(
j-1\right)  =\overline{0},\overline{U}\left(  j\right)  =\overline
{0},X>j,X_{j}^{g},j+\varsigma<X_{j}^{g}<j+\chi$

$\Rightarrow\left(  Y_{j}^{g},X_{j}^{g}\right)  \amalg A_{\Delta}^{g}\left(
j\right)  |\overline{L}\left(  j\right)  ,\overline{A}_{\Delta}^{g}\left(
j-1\right)  =\overline{0},X>j,X_{j}^{g},j+\varsigma<X_{j}^{g}<j+\chi$ (by
CD$^{g})$

$\Rightarrow\left(  Y_{m-\chi}^{g},X_{m-\chi}^{g}\right)  \amalg A_{\Delta
}^{g}\left(  m-\chi\right)  |\overline{L}\left(  m-\chi\right)  ,\overline
{A}_{\Delta}^{g}\left(  m-\chi-1\right)  =\overline{0},X>\left(
m-\chi\right)  ,X_{m-\chi}^{g},m>X_{m-\chi}^{g}>m-\chi+\varsigma.$

$\Rightarrow\left(  Y_{0}^{g},X_{0}^{g}\right)  \amalg A_{\Delta}^{g}\left(
m-\chi\right)  |\overline{L}\left(  m-\chi\right)  ,\overline{A}_{\Delta}%
^{g}\left(  m-\chi-1\right)  =\overline{0},X_{m-\chi}^{g}>\left(
m-\chi\right)  ,X_{m-\chi}^{g},m>X_{m-\chi}^{g}>m-\chi+\varsigma$

$\Rightarrow\left(  Y_{0}^{g},X_{0}^{g}\right)  \amalg A_{\Delta}^{g}\left(
m-\chi\right)  |\overline{L}\left(  m-\chi\right)  ,\overline{A}_{\Delta}%
^{g}\left(  m-\chi-1\right)  =\overline{0},X,m>X>m-\chi+\varsigma$ by the
$g-specific$ MLP assumption.

The proof of (82) fails because the event $\overline{L}\left(  j\right)
,\overline{A}_{\Delta}^{g}\left(  j-1\right)  =\overline{0},\overline
{U}\left(  j\right)  =\overline{0},X>j,X_{j}^{g},X_{j}^{g}<j+\varsigma$ is not
the same event as $\overline{L}\left(  j\right)  ,\overline{A}_{\Delta}%
^{g}\left(  j-1\right)  =\overline{0},\overline{U}\left(  j\right)
=\overline{0},X>j,X_{j}^{g},X_{j}^{g}<j+\varsigma$ under CD$^{g}$ because
$X_{j}^{g}<j+\varsigma$ does not imply $\overline{U}\left(  j\right)
=\overline{0}.$

\textbf{Proof that }$E\left[  Y_{0}^{\intercal}\right]  $\textbf{\ is
nonparametrically identified when a sufficiently long MLP exists }. In Section
3.3, we stated that $E\left[  Y_{0}^{\intercal}\right]  $ is nonparametrically
identified under the conditions of the previous theorem with the regime $g$ in
the theorem being the regime that always assigns exposure $zero.$ A proof follows.

Let $IN,A^{\intercal},\Xi^{\intercal},Y_{m}^{\intercal},X_{m}^{\intercal}$ be
as defined in Section 3.3 where we recall that because of the existence of the
MLP of length $\chi>\varsigma,$ all subjects with $\varsigma<X_{m}%
<m+\varsigma$ have $IN\left(  m\right)  =1.$ First in Eqs 78, 81, 82 we
replace $\left(  Y_{0}^{g},X_{0}^{g}\right)  $ by $\left(  Y_{0}^{\intercal
},X_{0}^{\intercal}\right)  ,A_{\Delta}^{g}\left(  m\right)  $ by
$A^{\intercal}\left(  m-\chi\right)  ,$ and redefine $L\left(  m\right)  $ as
$L\left(  m-\chi\right)  \ $with the component $\Xi\left(  m\right)  \ $of
$L\left(  m\right)  $ being replaced by $\Xi^{\intercal}\left(  m-\chi\right)
.$ Eq 82 now holds trivially because with probability one $m-\chi+\varsigma>X$
implies $IN\left(  m-\chi\right)  =1$ and thus $\Xi^{\intercal}\left(
m-\chi\right)  =0$ and $A^{\intercal}\left(  m-\chi\right)  =0.$ Furthermore
the proofs of Eqs 78 and 81 go through as above with only minor notational
changes. We therefore conclude that Eq 74 holds and thus that $E\left[
Y_{0}^{\intercal}\right]  $ is nonparametrically identified. The identifying
IPTW formula is explicitly given by
\begin{align*}
E\left[  Y_{0}^{\intercal}\right]   & =E\left[  YI\left\{  A^{\intercal
}\left(  K-\chi\right)  =\overline{0}\right\}  \mathbb{W}^{g,\ast}\right]  ,\\
\left\{  \mathbb{W}^{g,\ast}\right\}  ^{-1}  & =\\
&
{\displaystyle\prod\limits_{m=0}^{m=\left\lfloor X\right\rfloor }}
pr\left[  A^{\intercal}\left(  m-\chi\right)  =0|\overline{L}\left(
m-\chi\right)  ,\overline{A}^{\intercal}\left(  m-\chi-1\right)
=0,X>m\right]  \times\\
& \left\{
{\displaystyle\prod\limits_{m=\left\lfloor X+1\right\rfloor }^{K}}
pr\left[  A^{\intercal}\left(  m-\chi\right)  =0|\overline{L}\left(
m-\chi\right)  ,\overline{A}^{\intercal}\left(  m-\chi-1\right)  =0,X\right]
\right\}
\end{align*}

\bigskip

\section{Appendix 3: \textbf{Optimal Regime Models }:}

Suppose we now wish to estimate the regime $g_{opt}$ that maximizes $E\left[
Y_{0}^{g}\right]  $ over all regimes $g$ of the previous subsection. We will
do so by specifying an optimal regime stuctural nested mean model and
associated SNFTM.

To begin consider the dietary intervention $a\left(  k\right)  ,\underline
{g}_{opt,k+1}$ in which one follows there observed diet up to month $k,$ has a
$BMI$ increase of $a\left(  k\right)  $ over there maximum previous $BMI$ in
month $k,$ and follows the unknown optimal regime $g_{opt}$ thereafter. Let
$Y^{a\left(  k\right)  ,\underline{g}_{opt,k+1}},X^{a\left(  k\right)
,\underline{g}_{opt,k+1}}\ $be the associated counterfactuals. When $A\left(
k\right)  =a\left(  k\right)  ,$ write $\underline{g}_{opt,k+1}$ for the
regime $A\left(  k\right)  ,\underline{g}_{opt,k+1}.$ Note $X^{\underline
{g}_{opt,K+1}}=X.$

We will make the following assumptions:

\textbf{Optimal regime} \textbf{RC\ Assumption :} $A\left(  m\right)  $ is
statistically independent of $\left(  Y^{a\left(  m\right)  ,\underline
{g}_{opt,m+1}},X^{a\left(  m\right)  ,\underline{g}_{opt,m+1}}\right)  $ given
$\Xi\left(  m\right)  =1,\overline{L}(m),\overline{A}\left(  m-1\right)
\ $and $\overline{U}\left(  m\right)  =\overline{0}\left(  m\right)  $ for
each $a\left(  m\right)  \geq0$

\textbf{Optimal Regime CD Assumption}:
\begin{equation}
X^{\underline{g}_{opt,m}}>m+\zeta\Rightarrow\overline{U}\left(  m\right)
=\overline{0}\left(  m\right)  \
\end{equation}

We next recursively define the random variables $X^{a\left(  m\right)
,\underline{g}_{opt,m+1}}\left(  \psi\right)  \ $by the relationship that
$X^{\underline{g}_{opt,K+1}}\left(  \psi\right)  =X$ and, for $m=K,...,0.$
\begin{gather*}
X^{0\left(  m\right)  ,\underline{g}_{opt,m+1}}\left(  \psi\right)
=m+\exp\left\{  \omega\left(  a\left(  m\right)  ,\overline{A}\left(
m-1\right)  ,\overline{L}\left(  m\right)  ,\psi\right)  \right\}  \left(
X^{a\left(  m\right)  ,\underline{g}_{opt,m+1}}\left(  \psi\right)  -m\right)
\\
if\text{ }0<X^{a\left(  m\right)  ,\underline{g}_{opt,m+1}}\left(
\psi\right)  -m<1\\
X^{0\left(  m\right)  ,\underline{g}_{opt,m+1}}\left(  \psi\right)
=X^{a\left(  m\right)  ,\underline{g}_{opt,m+1}}\left(  \psi\right)  +\left\{
\exp\left\{  \omega\left(  a\left(  m\right)  ,\overline{A}\left(  m-1\right)
,\overline{L}\left(  m\right)  ,\psi\right)  \right\}  -1\right\} \\
if\text{ 1%
$<$%
}X^{a\left(  m\right)  ,\underline{g}_{opt,m+1}}\left(  \psi\right)  -m\\
X^{0\left(  m\right)  ,\underline{g}_{opt,m+1}}\left(  \psi\right)
=X^{a\left(  m\right)  ,\underline{g}_{opt,m+1}}\left(  \psi\right) \\
if\text{ }X^{a\left(  m\right)  ,\underline{g}_{opt,m+1}}\left(  \psi\right)
<m,
\end{gather*}

These equations recursively define $X^{a\left(  m\right)  ,\underline
{g}_{opt,m+1}}\left(  \psi\right)  $ in terms of the observed data, the regime
$\underline{g}_{opt,m+1}$ and the parameter vector $\psi$ as can be verified
by noting that these equations imply the following relationship between
$X^{a\left(  m\right)  ,\underline{g}_{opt,m+1}}\left(  \psi\right)  $ and
$X^{\underline{g}_{opt,m+1}}\left(  \psi\right)  .$
\begin{gather*}
X^{a\left(  m\right)  ,\underline{g}_{opt,m+1}}\left(  \psi\right)
=m+\frac{\exp\left\{  \omega\left(  A\left(  m\right)  ,\overline{A}\left(
m-1\right)  ,\overline{L}\left(  m\right)  ,\psi\right)  \right\}  }%
{\exp\left\{  \omega\left(  a\left(  m\right)  ,\overline{A}\left(
m-1\right)  ,\overline{L}\left(  m\right)  ,\psi\right)  \right\}  }\left(
X^{\underline{g}_{opt,m+1}}\left(  \psi\right)  -m\right) \\
\text{ if }0<X^{\underline{g}_{opt,m+1}}\left(  \psi\right)  -m<1,\text{
}0<X^{a\left(  m\right)  ,\underline{g}_{opt,m+1}}\left(  \psi\right)  <1
\end{gather*}

\begin{gather*}
X^{a\left(  m\right)  ,\underline{g}_{opt,m+1}}\left(  \psi\right) \\
=X^{\underline{g}_{opt,m+1}}\left(  \psi\right)  +\exp\left\{  \omega\left(
A\left(  m\right)  ,\overline{A}\left(  m-1\right)  ,\overline{L}\left(
m\right)  ,\psi\right)  \right\}  -\exp\left\{  \omega\left(  a\left(
m\right)  ,\overline{A}\left(  m-1\right)  ,\overline{L}\left(  m\right)
,\psi\right)  \right\} \\
if\text{ 1%
$<$%
}X^{a\left(  m\right)  ,\underline{g}_{opt,m+1}}\left(  \psi\right)  -m,\text{
1%
$<$%
}X^{\underline{g}_{opt,m+1}}\left(  \psi\right)  -m
\end{gather*}

\begin{gather*}
X^{a\left(  m\right)  ,\underline{g}_{opt,m+1}}\left(  \psi\right)
=m+\exp\left\{  \omega\left(  A\left(  m\right)  ,\overline{A}\left(
m-1\right)  ,\overline{L}\left(  m\right)  ,\psi\right)  \right\}  \left(
X_{m}^{\underline{g}_{opt,m+1}}\left(  \psi\right)  -m\right) \\
+1-\exp\left\{  \omega\left(  a\left(  m\right)  ,\overline{A}\left(
m-1\right)  ,\overline{L}\left(  m\right)  ,\psi\right)  \right\} \\
\text{ if }0<X_{m}^{\underline{g}_{opt,m+1}}\left(  \psi\right)  -m<1,\text{
}1<X^{a\left(  m\right)  ,\underline{g}_{opt,m+1}}\left(  \psi\right)  -m
\end{gather*}

\begin{gather*}
X^{a\left(  m\right)  ,\underline{g}_{opt,m+1}}\left(  \psi\right)
=m+\frac{\left\{  \left[  \exp\left\{  \omega\left(  A\left(  m\right)
,\overline{A}\left(  m-1\right)  ,\overline{L}\left(  m\right)  ,\psi\right)
\right\}  -1\right]  +\left(  X^{\underline{g}_{opt,m+1}}\left(  \psi\right)
-m\right)  \right\}  }{\exp\left\{  \omega\left(  a\left(  m\right)
,\overline{A}\left(  m-1\right)  ,\overline{L}\left(  m\right)  ,\psi\right)
\right\}  }\\
\text{ if }0<X^{a\left(  m\right)  ,\underline{g}_{opt,m+1}}-m\left(
\psi\right)  <1,\text{ }1<X^{\underline{g}_{opt,m+1}}-m\left(  \psi\right)
\end{gather*}
We next assume an optimal regime SNFTM given by
\begin{equation}
X^{a\left(  m\right)  ,\underline{g}_{opt,m+1}}\left(  \psi^{\ast}\right)
=X^{a\left(  m\right)  ,\underline{g}_{opt,m+1}}wp1
\end{equation}
for an unknown value $\psi^{\ast}$ of the vector $\psi.$

We also assume the optimal regime SNMM%

\begin{gather}
\gamma_{m}\left[  a\left(  k\right)  ,\overline{a}\left(  m-1\right)
,\overline{l}\left(  m\right)  ,x,\beta^{\ast}\right]  \equiv\\
E\left[  Y^{a\left(  k\right)  ,\underline{g}_{opt,k+1}}|\overline{L}%
_{m}=\overline{l}_{m},\overline{A}_{m}=\overline{a}_{m},X^{0\left(  k\right)
,\underline{g}_{opt,k+1}}\left(  \psi^{\ast}\right)  =x\right] \nonumber\\
-E\left[  Y^{0\left(  k\right)  ,\underline{g}_{opt,k+1}}|\overline{L}%
_{m}=\overline{l}_{m},\overline{A}_{m}=\overline{a}_{m},X^{0\left(  k\right)
,\underline{g}_{opt,k+1}}=x\right] \nonumber
\end{gather}
Above $\omega\left(  a\left(  t\right)  ,\overline{a}\left(  t-1\right)
,\overline{l}\left(  t\right)  ,\psi\right)  $ and $\gamma_{m}\left(  a\left(
m\right)  ,\overline{a}\left(  m-1\right)  ,\overline{l}\left(  m\right)
,\psi\right)  $ are known functions $\gamma_{m}\left[  a\left(  k\right)
,\overline{a}\left(  m-1\right)  ,\overline{l}\left(  m\right)  ,x,\beta
\right]  $ satisfying $\omega\left(  a\left(  t\right)  ,\overline{a}\left(
t-1\right)  ,\overline{l}\left(  t\right)  ,\psi\right)  =0$ if $a\left(
t\right)  =0$ or $\psi=0\ $and $\gamma_{m}\left(  a\left(  m\right)
,\overline{a}\left(  m-1\right)  ,\overline{l}\left(  m\right)  ,\beta\right)
=0$ if $a\left(  m\right)  =0$ or $\beta=0.$

Recall the optimal regime itself remains unknown. However we show below that
the following algorithm evaluated at the true $\left(  \beta^{\ast},\psi
^{\ast}\right)  $ would find the optimal regime $g_{opt}\ $under the following
additional condition, we we henceforth assume to hold.

\textbf{Additional} \textbf{Condition }: For each $\overline{a}\left(
m-1\right)  ,\overline{l}\left(  m\right)  ,x,\beta,m$ the function
$\gamma_{m}^{opt}\left[  a\left(  k\right)  ,\overline{a}\left(  m-1\right)
,\overline{l}\left(  m\right)  ,x,\beta\right]  $ is either everywhere zero or
is strictly concave in a$\left(  k\right)  $ on the support of $A\left(
k\right)  .$

\textbf{Optimal Regime Algorithm:}

Given any $\left(  \beta,\psi\right)  ,$ calculate $g_{opt,\left(  \beta
,\psi\right)  }=\left\{  g_{opt\left(  \beta,\psi\right)  ,m}\left[
\overline{a}\left(  m\right)  ,\overline{l}\left(  m\right)  )\right]
;m=K,...,0\right\}  $ as follows.

Calculate\ $X^{0\left(  K\right)  ,\underline{g}_{opt,K+1}}\left(
\psi\right)  .$ Define%

\begin{align*}
& g_{opt\left(  \beta,\psi\right)  ,K}^{\ast}\left[  \overline{A}\left(
K-1\right)  ,\overline{L}\left(  K\right)  )\right] \\
& =I\left(  X\leq K\right)  \arg\max_{a\left(  K\right)  }\left[  \gamma
_{K}\left\{  a(K),\overline{A}\left(  K-1\right)  ,\overline{L}\left(
K\right)  ,X,\beta\right\}  \right]  +I\left(  X>K\right)  \times\\
& \arg\max_{a\left(  K\right)  }E\left[  \gamma_{K}\left\{  a\left(  K\right)
,\overline{A}\left(  K-1\right)  ,\overline{L}\left(  K\right)  ,X^{0\left(
K\right)  ,\underline{g}_{opt,K+1}}\left(  \psi\right)  ,\beta\right\}
|\overline{A}\left(  K-1\right)  ,\overline{L}\left(  K\right)  ,X>k\right]
\end{align*}

Calculate $g_{opt\left(  \beta,\psi\right)  ,K}\left[  \overline{A}\left(
K-1\right)  ,\overline{L}\left(  K\right)  )\right]  =\min\left\{  A\left(
K\right)  ,g_{opt\left(  \beta,\psi\right)  ,K}^{\ast}\left[  \overline
{A}\left(  K-1\right)  ,\overline{L}\left(  K\right)  )\right]  \right\}  $

Calculate $X^{\underline{g}_{opt\left(  \beta,\psi\right)  ,K}}\left(
\psi\right)  =X^{g_{opt\left(  \beta,\psi\right)  ,K}\left[  \overline
{A}\left(  K\right)  ,\overline{L}\left(  K\right)  \right]  ,\underline
{g}_{opt,K+1}}\left(  \psi\right)  $ $.$

Recursively for $m=K-1,...,0,$ calculate

$X^{0\left(  m\right)  ,\underline{g}_{opt,m+1}\left(  \beta,\psi\right)
}\left(  \psi\right)  ,$%

\begin{align*}
& g_{opt\left(  \beta,\psi\right)  ,m}^{\ast}\left[  \overline{A}\left(
m-1\right)  ,\overline{L}\left(  m\right)  )\right] \\
& =I\left(  X\leq m\right)  \arg\max_{a\left(  m\right)  }E\left[  \gamma
_{m}\left\{  a\left(  m\right)  ,\overline{A}\left(  m-1\right)  ,\overline
{L}\left(  m\right)  ,X,\beta\right\}  \right]  +I\left(  X>m\right)  \times\\
& \arg\max_{a\left(  m\right)  }E\left[  \gamma_{m}\left\{  a\left(  m\right)
,\overline{A}\left(  m-1\right)  ,\overline{L}\left(  m\right)  ,X^{0\left(
m\right)  ,\underline{g}_{opt\left(  \beta,\psi\right)  ,m+1}}\left(
\psi\right)  ,\beta\right\}  |\overline{A}\left(  m-1\right)  ,\overline
{L}\left(  m\right)  ,X>m\right]
\end{align*}

Calculate $g_{opt\left(  \beta,\psi\right)  ,m}\left[  \overline{A}\left(
m\right)  ,\overline{L}\left(  m\right)  )\right]  =\min\left\{  A\left(
m\right)  ,g_{opt\left(  \beta,\psi\right)  ,m}^{\ast}\left[  \overline
{A}\left(  m-1\right)  ,\overline{L}\left(  m\right)  )\right]  \right\}  .$

Calculate $X^{\underline{g}_{opt\left(  \beta,\psi\right)  ,m}}\left(
\psi\right)  =X^{g_{opt\left(  \beta,\psi\right)  ,m}\left[  \overline
{A}\left(  m\right)  ,\overline{L}\left(  m\right)  )\right]  ,\underline
{g}_{opt,m+1}}\left(  \psi\right)  $

Note to carry out this algorithm we will need to be able to estimate\newline%
$E\left[  \gamma_{m}\left\{  a\left(  m\right)  ,\overline{A}\left(
m-1\right)  ,\overline{L}\left(  m\right)  ,X^{0\left(  m\right)
,\underline{g}_{opt\left(  \beta,\psi\right)  ,m+1}}\left(  \psi\right)
,\beta\right\}  |\overline{A}\left(  m-1\right)  ,\overline{L}\left(
m\right)  ,X>m\right]  $ for all possible values of $a\left(  m\right)  $ in
the support of $A\left(  m\right)  .$ One possibility is to specify and fit an
appropriate multivariate regression model with the possible values of
$a\left(  m\right)  $ indexing the multivariate outcomes at time $m.$

To understand why this is the correct algorithm, we first note that any regime
at $m$ can be a function of $X$ only if $X\leq m,$ so that $X$ is known by
$m.$ When $X>m,$ we must average over $X^{0\left(  m\right)  ,\underline
{g}_{opt\left(  \beta,\psi\right)  ,m+1}}\left(  \psi\right)  $ because
$X^{0\left(  m\right)  ,\underline{g}_{opt\left(  \beta,\psi\right)  ,m+1}%
}\left(  \psi\right)  $ is a function of $X.$ When $X>m,$ $X^{a\left(
m\right)  ,\underline{g}_{opt\left(  \beta,\psi\right)  ,m+1}}\left(
\psi\right)  $ will be the value of $X^{\underline{g}_{opt\left(  \beta
,\psi\right)  ,m}}\left(  \psi\right)  $ if the regime $g_{opt\left(
\beta,\psi\right)  }$ dictates the exposure $a\left(  m\right)  .$ The optimal
regime will choose the $a\left(  m\right)  $ that optimizes the contribution
to the utility at time $m.$ But the optimizing $a\left(  m\right)  $ depends
on the $a\left(  k\right)  $ chosen for the regime for $k>m.$ Thus we need to
use backward recursion to estimate the optimal regime.

To be more specific consider the subgroup of subjects with a history\newline%
$\left(  \overline{A}\left(  K-1\right)  ,\overline{L}\left(  K\right)
,X\right)  $ with $X<K$ so $X=X^{0\left(  K\right)  ,\underline{g}_{opt,K+1}%
}\left(  \psi^{\ast}\right)  \in\overline{L}\left(  K\right)  .$ Then
$a\left(  K\right)  =g_{opt\left(  \beta,\psi\right)  ,K}^{\ast}\left[
\overline{A}\left(  K-1\right)  ,\overline{L}\left(  K\right)  )\right]  $
that maximizes\newline$\gamma_{K}\left\{  a(K),\overline{A}\left(  K-1\right)
,\overline{L}\left(  K\right)  ,X,\beta^{\ast}\right\}  $ is clearly the
optimal treatment choice at $K.$ However, we are only considering regimes
(interventions) that do not force subjects to gain weight. We now argue that
for any subject with $A\left(  K\right)  $ less than $g_{opt\left(  \beta
,\psi\right)  ,K}^{\ast}\left[  \overline{A}\left(  K-1\right)  ,\overline
{L}\left(  K\right)  )\right]  ,$ the optimal decison is not to intervene at
all, so the subject receives his observed treatment $A\left(  K\right)  $. The
subject with $A\left(  K\right)  $ less than $g_{opt\left(  \beta,\psi\right)
,K}^{\ast}\left[  \overline{A}\left(  K-1\right)  ,\overline{L}\left(
K\right)  )\right]  $ could still have received any treatment between $0$ and
$A\left(  K\right)  .$ However among this set of treatments, the treatment
$A\left(  K\right)  $ is optimal by Condition a) above.

Next consider the subgroup of subjects with a history $\left(  \overline
{A}\left(  K-1\right)  ,\overline{L}\left(  K\right)  ,X\right)  $ with $X>K$
so $X\notin\overline{L}\left(  K\right)  \ $and $X^{0\left(  K\right)
,\underline{g}_{opt,K+1}}\left(  \psi^{\ast}\right)  >K.$ To find the optimal
treatment we average over $X^{0\left(  K\right)  ,\underline{g}_{opt,K+1}%
}\left(  \psi^{\ast}\right)  $. Since the average over $X^{0\left(  K\right)
,\underline{g}_{opt,K+1}}\left(  \psi^{\ast}\right)  $ of a function that is
concave in $a\left(  K\right)  $ for every possible value of $X^{0\left(
K\right)  ,\underline{g}_{opt,K+1}}\left(  \psi^{\ast}\right)  $ remains a
concave function of $a\left(  K\right)  ,$ we again take $g_{opt\left(
\beta,\psi\right)  ,K}\left[  \overline{A}\left(  K-1\right)  ,\overline
{L}\left(  K\right)  )\right]  =\min\left\{  A\left(  K\right)  ,g_{opt\left(
\beta,\psi\right)  ,K}^{\ast}\left[  \overline{A}\left(  K-1\right)
,\overline{L}\left(  K\right)  )\right]  \right\}  .$

That the same argument holds for each $m$ is a standard dynamic programming
argument as discussed in Robins (2004).

Since $\left(  \beta^{\ast},\psi^{\ast}\right)  $ are unknown we must estimate
them by g-estimation. Define
\begin{align*}
X_{m}^{g_{opt\left(  \beta,\psi\right)  }}\left(  \psi\right)   &
=X^{\underline{g}_{opt,m}}\left(  \psi\right) \\
\ Y_{m}^{g_{opt\left(  \beta,\psi\right)  }}\left(  \beta,\psi\right)   &
=Y-\sum_{m}^{K}\gamma_{m}\left[  A(m),\overline{A}\left(  m-1\right)
,\overline{L}\left(  m\right)  ,X_{m}^{g_{opt\left(  \beta,\psi\right)  }%
}\left(  \psi\right)  ,\beta\right]  .
\end{align*}

Note these equations are much more complex than the equations for $\psi$ using
a SNFTM and SNMM for a fixed $g$ in that $g_{opt}$ is now not known but
depends on the parameters $\left(  \beta,\psi\right)  $ through the above
algorithm for $g_{opt,\left(  \beta,\psi\right)  }.$ Thus we can no longer
estimate $\psi^{\ast}$ independently of $\beta^{\ast}$ since $X_{m}%
^{g_{opt\left(  \beta,\psi\right)  }}\left(  \psi\right)  $ is now a function
of $\beta$ as well as $\psi$ through its dependence on $g_{opt\left(
\beta,\psi\right)  }.$ Rather, we must solve both pairs of g-estimation
equations simultaneously.

Specifically, given the optimal regime RC\ and CD assumptions, to obtain CAN
estimators of the unknown parameters, we find jointly $\left(  \widetilde
{\beta},\widetilde{\psi}\right)  $ so that both the score test for the
covariate vector depending on $X_{m}^{g_{opt\left(  \beta,\psi\right)  }%
}\left(  \psi\right)  $ is precisely zero and the score test for the covariate
vector depending on $Y_{m}^{g_{opt\left(  \beta,\psi\right)  }}\left(
\beta,\psi\right)  $ is precisely zero (both tests restricted to subjects with
$X_{m}^{g_{opt\left(  \beta,\psi\right)  }}\left(  \psi\right)  >\zeta$ and
$\Xi\left(  m\right)  =1.)$ This turns out to be a very difficult
computational problem. Robins (4) describes a number of computational
simplifications, but they are beyond the scope of the current paper. Finally
we obtain $g_{opt\left(  \widetilde{\beta},\widetilde{\psi}\right)  }$ as our
estimate of the optimal regime $g_{opt\left(  \beta^{\ast},\psi^{\ast}\right)
}$ and n$^{-1}\sum_{i}^{n}Y_{0}^{g_{opt\left(  \widetilde{\beta}%
,\widetilde{\psi}\right)  }}\left[  \left(  \widetilde{\beta},\widetilde{\psi
}\right)  \right]  $ as our estimate of the expected utility $E\left[
Y_{0}^{g_{opt}}\right]  $ under the optimal regime.

Both estimation of $E\left[  Y_{0}^{g}\right]  $ for a known $g$ and of
$E\left[  Y_{0}^{g_{opt}}\right]  $ can be modified to allow for censoring at
end of follow-up at $K+1$ and for intactable unmeasured confounding in certain
subgroups using methods exactly analogous to the methods for estimation of
$E\left[  Y_{0}\right]  $.

\end{document}